\documentclass[useAMS,usenatbib]{mn2e}

\usepackage[english]{babel}
\usepackage{amsmath}
\usepackage{txfonts}
\usepackage{graphicx}
\usepackage{url}
\usepackage{color}

\usepackage{amssymb}

\newcommand{\pd}{\partial}
\newcommand{\ud}{\ensuremath{\mathrm{d}}}

\title[Dynamics of Neutrino-Driven Supernova Explosions]{The Dynamics of Neutrino-Driven Supernova Explosions after Shock Revival in 2D and 3D} 

\author[B.~M\"uller]{B.~M\"uller
\thanks{E-mail: bernhard.mueller@monash.edu}
\\ Monash Centre for Astrophysics, School of
Physics and Astronomy, 19 Rainforest Walk, Monash University, Victoria 3800,
Australia }

\begin{document}

\maketitle

\begin{abstract}
We study the growth of the explosion energy after shock revival in
neutrino-driven explosions in two and three dimensions (2D/3D) using
multi-group neutrino hydrodynamics simulations of an $11.2 M_\odot$
star.  The 3D model shows a faster and steadier growth of the
explosion energy and already shows signs of subsiding accretion after
one second. By contrast, the growth of the explosion energy in 2D is
unsteady, and accretion lasts for several seconds as confirmed by
additional long-time simulations of stars of similar
masses. Appreciable explosion energies can still be reached, albeit at
the expense of rather high neutron star masses. In 2D, the binding
energy at the gain radius is larger because the strong excitation of
downward-propagating $g$-modes removes energy from the freshly
accreted material in the downflows.  Consequently, the mass outflow
rate is considerably lower in 2D than in 3D. This is only partially
compensated by additional heating by outward-propagating acoustic
waves in 2D. Moreover, the mass outflow rate in 2D is reduced because
much of the neutrino energy deposition occurs in downflows or bubbles
confined by secondary shocks without driving outflows. Episodic
constriction of outflows and vertical mixing of colder shocked
material and hot, neutrino-heated ejecta due to Rayleigh-Taylor
instability further hamper the growth of the explosion energy in 2D.
Further simulations will be necessary to determine whether these
effects are generic over a wider range of supernova progenitors.
\end{abstract}

\begin{keywords}
supernovae: general -- hydrodynamics -- instabilities -- neutrinos --
radiative transfer
\end{keywords}

\section{Introduction}
\label{sec:intro}
After decades of research, the mechanism powering core-collapse
supernovae is still one of the outstanding problems in theoretical
astrophysics. The so-called delayed neutrino-driven mechanism
currently remains the most popular and best explored explanation (see
\citealt{janka_12,burrows_13} for current reviews) for ``ordinary''
supernova explosions not exceeding $\mathord{\sim} 10^{51} \ \mathrm{erg}$.  In its
modern guise, the neutrino-driven mechanism relies on additional
support for shock expansion in the form of hydrodynamic instabilities
like convection \citep{herant_92,burrows_92,herant_94,burrows_95,janka_96,mueller_97} and the standing accretion shock
instability (SASI, \citealp{blondin_03,laming_07,foglizzo_07,guilet_12}). Indeed, many successful
multi-dimensional simulations of neutrino-driven shock revival (mostly
in 2D, i.e.\ under the assumption of axisymmetry) have strengthened
our confidence in the neutrino-driven mechanism over the recent years
\citep{buras_06b,marek_09,yakunin_10,suwa_10,mueller_12a,mueller_12b,janka_12b,bruenn_13,suwa_13,pan_15}. However,
both the long-time evolution of the successful 2D models as well as
the advent of three-dimensional (3D) simulations have revealed two
serious challenges for the neutrino-driven paradigm: With the
exception of \citet{bruenn_13}, the majority of successful 2D
simulations tended to produce explosions that are probably
underenergetic. Moreover, the most ambitious self-consistent 3D
simulations with multi-group neutrino transport have so far either
failed to yield explosions \citep{hanke_13,tamborra_14a,tamborra_14b},
or, in the few successful cases, showed a considerable delay in shock
revival \citep{melson_15b,lentz_15} and significantly smaller
explosion energies \citep{takiwaki_14}. Only the explosion of a $9.6
M_\odot$ star recently simulated by \citet{melson_15a} is an exception
from this trend. Whether the neutrino-driven mechanism provides a
robust explanation for shock revival and can account for the observed
explosion properties of core-collapse supernovae may appear doubtful
in the light of these results.

There is now an emerging consensus about the reasons underlying the
more fundamental problem of missing
or delayed
explosions in 3D.  Both simple
light-bulb and leakage-based simulations
\citep{hanke_12,couch_12a,couch_12b,couch_14a,couch_15} as well as
models with multi-group transport \citep{hanke_13,takiwaki_14} find an
artificial accumulation of turbulent kinetic energy on large spatial
scales in 2D due to the action of the inverse turbulent cascade
\citep{kraichnan_76}. Effectively, the forward cascade in 3D
provides for more efficient damping/dissipation of the
turbulent motions in the post-shock region, resulting in smaller
turbulent kinetic energies. Since the turbulent kinetic energy is
directly related to the Reynolds stresses (i.e.\ loosely speaking, the
``turbulent pressure'') that have been identified as the primary agent
fostering shock expansion in multi-D
\citep{burrows_95,murphy_12,couch_15,mueller_15}, this affects
the critical neutrino luminosity $L_\mathrm{crit}$
\citep{burrows_93,murphy_08b} required to power an explosive runaway.
However, even though the higher explosion threshold in 3D has emerged as a
systematic effect, it nonetheless remains a small effect: Current
light-bulb models \citep{hanke_12,couch_12b,dolence_13} invariably show
a similar reduction of $20\ldots 30\%$ in critical luminosity in 2D/3D
compared to 1D, with \citet{dolence_13} still finding a slightly lower
explosion threshold in 3D. Likewise, neutrino hydrodynamics simulation
\citep{hanke_13,takiwaki_14} show very similar heating conditions in
2D and 3D prior to shock revival (and even transient phases with
better heating conditions in 3D in \citealt{hanke_13}), although the
small differences between 2D and 3D eventually thwart shock revival in
the 3D models of \citet{hanke_13} and \citet{tamborra_14a,tamborra_14b}.  Even
though the adverse effects of the forward cascade in 3D may still be
underestimated by the relatively crude grid resolution that current
models can afford \citep{hanke_12,couch_12b,abdikamalov_14}, it thus
emerges that 3D models of neutrino-driven supernova explosions are
very close to the explosion threshold. Consequently, relatively small
refinements in the simulations and the initial models may be
sufficient to obtain explosions, e.g.\ \emph{moderate} rotation
\citep{nakamura_14}, magnetic fields \citep{obergaulinger_14},
asphericities in the progenitor core
\citep{couch_13,couch_15b,mueller_15}, or modifications to
the standard neutrino interaction rates \citep{melson_15b}. The emergence of the
spiral mode of the SASI could even allow strongly SASI-dominated
models to explode earlier in 3D than in 2D \citep{fernandez_15}.  

\begin{figure}
\includegraphics[width=\linewidth]{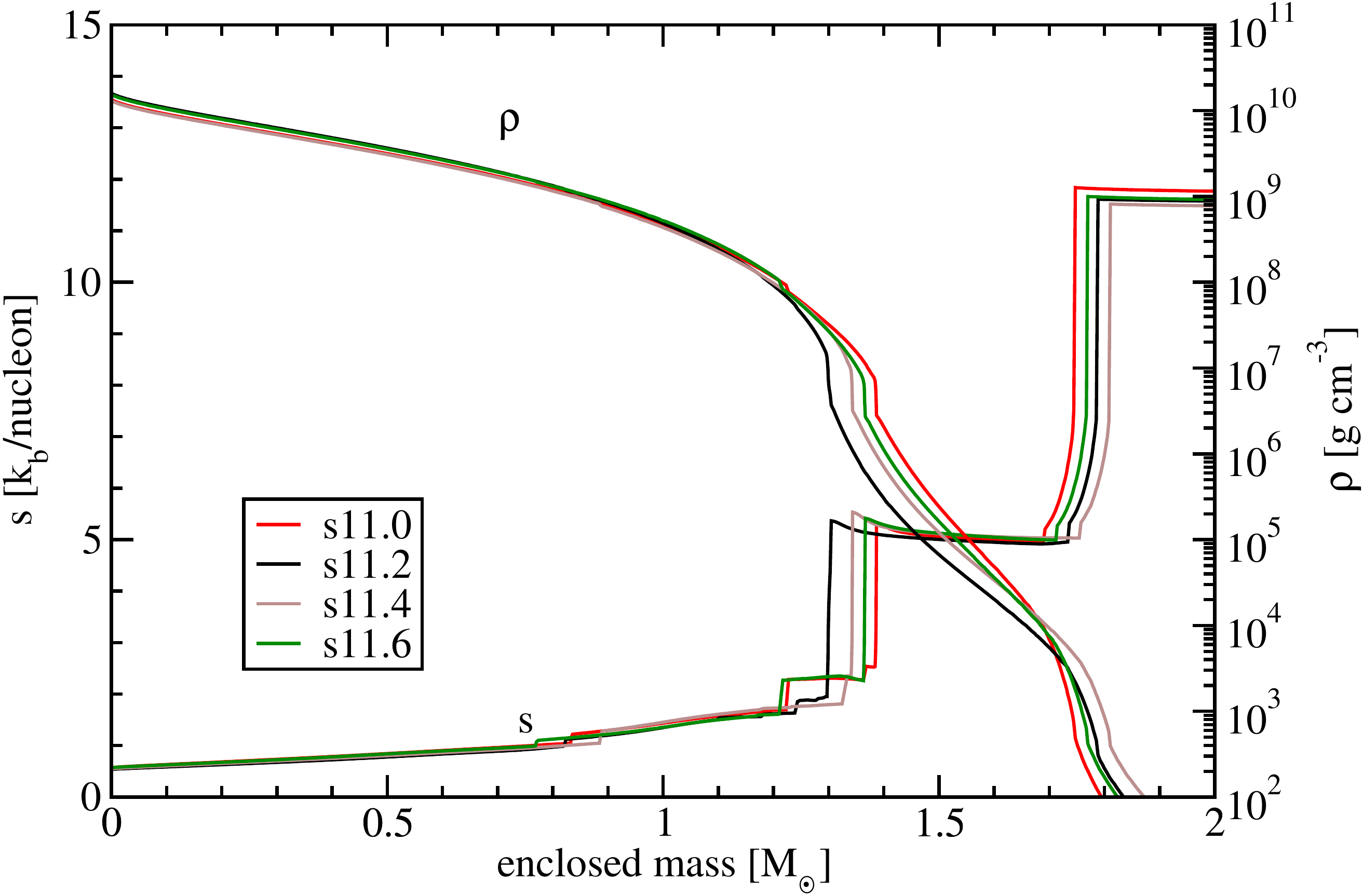}
\caption{Density $\rho$ and entropy $s$ for the four progenitors s11.0
  (red), s11.2 (black), s11.4 (light brown), and s11.6 (green) as a
  function of enclosed mass.
\label{fig:comp_progs}}
\end{figure}

By contrast, the problem of underenergetic neutrino-driven explosions
in 2D has so far gone without a convincing theoretical explanation,
and fewer suggestions have been made to remedy it, although it may be
more serious in the sense that it affects even models with
\emph{successful} shock revival: While recent analyses of
well-observed supernovae
\citep{tanaka_09,utrobin_11,poznanski_13,chugai_14,pejcha_15} suggest
a broader range of explosion energies for type II-P supernovae between
$\mathord{\sim} (0.1 \ldots 2 )\times 10^{51} \ \mathrm{erg}$ instead
of a single ``canonical'' value of $10^{51} \ \mathrm{erg}$, there is
arguably still a discrepancy, since almost none of the successful 2D
and 3D models with multi-group neutrino transport show explosion
energies considerably above $10^{50} \ \mathrm{erg}$, e.g.\ the final
values at the end of the simulations are a few $10^{49}
\ \mathrm{erg}$ in \citet{marek_09} for progenitors with $11.2
M_\odot$ and $15 M_\odot$, $\lesssim 10^{50} \ \mathrm{erg}$ for a
$13 M_\odot$ progenitor in \citet{suwa_10}, and $4\times 10^{49}
\ \mathrm{erg}$ ($11.2 M_\odot$ progenitor), $1.3 \times 10^{50}
\ \mathrm{erg}$ ($15 M_\odot$), and $1.3 \times 10^{50}
\ \mathrm{erg}$ ($27 M_\odot$) in \citet{janka_12b}.  Moreover, these
estimates are not corrected for the ``overburden'', i.e.\ the binding
energy of the layers outside the shock, so that it remains unclear
whether the explosions become energetic enough to shed the envelope at
all.  At first glance, the low explosion energies may simply be due to
the fact that the simulations typically terminate before the explosion
energy reaches its asymptotic value.  While it can be argued that the
final explosion energies may yet be higher by a factor of several because
continued accretion sustains strong neutrino emission after shock
revival that can power outflows from the gain region for
$\mathord{\gtrsim} 0.5 \ \mathrm{s}$, this assumption creates several
other problems: Sustained accretion over $\mathord{\gtrsim} 0.5
\ \mathrm{s}$ might shift the resulting remnant mass distribution well
above the average birth mass of neutron stars $M_\mathrm{grav}\approx
1.35 M_\odot$ \citep{schwab_10} inferred from observations (which may,
however, suffer from a selection bias).  Only the 2D models of
\citet{bruenn_14} form an exception from this trend; they obtain
explosion energies in the range of $(3.4\ldots 8.8) \times 10^{50}
\ \mathrm{erg}$ for progenitors between $12 M_\odot$ and $25 M_\odot$
as well as reasonable Nickel masses.

While this is encouraging, explosion energies above $10^{51}
\ \mathrm{erg}$ still remain unexplained, and the problem of
underenergetic supernova explosion will arguably still linger as long
as the discrepancies between the different simulation codes are not
resolved.  Interestingly, \citet{melson_15a} recently reported that 3D
effects, while apparently detrimental for shock revival in more
massive progenitors, actually boost the explosion energy in a $9.6
M_\odot$ progenitor by $\mathord{\sim} 10\%$ by reducing the infall
velocity in the accretion downflow and hence the cooling below the
gain layer. Here, we further investigate their intriguing premise that
3D effects, while hurtful for shock revival, can prove beneficial in
other situations and contribute to solving the problem of low
explosion energies.

In this paper, we present a successful 3D multi-group neutrino
hydrodynamics simulation of an $11.2 M_\odot$ progenitor with the
\textsc{CoCoNuT-FMT} code \citep{mueller_10,mueller_15} as further
evidence that 3D turbulence plays a positive role \emph{after} the
onset of the explosion. By comparing the dynamics and energetics of
the 3D explosion model to several long-time simulations of 2D
progenitors in the mass range between $11 M_\odot$ and $11.6 M_\odot$,
we demonstrate how 3D effects can lead to a faster, more robust growth
of the explosion energy provided that shock revival can be achieved.
So far, the long-time effects of the dimensionality during the first
hundreds of milliseconds to seconds after shock revival have received
less attention than the role of the third dimension in shock revival:
Successful first-principle models are still scarce and cannot be
evolved for a sufficiently long time yet to address this phase in
detail, while light-bulb-based studies of supernova energetics in 2D
and 3D \citep{handy_14} cannot adequately account for the feedback of
the subsiding accretion onto the neutrino heating and do not show the
drawn-out long-time accretion characteristic of first-principle
models. In this paper, we show that this phase deserves a closer look.

Our paper is structured as follows: In Section~\ref{sec:setup}, we
review the numerical methods for hydrodynamics and neutrino transport
used in our version of the \textsc{CoCoNuT} code, including a brief
sketch of the modifications used in the 3D version.  In
Section~\ref{sec:overview}, we first give a descriptive overview of the
differences in shock propagation and explosion properties between the
2D and 3D models. Since the question of shock revival in 3D
is not the objective of our current study, we only provide
a brief, cautionary assessment of shock revival in the 3D
model against the backdrop of recent first-principle
models in Section~\ref{sec:assessment}. In Section~\ref{sec:analysis},
we then analyze the physical effects underlying
these differences by combining a separate analysis of the outflows and
downflows in the spirit of \citet{melson_15a}.  Finally, we
summarize our results and discuss their implications in
Section~\ref{sec:conclusions}.

\begin{figure}
\includegraphics[width=\linewidth]{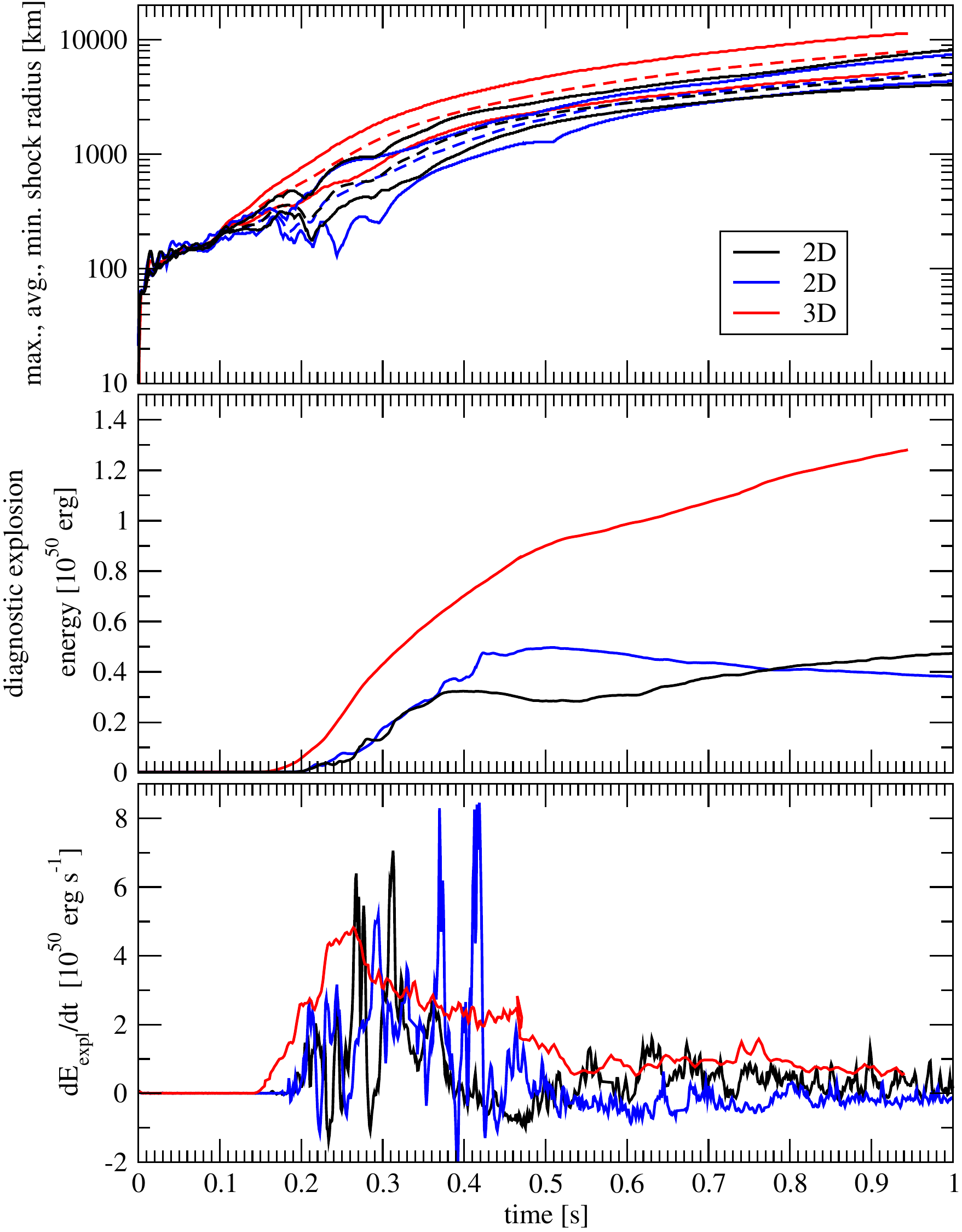}
\caption{Shock propagation and diagnostic explosion energy $E_\mathrm{expl}$  for
the $11.2 M_\odot$ progenitor in 2D and 3D:
The top panel shows the maximum, minimum (solid), and average
(dashed) shock radius for model s11.2\_2Da (black),
s11.2\_2Db (blue), and s11.2\_3D (red). The middle and bottom
panel show the diagnostic explosion energy $E_\mathrm{expl}$
and its time derivative $\ud E_\mathrm{expl}/\ud t$.
\label{fig:shock_2d_3d}
}
\end{figure}

\begin{figure}
\includegraphics[width=\linewidth]{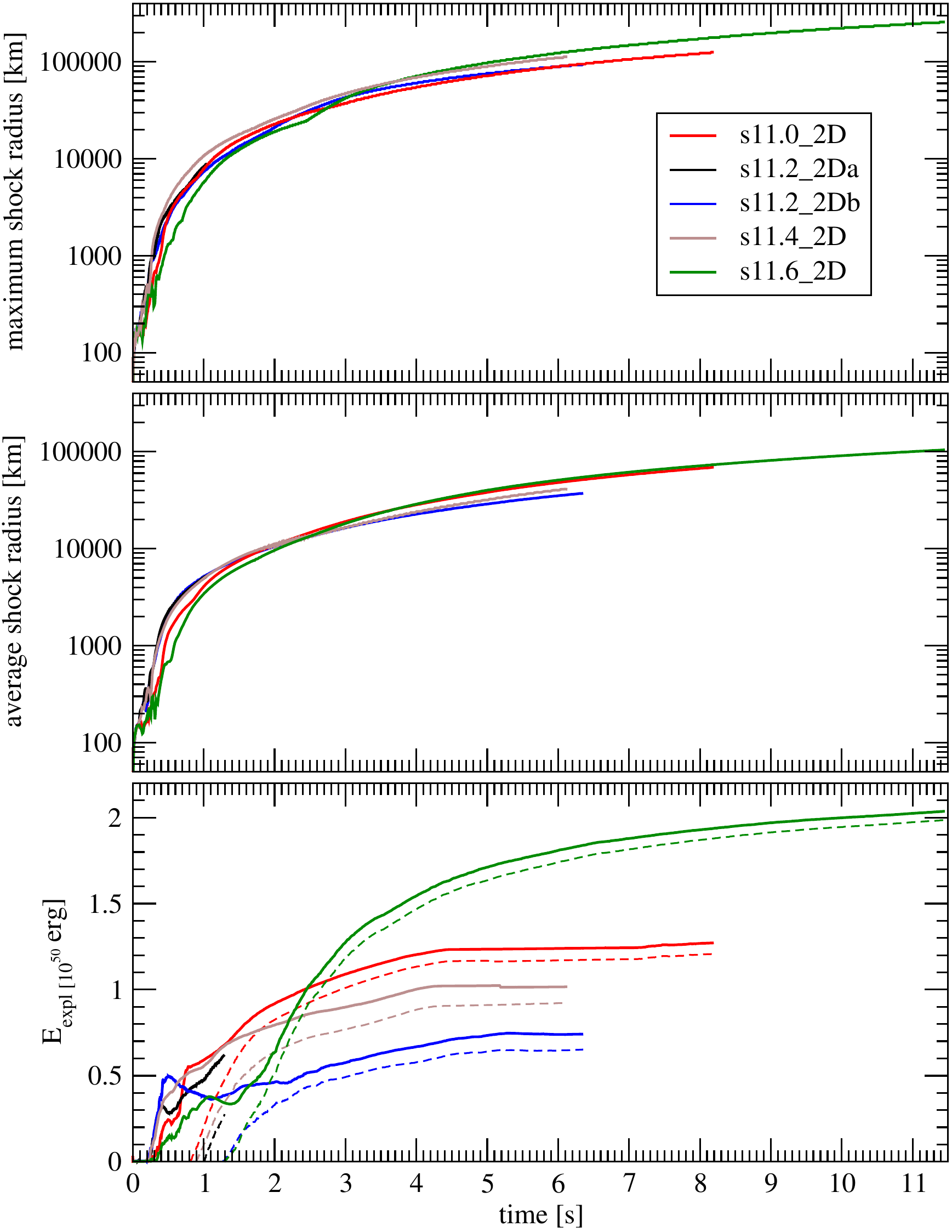}
\caption{Shock propagation and diagnostic explosion energy
  $E_\mathrm{expl}$ for the different progenitors in 2D: The top and
  middle panel show the maximum and average shock radius,
  respectively. The bottom panel shows the diagnostics
    explosion energy $E_\mathrm{expl}$ as a function of time (solid
    lines). Dashed lines show the time evolution
      $E_\mathrm{expl}-E_\mathrm{ov}$, i.e.\ the diagnostic energy
      corrected for the binding energy (overburden) $E_\mathrm{ov}$ of
      the material ahead of the shock. Red, black, blue, light brown,
    and green curves are used for models s11.0\_2D, s11.2\_2Da,
    s11.2\_2Db, s11.4\_2D, s11.6\_2D.
\label{fig:shock_prog}
}
\end{figure}

\section{Numerical Methods and Model Setup}
\label{sec:setup}

\subsection{Progenitor Models}
We simulate the collapse and post-bounce phase of the (non-rotating)
$11.2 M_\odot$ solar-metallicity progenitor s11.2 of
\citet{woosley_02} in 2D and 3D.  In order to gauge the effect of
stochastic model variations, we perform two different 2D simulations
(s11.2\_2Da and s11.2\_2Db\footnote{Scattering on nuclei (including
  $\alpha$-particles) was switched off after bounce for model
  s11.2\_2Db, which leads to minor changes in early shock propagation,
  but is inconsequential for the long-time evolution. 
    Since the energy exchange due to recoil in neutrino-nucleon
    scattering was taken to be proportional to the total scattering
    opacity instead of the neutrino-nucleon scattering opacity only
    (see Equation~A31 in \citealt{mueller_15}) in model s11.2\_2Da and
    all other models, the runs with neutrino scattering on nuclei
    overestimate pre-heating from heavy flavor neutrinos outside the
    shock during the early post-bounce phase (whereas nuclei actually
    receive negligible recoil in scattering reactions), which leads to
    a slight reduction of the heavy flavor neutrino luminosity and a
    slightly slower contraction of the proto-neutron star compared to
    s11.2\_2Db.})  for this progenitor, and we also conduct
simulations for three other solar metallcity progenitors of
\citet{woosley_02} with similar zero-age main sequence (ZAMS) mass and
density structure ($11 M_\odot$, $11.4 M_\odot$, $11.6 M_\odot$). We
randomly perturb the radial velocity $v_r$ at the beginning of
collapse using a perturbation amplitude $\delta v_r/v_r=10^{-5}$.

In Figure~\ref{fig:comp_progs}, we show density and entropy profiles for the
four progenitors simulated in the different 2D and 3D runs.
Despite small differences in the location of the interfaces between
the different shells, the models are very similar in terms of
structure with a pronounced density jump between the silicon
shell and the convective shell above the active oxygen burning
zone. As we shall see, the 2D models of the different progenitors
are qualitatively very similar in terms of explosion dynamics
and energetics and should thus illustrate the generic behaviour
of supernova explosions originating from progenitors in this mass range.

\begin{figure*}
\includegraphics[width=0.32 \textwidth]{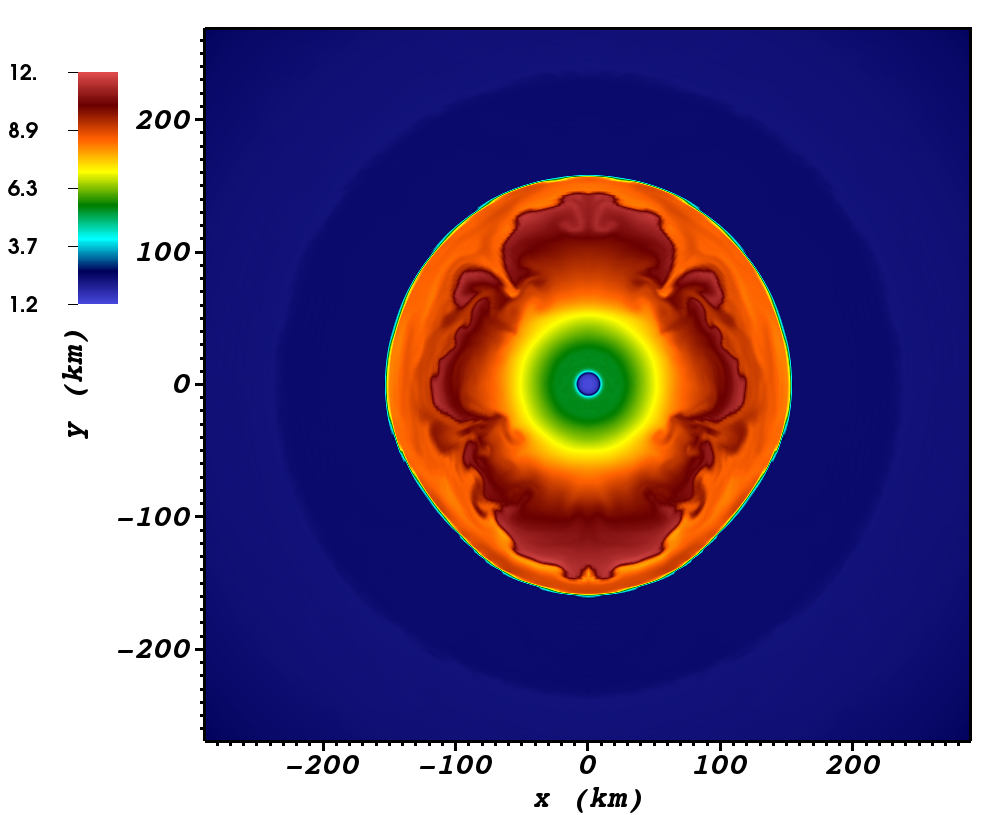}
\includegraphics[width=0.32 \textwidth]{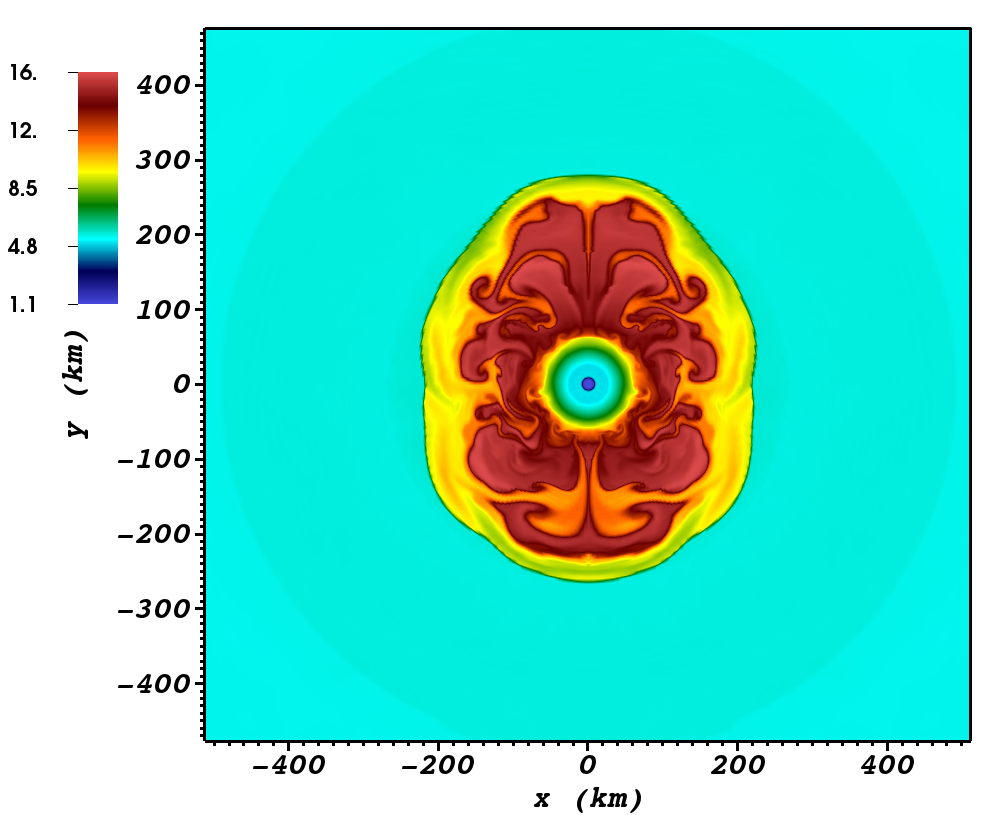}
\includegraphics[width=0.32 \textwidth]{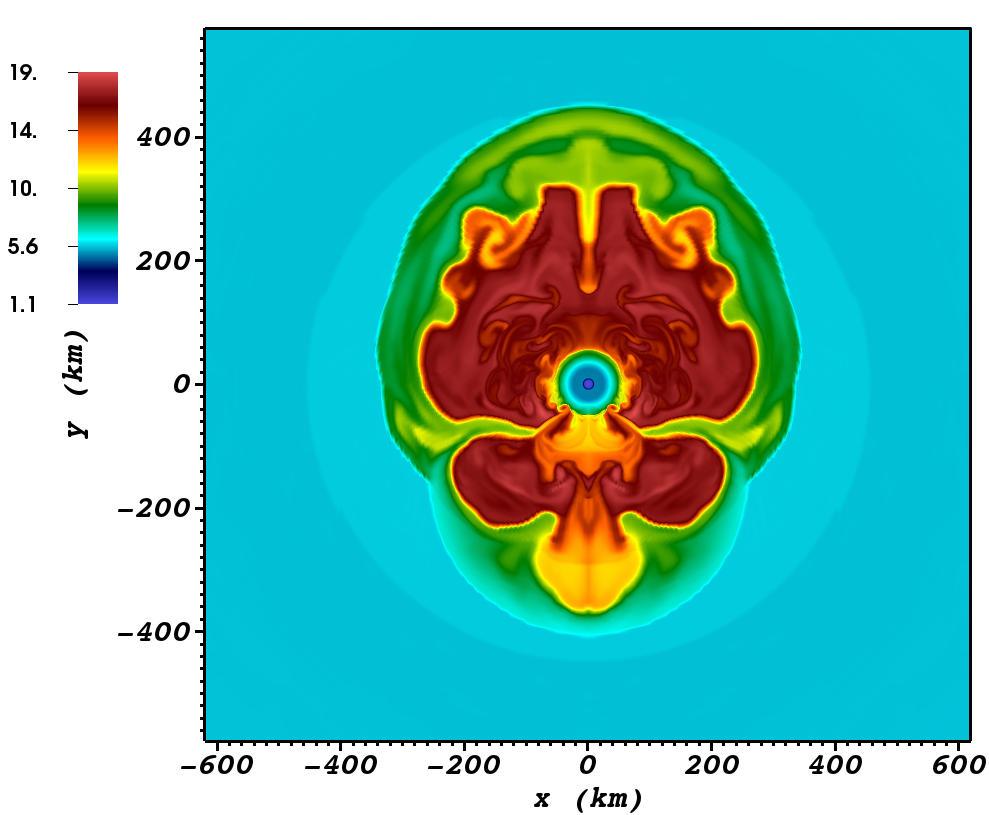}
\\
\includegraphics[width=0.32 \textwidth]{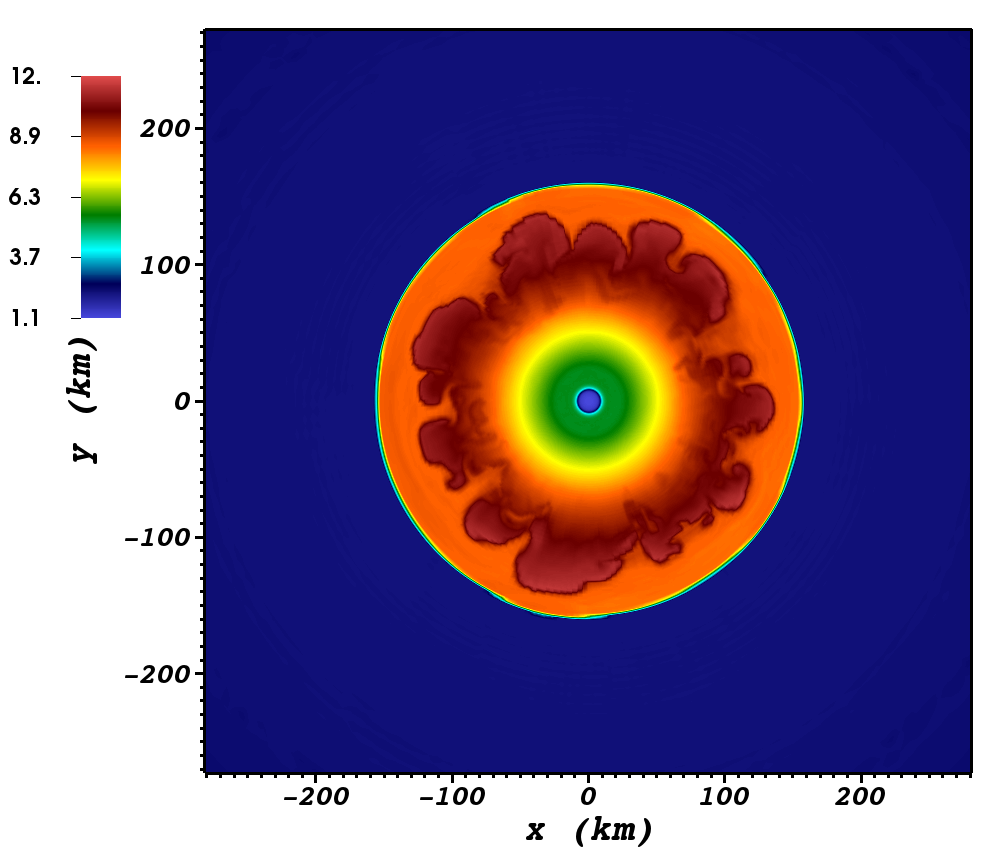}
\includegraphics[width=0.32 \textwidth]{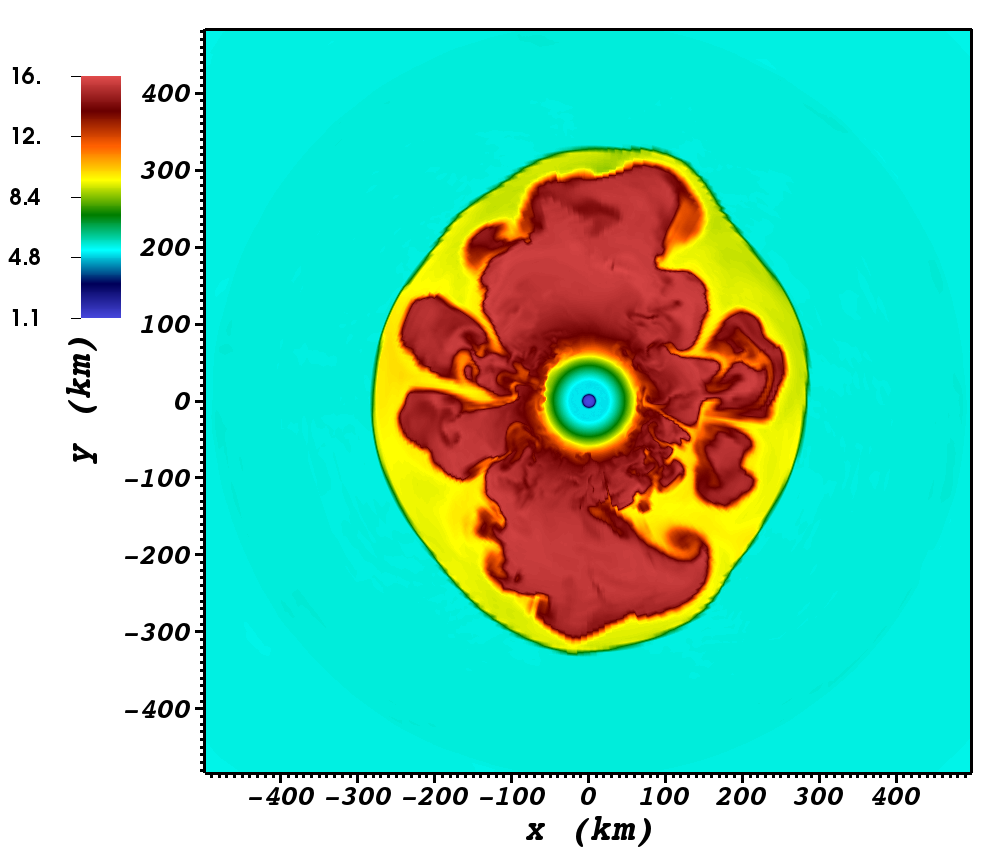}
\includegraphics[width=0.32 \textwidth]{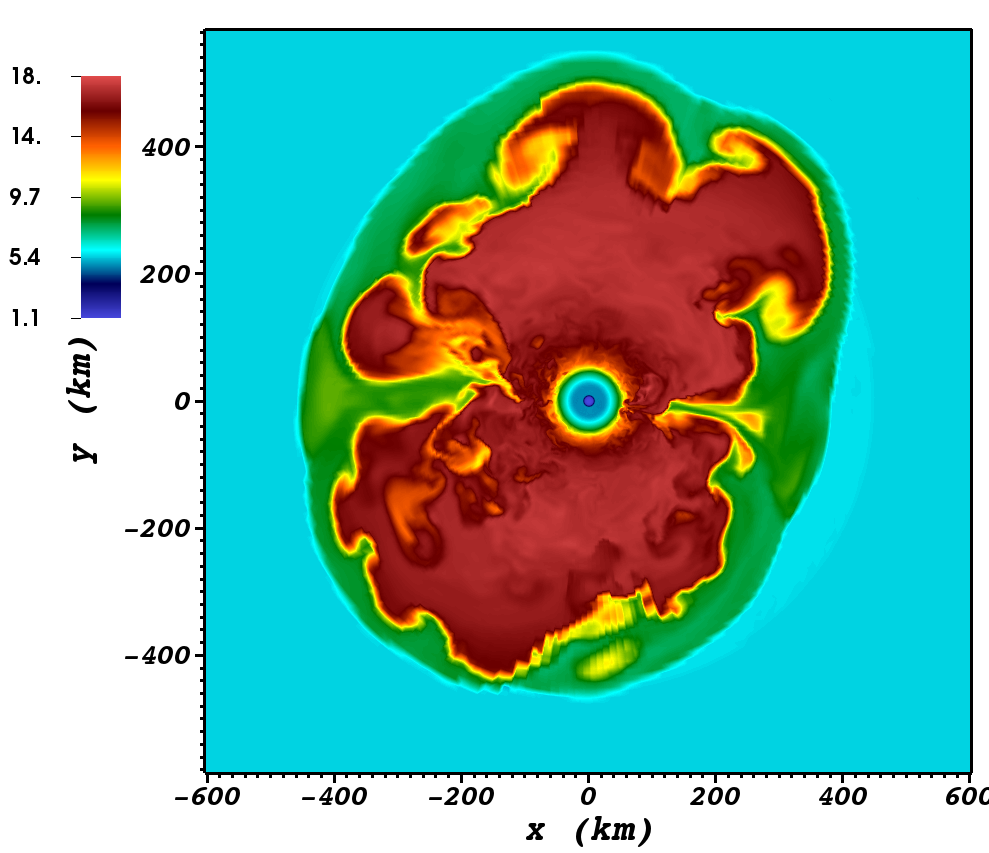}
\caption{Specific entropy for model s11.2\_2Da (top row) and
model s11.2\_3D (in a slice almost perpendicular to the equatorial plane,
bottom row)
at post-bounce times of
$80 \ \mathrm{ms}$,
$140 \ \mathrm{ms}$,
and
$181 \ \mathrm{ms}$ (left to right).
 Note that a different color scale for
the entropy is used
for  each of these snapshots.
\label{fig:snap_2d_3d_1}
}
\end{figure*}

\begin{figure*}
\includegraphics[width=0.32 \textwidth]{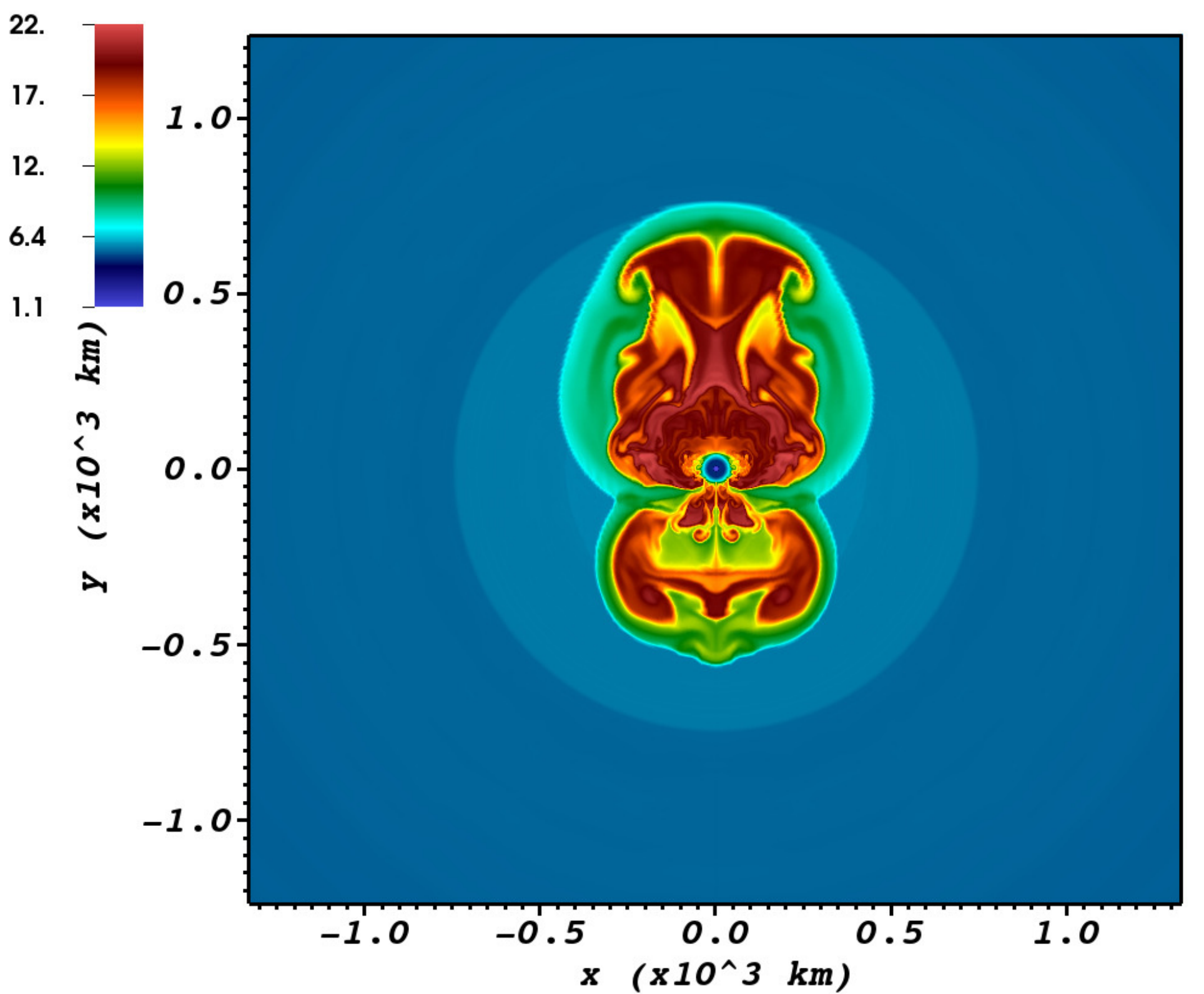}
\includegraphics[width=0.32 \textwidth]{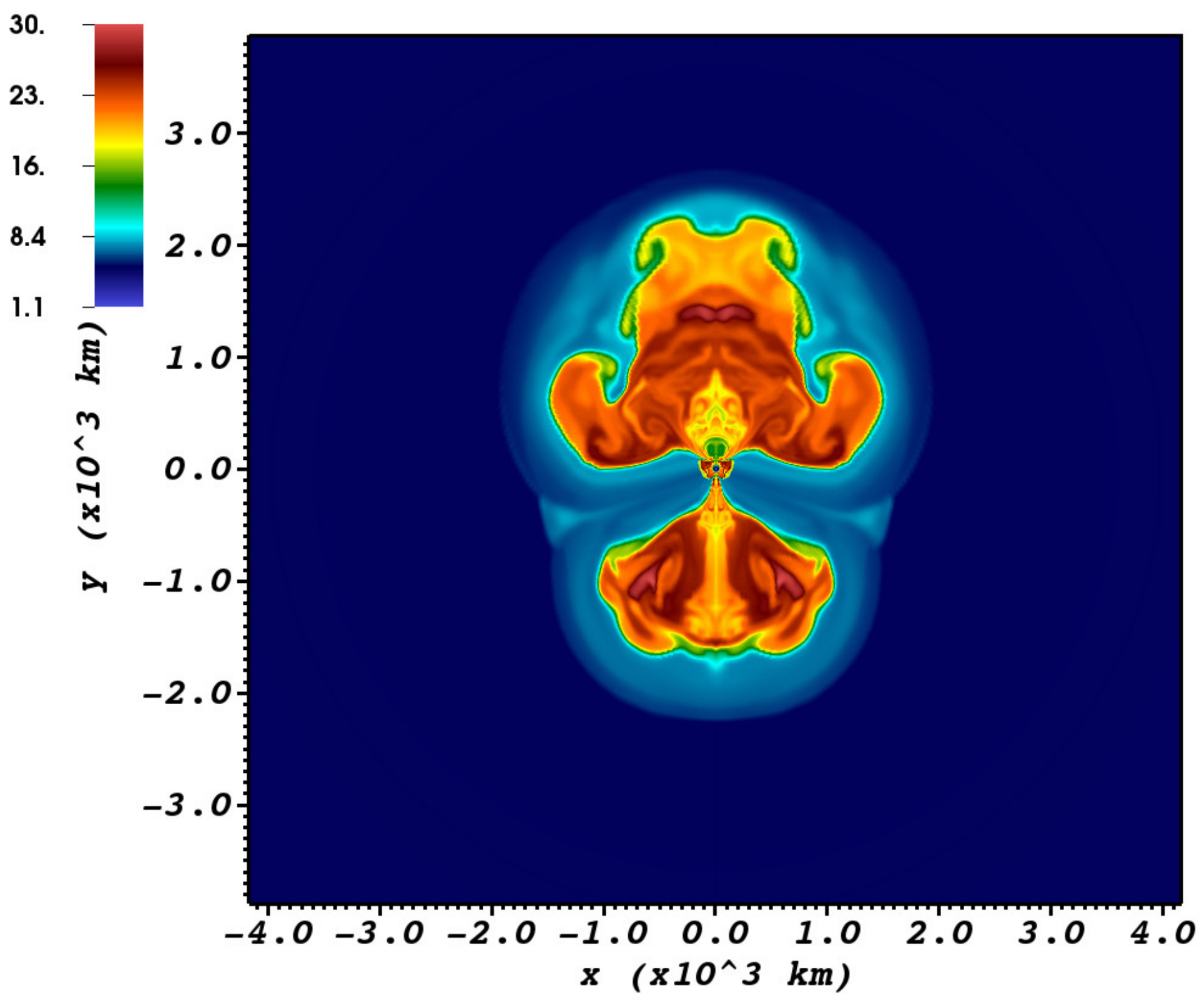}
\includegraphics[width=0.32 \textwidth]{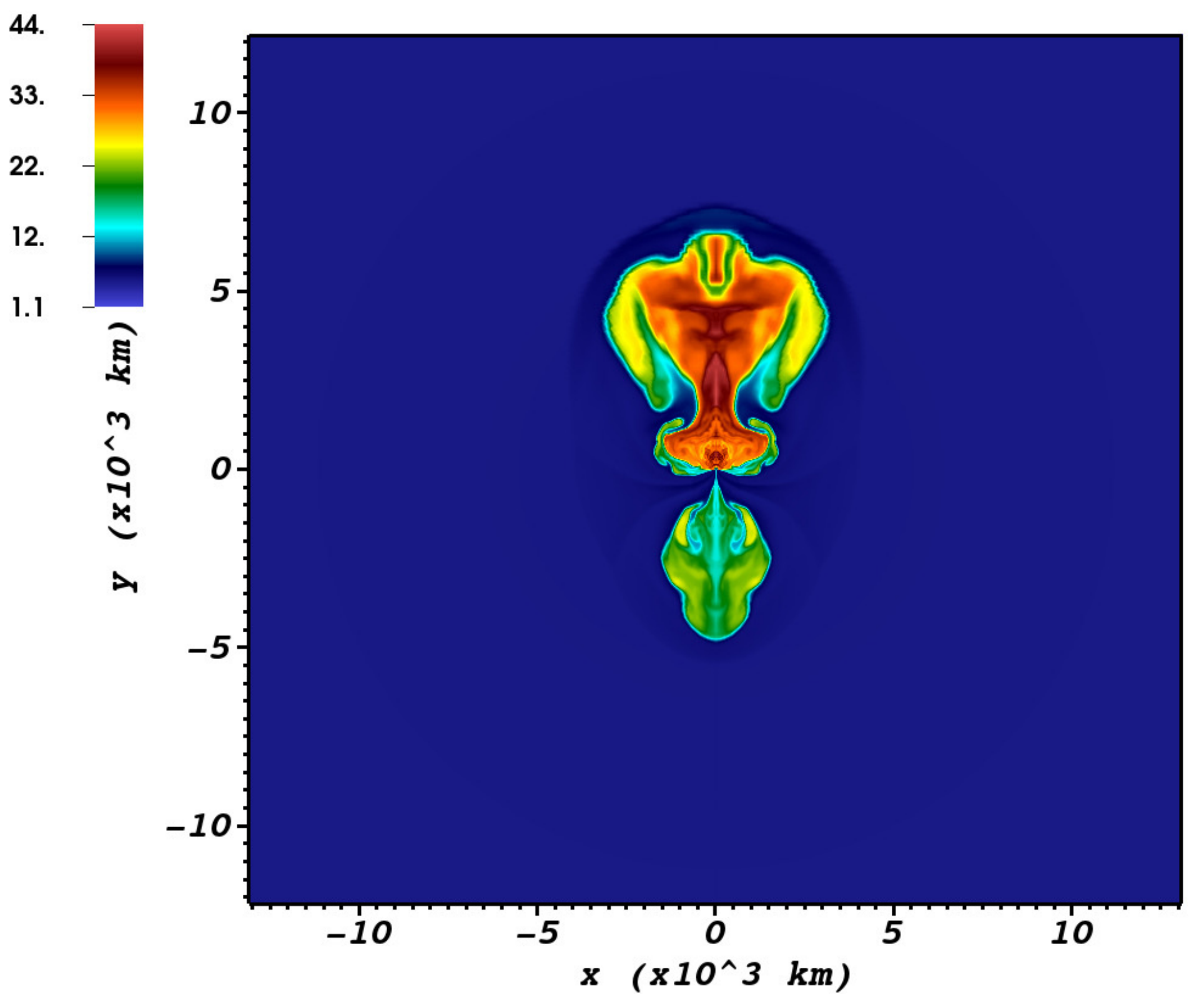}
\\
\includegraphics[width=0.32 \textwidth]{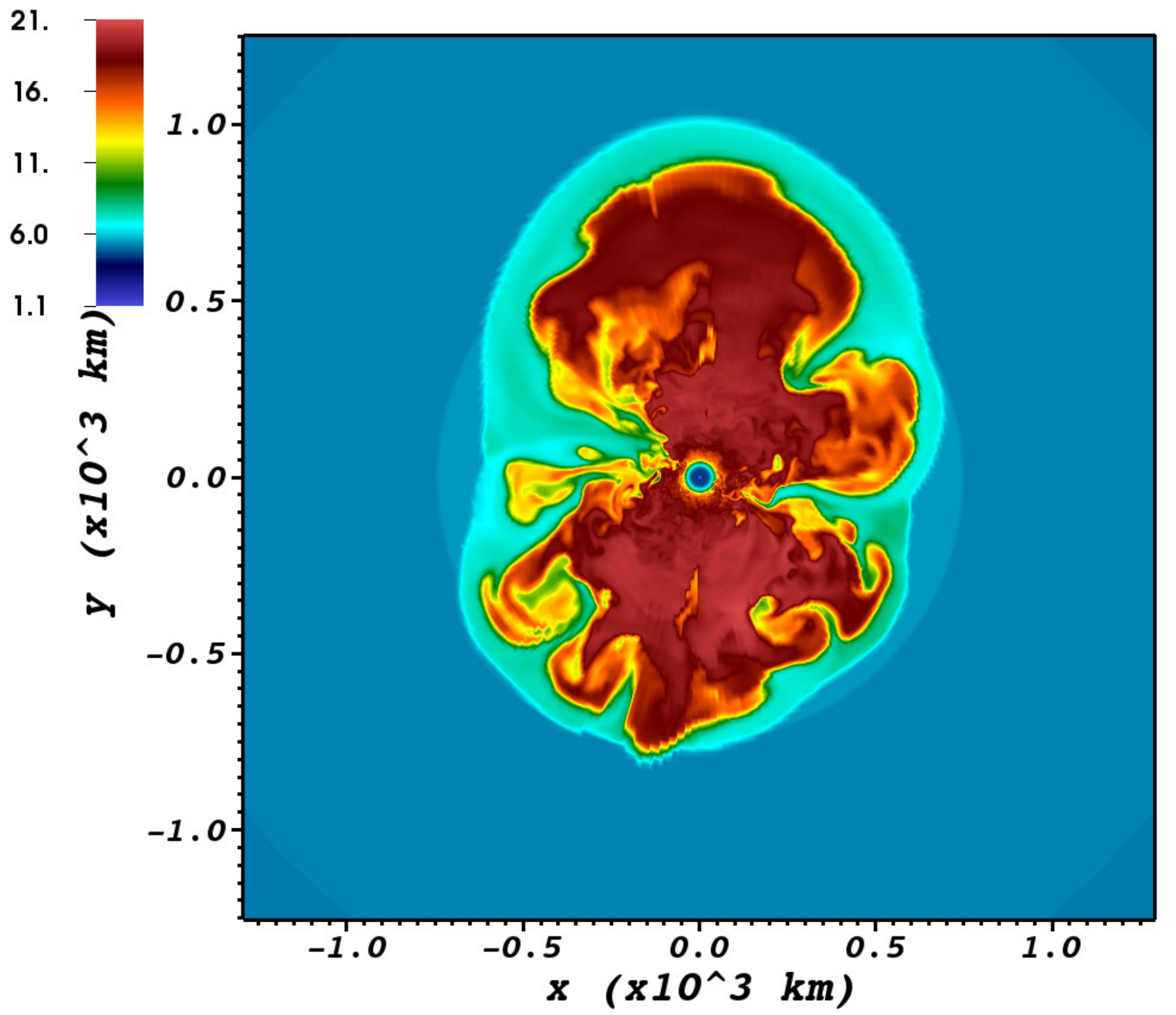}
\includegraphics[width=0.32 \textwidth]{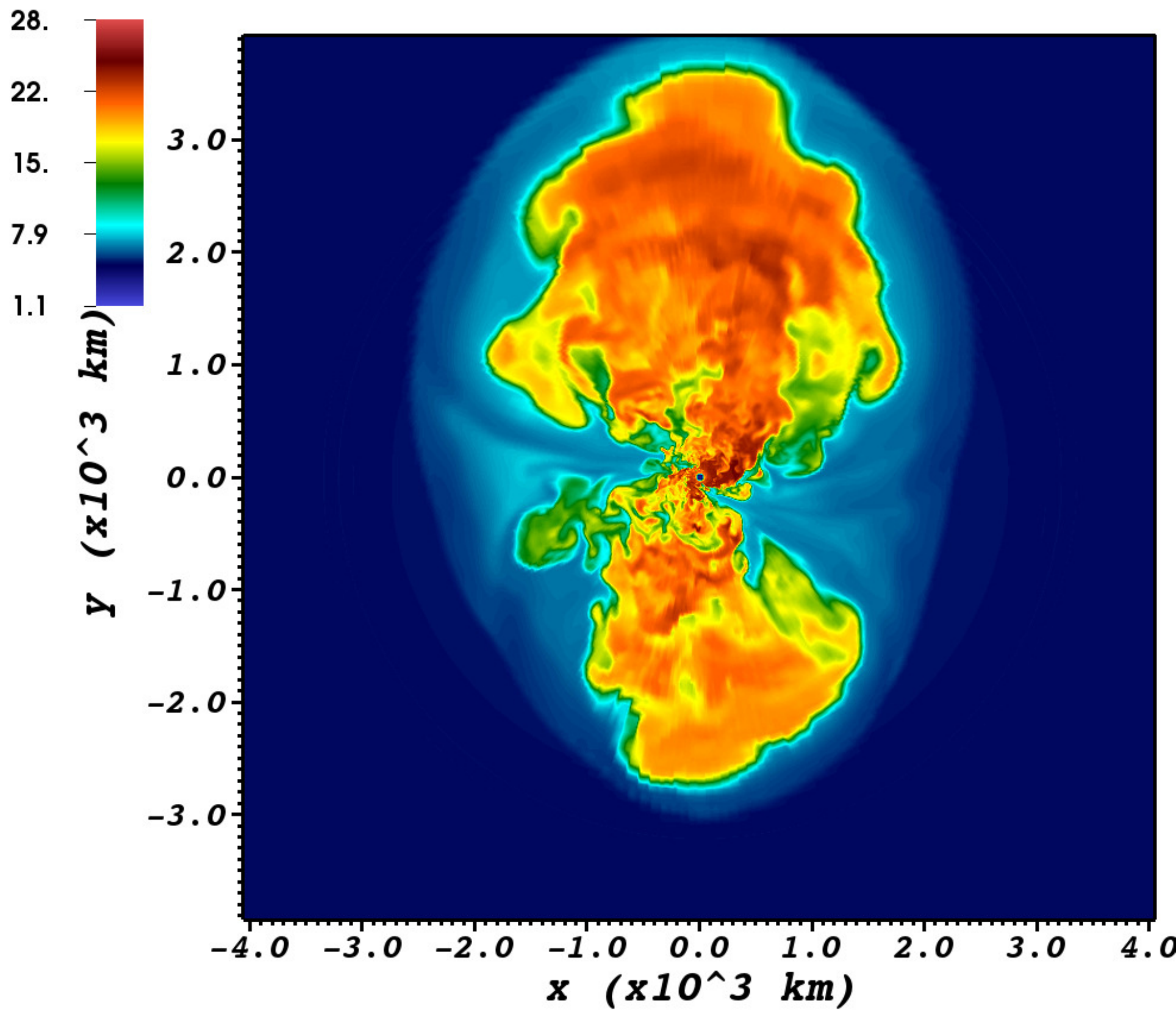}
\includegraphics[width=0.32 \textwidth]{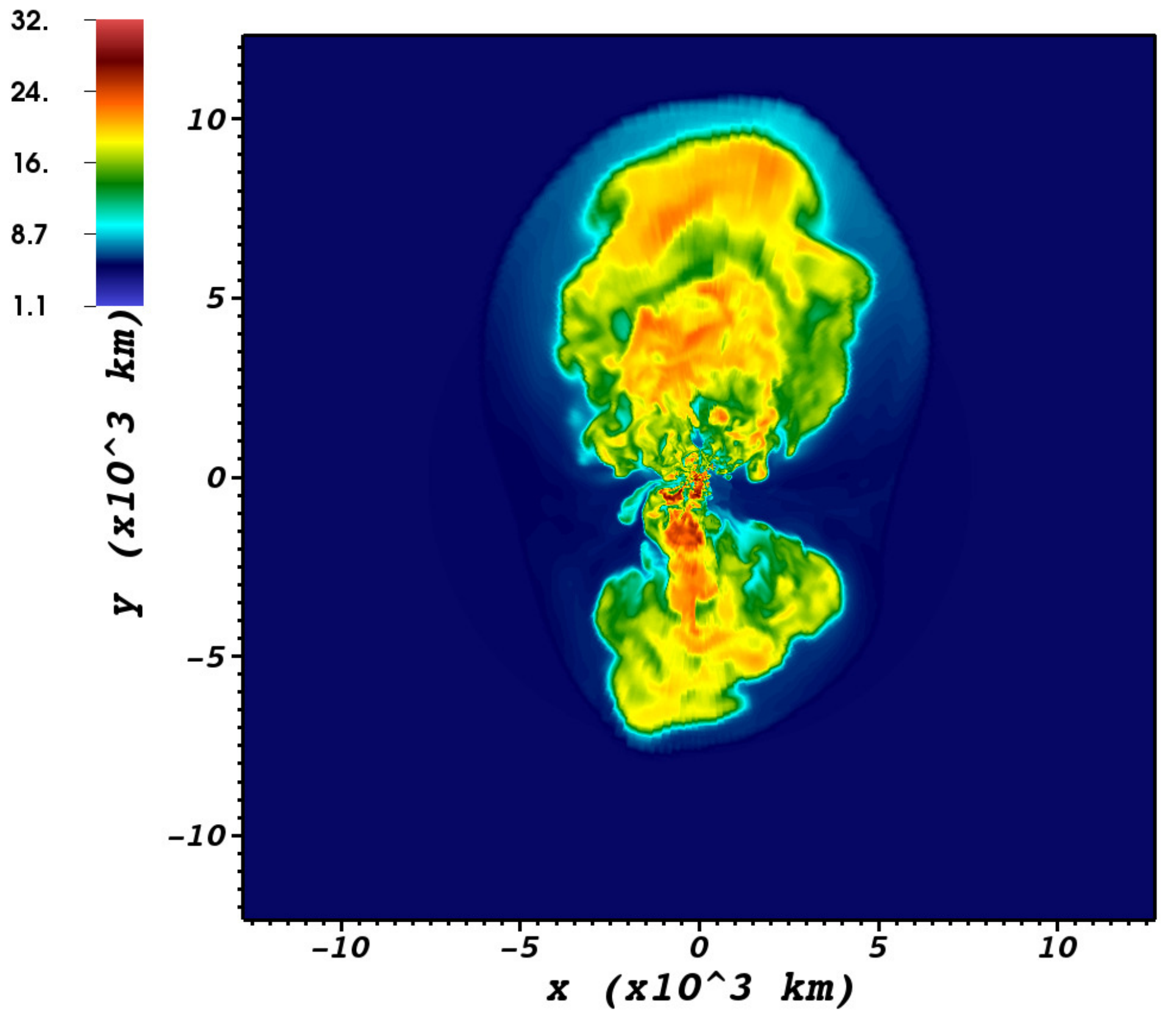}
\caption{Specific entropy for model s11.2\_2Da (top row) and
model s11.2\_3D (in a slice almost perpendicular to the equatorial plane,
bottom row)
at post-bounce times of
$241 \ \mathrm{ms}$,
$471 \ \mathrm{ms}$,
and
$944 \ \mathrm{ms}$ (left to right). Note that a different color scale for
the entropy is used
for  each of these snapshots.
\label{fig:snap_2d_3d_2}
}
\end{figure*}

\begin{figure*}
\includegraphics[width=0.48\linewidth]{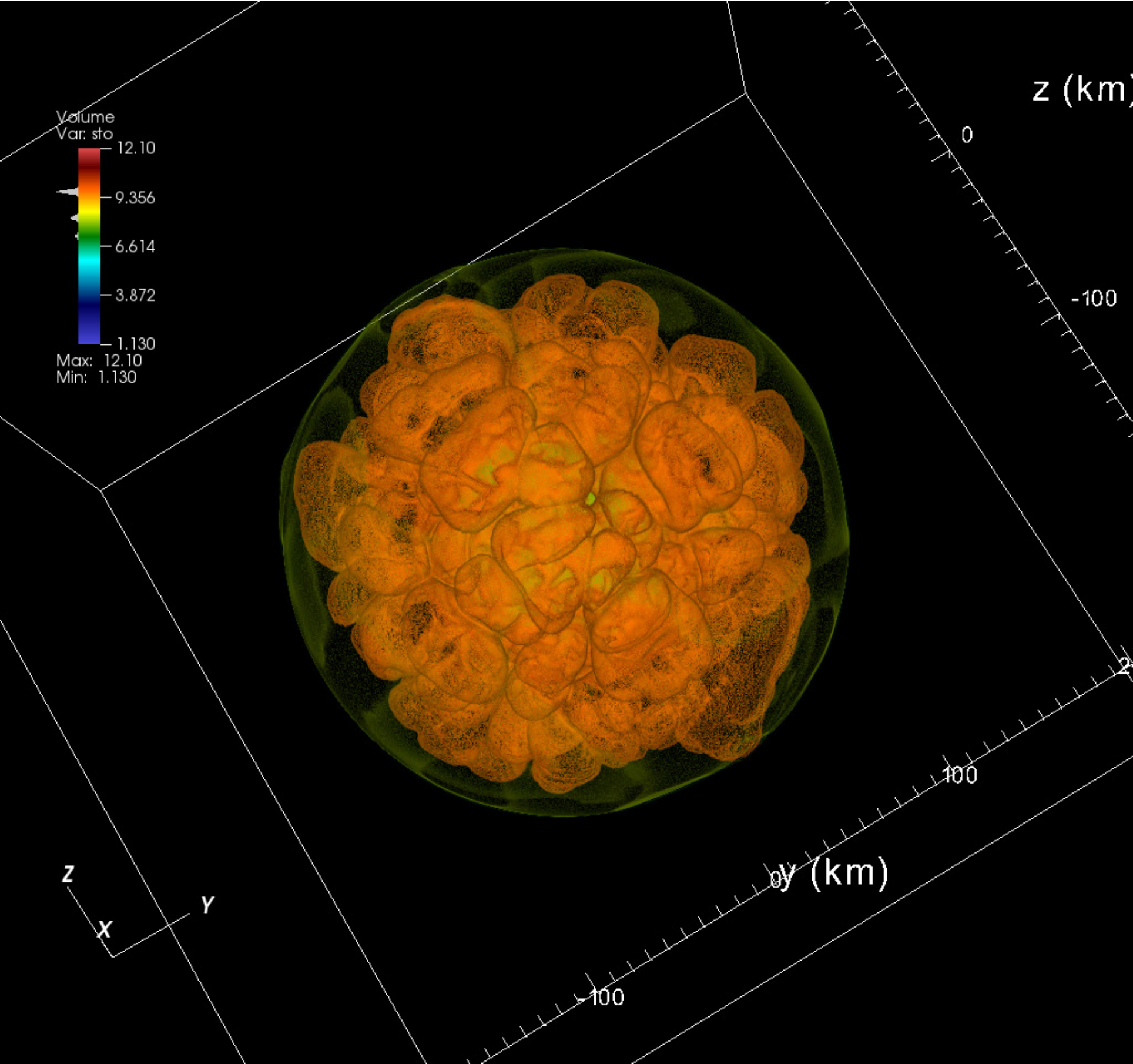}
\includegraphics[width=0.48\linewidth]{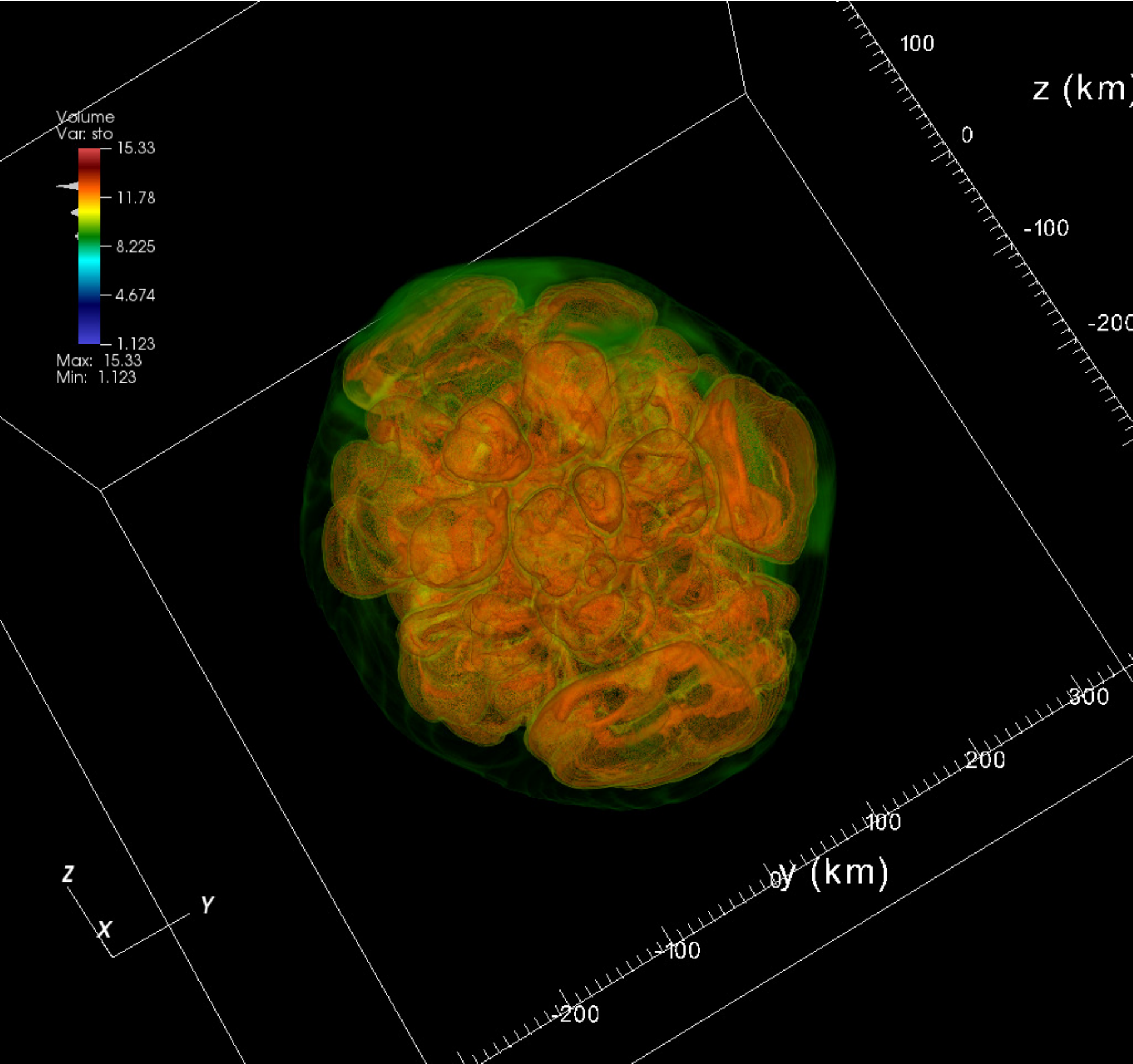}\\
\vspace{0.06cm}
\includegraphics[width=0.48\linewidth]{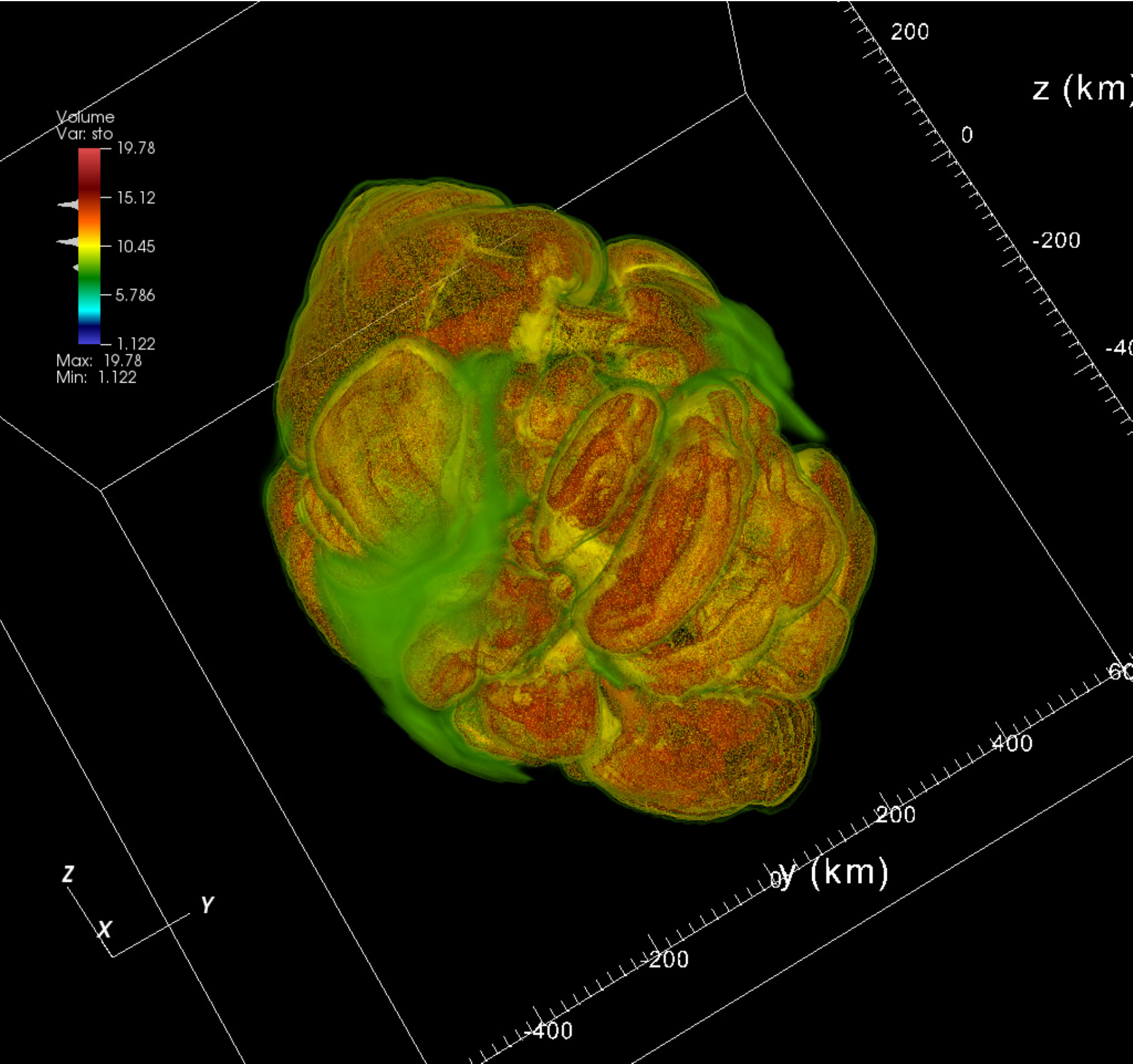}
\includegraphics[width=0.48\linewidth]{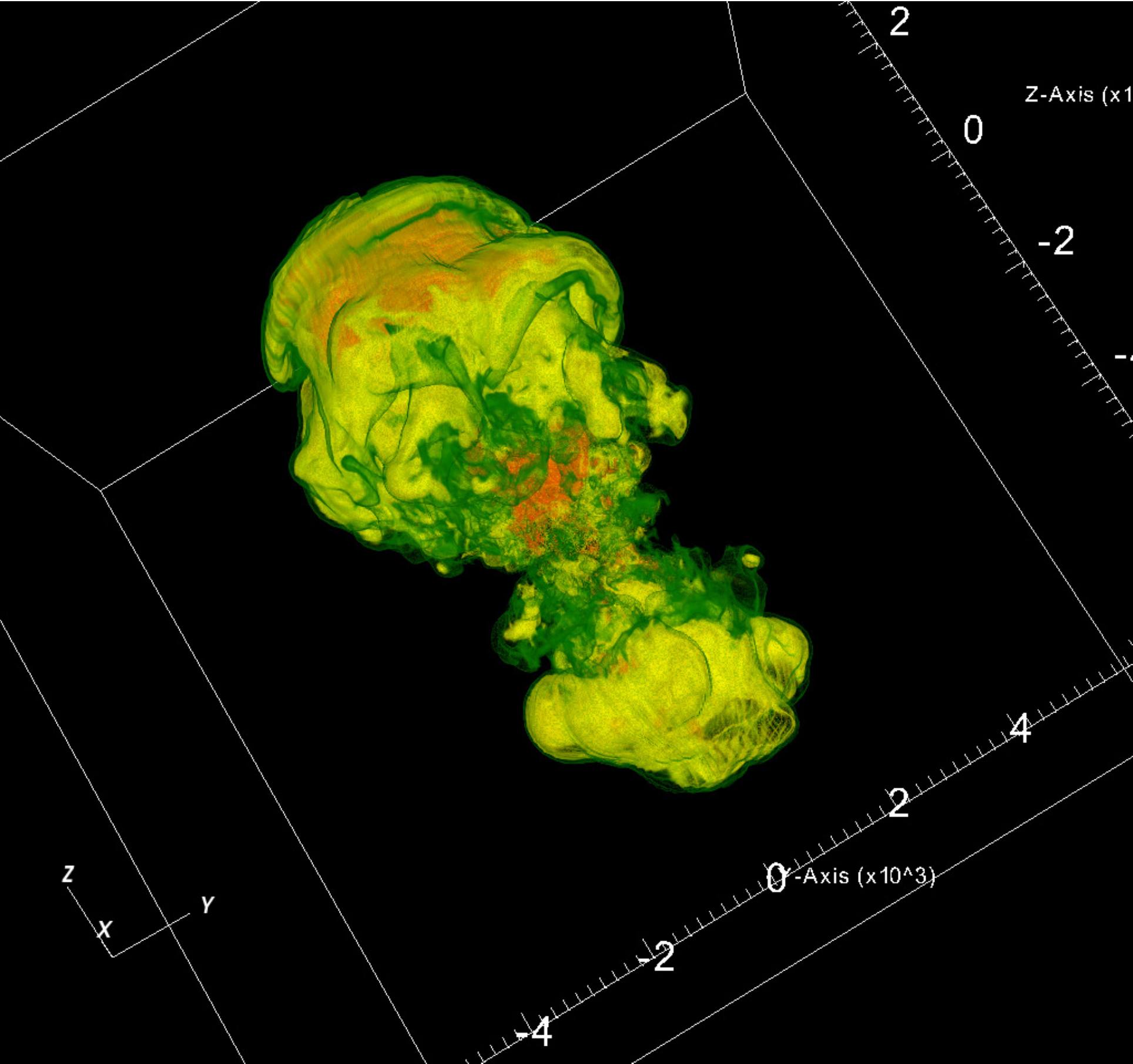}
\caption{Volume rendering of the entropy
in model s11.2\_3D at post-bounce times
of $89 \ \mathrm{ms}$ (top left), 
$134 \ \mathrm{ms}$ (top right), $210\ \mathrm{ms}$ (bottom left), and $580\ \mathrm{ms}$
(bottom right).
\label{fig:volume}
}
\end{figure*}

\subsection{Numerical Methods}

The simulations are performed with the general relativistic (GR)
neutrino hydrodynamics code \textsc{CoCoNuT}
\citep{dimmelmeier_02_a,mueller_10,mueller_15}. Our version of
\textsc{CoCoNuT} uses piecewise parabolic (PPM) reconstruction
\citep{colella_84} combined with a hybrid HLLC/HLLE Riemann solver
\citep{mignone_05_a} to obtain higher-order spatial accuracy.
\textsc{CoCoNuT} employs spherical polar coordinates
$(r,\theta,\varphi)$, which leads to strong time-step constraints near
the polar axis in 3D due to the converging grid geometry. We
circumvent this problem using an adapted version of the mesh
coarsening scheme of \citet{cerda_09}: While the equations of
hydrodynamics are solved on a fine grid with constant spacing $\delta
\varphi$ in longitude everywhere, a filter is applied to the solution
after each time step to remove short wavelength noise in the
$\varphi$-direction by projecting the conserved variables onto
piecewise linear/quadratic\footnote{For example, we use linear
  functions for the Eulerian density $D$ and the mass fractions $X_i$
  so that the conserved partial masses $D X_i$ are represented by
  quadratic functions.}  functions in ``supercells'' that contain
$2^{n(\theta)}$ fine cells in the $\varphi$-direction. The projection
algorithm is implemented conservatively, and the slopes for the
filtered solution are obtained using the monotonized-central (MC)
limiter of \citet{van_leer_77}. The supercell size $2^{n(\theta)}$ is
chosen such that $n \sin \theta>1/2$ is maintained at any
latitude. This ensures that the allowed CFL time step at high
latitudes is at most shorter by a factor of $\mathord{\sim} 2$
compared to the equatorial region, and limits the filtering to a
region of $30^\circ$ around the pole, which corresponds to $13.3\%$ of
the total volume. Similar techniques have long been used in numerical
meteorology, cp.\ \citet{kageyama_04} and Chapter~18 in \citet{boyd}.
Polar filtering allows us to maintain the same effective angular
resolution of $1.4^\circ$ in 2D and 3D with grids
of  $N_r\times N_\theta =550 \times 128 $ zones (2D)
and
 $N_r\times N_\theta \times N_\varphi=550\times128 \times 256$ zones
(3D, fine grid) covering the innermost
$10^5 \ \mathrm{km}$ of the star, respectively.

Like any other solution to avoid the coordinate singularity and the
excessive time step constraint near the axis such as Cartesian
coordinates, overset grids
\citep{kageyama_04,wongwathanarat_10a,melson_15a} or cubed-sphere grids
\citep{ronchi_96,koldoba_02,zink_08,fragile_09}, this polar filtering
procedure has specific advantages and disadvantages: Unlike Cartesian
codes, polar filtering allows us to maintain spherical symmetry in the
initial conditions and explicit symmetry-breaking terms can be
avoided.  Different from cubed-sphere grids, the grid remains
orthogonal; and global conservation laws are easier to enforce than
for overlapping overset grids. On the other hand, projecting the
solution to piecewise linear function effectively introduces an
anisotropy in the numerical viscosity and diffusivity (an unwelcome
effect that is minimized by overset or cubed-sphere grids but also
manifests itself for Cartesian grids that are prone to the development
of $m=4$ modes).

The space-time metric is computed using the extended conformal
flatness condition (xCFC, \citealp{cordero_09}). Because the asphericities
in the gravitational field are small for non-rotating core-collapse
supernovae we use the monopole approximation for the gravitational
field, i.e.\ the lapse function $\alpha$, the conformal factor $\phi$,
and the radial component $\beta_r$ of the shift vector only depend on
$r$, and the non-radial components $\beta_\theta$ and $\beta_\varphi$
of the shift vector are set to zero.

For the neutrinos, we use the fast multi-group transport (FMT) scheme
of \citet{mueller_15}, which is based on a stationary two-stream
solution of the relativistic transfer equation that is combined with
an analytic variable Eddington factor closure at low optical
depths. This schemes includes general relativistic effects under
the assumption of a stationary metric, but neglects velocity-dependent
effects like Doppler shift and aberration. The neutrino rates include
emission, absorption, and elastic scattering by nuclei and free
nucleons (along the lines of \citealp{bruenn_85}) as well as an
effective one-particle rate for nucleon-nucleon bremsstrahlung and an
approximate treatment of the energy exchange in neutrino-nucleon
scattering reactions. Comparisons of the FMT scheme with the more
sophisticated relativistic neutrino transport solver
\textsc{Vertex} \citep{rampp_02,mueller_10} showed excellent
qualitative and good quantitative agreement. For more details,
we refer the reader to \citet{mueller_15}.

In order to further alleviate the time-step constraint, 
the innermost part of the computational domain
(where densities exceed  $\mathord{\sim} 5 \times 10^{11} \ \mathrm{g} \ \mathrm{cm}^{-3}$)
is calculated in spherical symmetry using a conservative
implementation of mixing-length theory for proto-neutron star
convection, a procedure that has been used in the context
of supernova simulations before (e.g.\ \citealp{wilson_88,huedepohl_phd}).
The transition density is adjusted such that it lies inside the convectively
stable cooling layer.

In the high-density regime, we use the equation of state (EoS) of
\citet{lattimer_91} with a bulk incompressibility modulus of nuclear
matter of $K=220 \ \mathrm{MeV}$. At low densities, we employ an EoS
accounting for photons, electrons and positrons of arbitrary
degeneracy, an ideal gas contribution from baryons (nucleons, protons,
$\alpha$-particles and 14 other nuclear species), Nuclear reactions
are treated using ``flashing'' as described in \citet{rampp_02}.

\begin{table*}
  \caption{Overview of Simulations.
The extrapolation of the final remnant masses (last column) is discussed in Section~\ref{sec:explosion_properties}.
\label{tab:models}
}
  \begin{center}
    \begin{tabular}{ccccccc}
      Model  & Progenitor & Dimensionality &  Post-Bounce  & Diagnostic Energy & Baryonic Neutron Star & Extrapolated Baryonic\\
             &            &                &  Time Reached [s] & Reached [erg] & Mass Reached $[M_\odot]$ & Remnant Mass $[M_\odot]$ \\
\hline
\hline
      s11.0\_2D & s11.0 &  2               & 8.195  & $1.3 \times 10^{50}$ & 1.62 & 1.62\\
      s11.2\_3D & s11.2 &  3               & 0.944  & $1.3 \times 10^{50}$ & 1.33 & 1.48 \\
      s11.2\_2Da & s11.2 &  2              & 1.044  & $5.0 \times 10^{49}$ & 1.37 & --- \\
      s11.2\_2Db & s11.2 &  2              &  6.003 & $7.8 \times 10^{49}$ & 1.47 & 1.69 \\
      s11.4\_2D & s11.4 &  2               &  6.129 & $1.0 \times 10^{50}$ & 1.56 & 1.63 \\
      s11.6\_2D & s11.6 &  2               & 11.453 & $2.1 \times 10^{50}$ & 1.62 & 1.63\\
    \end{tabular}
  \end{center}
\end{table*}

\begin{figure}
\includegraphics[width=\linewidth]{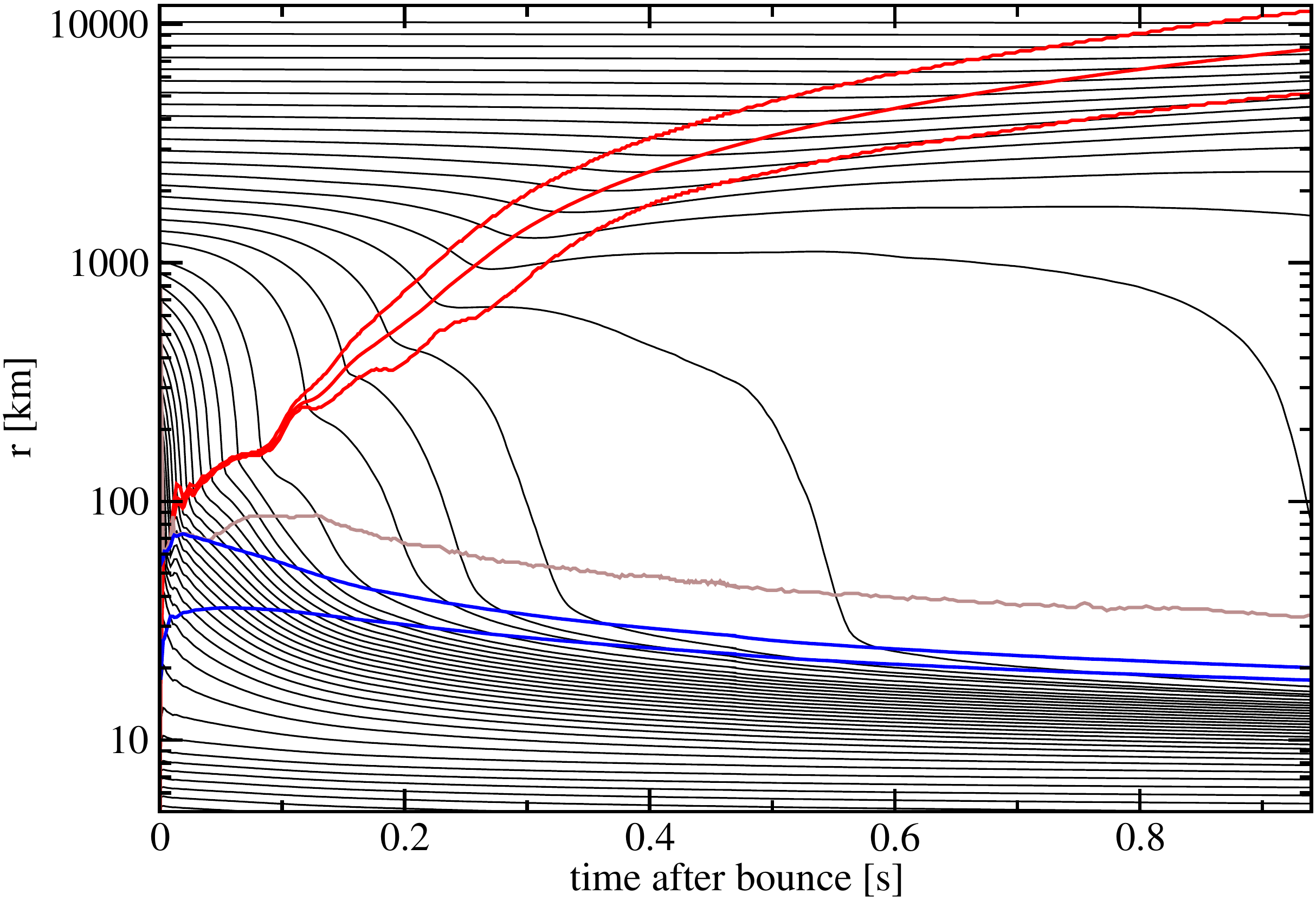}
\caption{Selected mass shell trajectories (black) for model
s11.2\_3D computed from spherically averaged density profiles.
The trajectories start with roughly equal spacing in $\log r$
shortly before bounce. The plot also shows the maximum,
average, and minimum shock radius (red), the gain radius
(light brown), and the radii corresponding to
densities of $10^{11} \ \mathrm{g} \ \mathrm{cm}^{-3}$
and $10^{12} \ \mathrm{g} \ \mathrm{cm}^{-3}$ (blue).
\label{fig:mass_shells}
}
\end{figure}

\begin{figure}
\includegraphics[width=\linewidth]{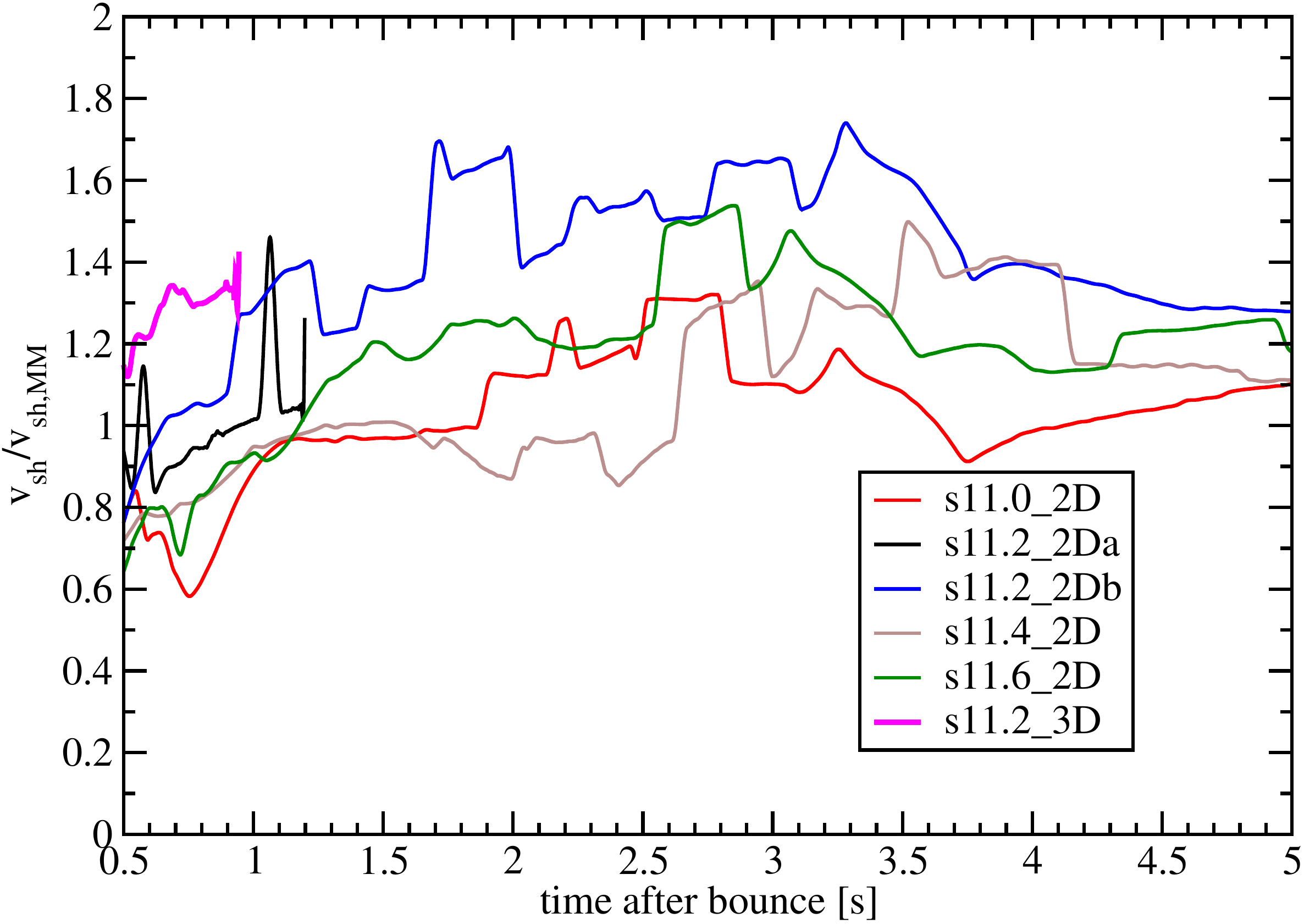}
\caption{Ratio $v_\mathrm{sh}/v_\mathrm{sh,MM}$ between the
  angle-averaged shock velocity $v_\mathrm{sh}= \ud r_\mathrm{sh,avg}/
  \ud t$ and the shock velocity $v_\mathrm{sh,MM}=0.794
  (E_\mathrm{expl}/M_\mathrm{expl})^{1/2}
      [M_\mathrm{expl}/(\rho_\mathrm{pre} r^3)]^{0.19}$ predicted by
      the model of \citet{matzner_99}.  Note that $r_\mathrm{sh,avg}$
      and its numerical derivatives need to be smoothed considerably
      to allow for a useful comparison.
      $v_\mathrm{sh}/v_\mathrm{sh,MM}$ is only shown until $5
      \ \mathrm{s}$ after bounce because the automatic smoothing
      procedure becomes ineffective toward the end of simulations
      s11.0\_2D, s11.2\_2Db, and s11.4\_2D and
      $v_\mathrm{sh}/v_\mathrm{sh,MM}$ becomes highly oscillatory.
\label{fig:matzner}
}
\end{figure}

\begin{figure}
\includegraphics[width=\linewidth]{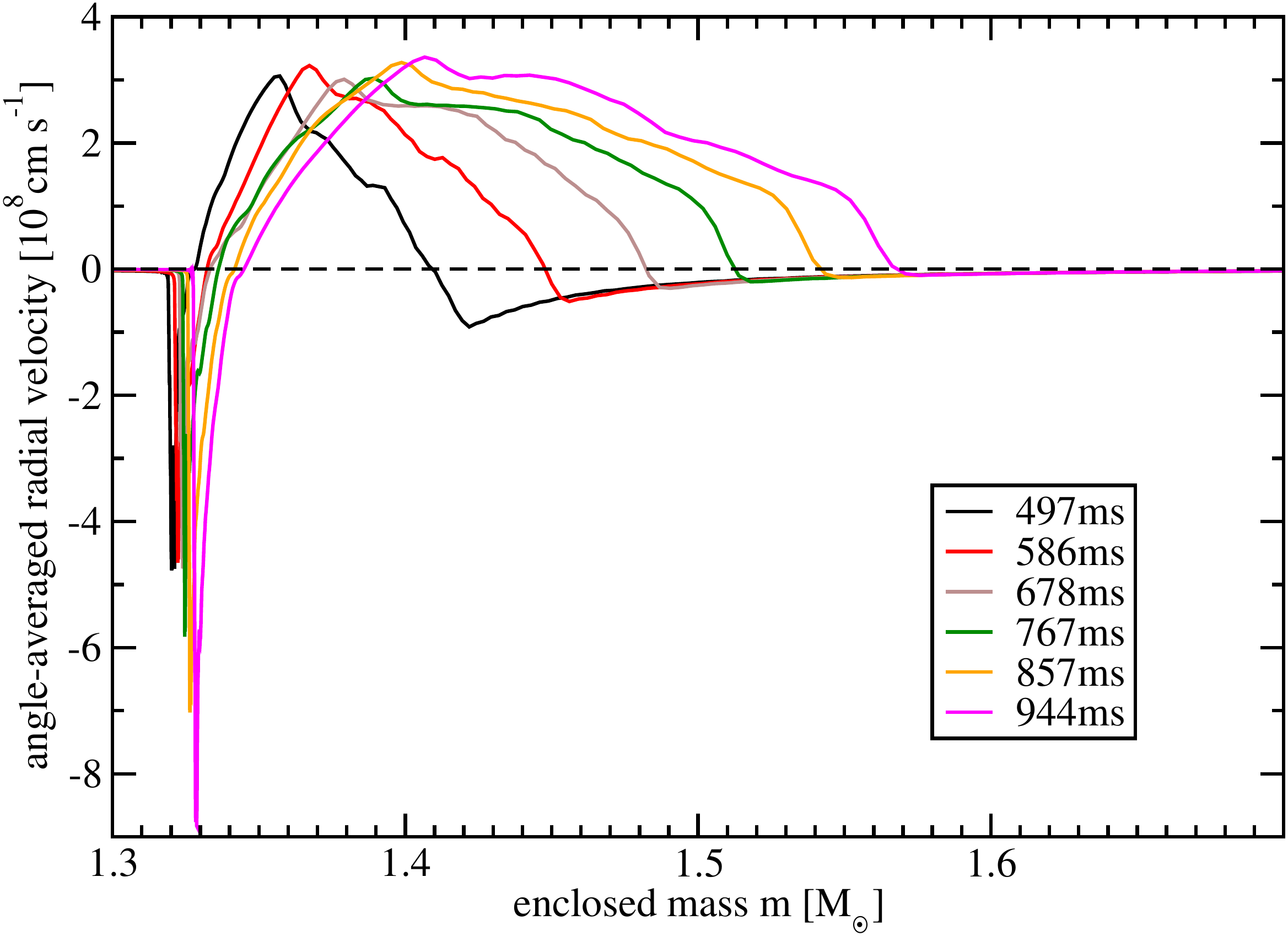}
\caption{Angle-averaged, density-weighted velocity profiles for
model s11.2\_3D at different post-bounce times. At the
end of the simulation, the angle-averaged velocity
is positive outside a mass coordinate of $1.35 M_\odot$, but the
zero point is still moving outward in mass. Note that the angle-average
extends over the post-shock and pre-shock region and cannot be used
to infer the post-shock velocity directly.
\label{fig:velocity_profiles}
}
\end{figure}

\begin{figure}
\includegraphics[width=\linewidth]{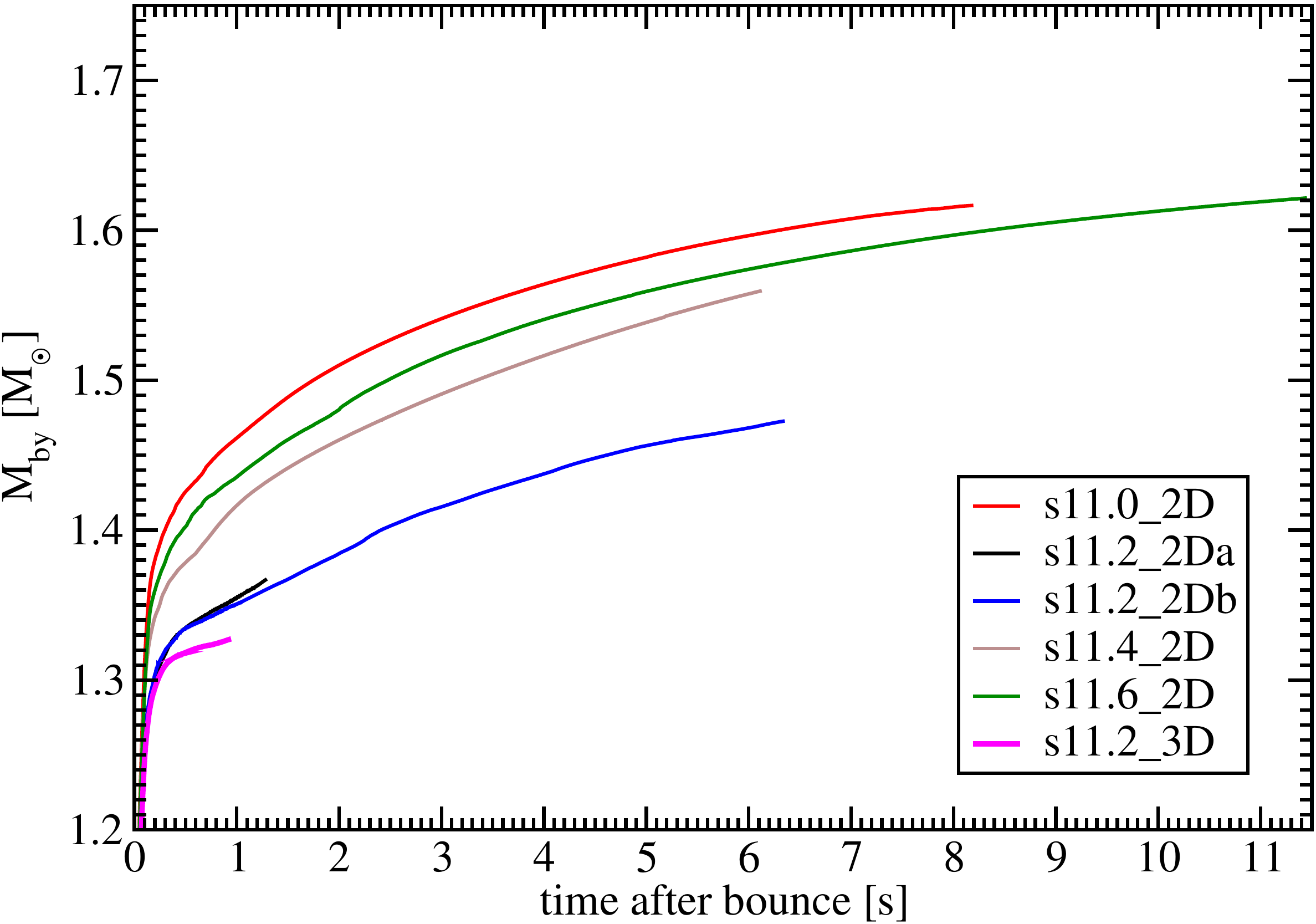}
\caption{Baryonic neutron star masses (comprising all matter at
  densities higher than $10^{11} \ \mathrm{g} \ \mathrm{cm}^{-3}$) for
  the different 2D and 3D simulations as a function of time.
\label{fig:pns_masses}
}
\end{figure}

\section{Overview of Simulation Results}
\label{sec:overview}
In all our simulations, runaway shock expansion sets in when the
Si/SiO interface reaches then shock and the mass accretion rate drops
rapidly.  Figures~\ref{fig:shock_2d_3d} (all 2D/3D $11.2 M_\odot$
models) and \ref{fig:shock_prog} (long-time evolution of all 2D
models) provide an overview over the propagation of the shock and the
growth of the explosion energies for the different models; they show the
maximum, minimum (only Figure~\ref{fig:shock_2d_3d}), and
angled-averaged shock radius, as well as the ``diagnostic explosion
energy'' $E_\mathrm{expl}$, which we define as the total net energy
(i.e.\ gravitational+internal+kinetic energy) of all the material that
is nominally unbound and is moving outward with positive radial
velocity at a given time (cp.\ \citealt{mueller_12a,bruenn_14}). The
nucleon rest masses are not included in the internal energy,
i.e.\ nucleon recombination only contributes to the diagnostic energy
once it actually takes places.  Figure~\ref{fig:shock_2d_3d} also
shows the time derivative of the diagnostic energy. Key results of the
simulations, including the diagnostics energy and the baryonic remnant
mass at the end of the simulations as well as estimates for the final
remnant mass (see Section~\ref{sec:explosion_properties} below), are
given in Table~\ref{tab:models}.

\subsection{Differences Between 2D and 3D During the First Second}
For the $11.2 M_\odot$ progenitor, the first second after bounce is
shown in detail in Figure~\ref{fig:shock_2d_3d} both in 2D and 3D.  In
addition, Figures~\ref{fig:snap_2d_3d_1} and \ref{fig:snap_2d_3d_2}
illustrate the multi-dimensional flow morphology for models s11.2\_2Da
and s11.2\_3D on meridional slices, and 3D ray-cast images of
neutrino-heated convective bubbles in model s11.2\_3D before and after
shock revival are shown in Figure~\ref{fig:volume}.

Prior to the infall of the Si/SiO interface, we find very similar
shock trajectories independent of dimensionality. However, prompt
convection develops slightly differently in 2D and 3D, and its
residual effect on the entropy and lepton number profiles leads to a
slight divergence between the 2D and 3D models already at early
times in many quantities (neutron star radius, gain radius, cooling
profiles, etc.). This effect is not unphysical \emph{per se}, but is
most probably exaggerated in our models because the FMT scheme tends
to overestimate the strength of prompt convection. In view of the
large systematic effects that we shall discuss later, it is also
inconsequential, but needs to be borne in mind when comparing the
different models.

After the infall of the Si/SiO interface, the shock expands slightly
faster in 3D than in 2D, and the explosion energy starts to reach
appreciable positive values several tens of milliseconds earlier.
Snapshots of the entropy for models s11.2\_3D and s11.2\_2Da during
this phase can be seen in the middle and right columns of
Figure~\ref{fig:snap_2d_3d_1}, which show the development of large
convective plumes in both cases.  The reader will note that the plumes
are somewhat aligned with the coordinate axis in 3D, which is clearly
a result of the coordinate choice but need not be considered harmful
as discussed in Section~\ref{sec:assessment}.  At later times, the
morphology of the 3D model is quite different; instead of the broad,
laminar downflows characteristic for 2D explosions, the interface
between the downflows and the hot, neutrino-heated ejecta eventually
becomes turbulent during the infall, resulting in corrugated downflows
and partial mixing with the neutrino-heated ejecta, as can be seen
most perspicuously in the middle column of Figure~\ref{fig:snap_2d_3d_2}.

Soon after shock revival, the 2D models start to go through episodes
of halting shock expansion or even transient shock recession. While
the growth rate $\ud E_\mathrm{expl}/\ud t$ of the diagnostic explosion energy
reaches values comparable to 3D for $100 \ldots 200 \ \mathrm{ms}$, the 
explosion energy grows much less steadily in the long term and has reached
only $(4\ldots 5) \times 10^{49} \ \mathrm{erg}$ after $1 \ \mathrm{s}$. By contrast,
the 3D model shows a steady growth of the explosion energy
($1.3 \times 10^{50} \ \mathrm{erg}$ by the end of the simulation), and
considerably faster shock expansion. As illustrated
by the mass shell trajectories in Figure~\ref{fig:mass_shells},
the spherically averaged radial velocity behind the shock
becomes positives about $300 \ \mathrm{ms}$ after bounce,
and the mass shells shocked later than $500 \ \mathrm{ms}$
after bounce appear to move outward steadily instead of eventually
falling back onto the proto-neutron stars.

\subsection{Shock Propagation During the First Seconds}
\label{sec:matzner}
Before we attempt to extrapolate the final remnant masses, it is
useful to point out a simple analytic relation between the diagnostic
energy and the shock velocity. During the later phases of the
explosion when the explosion energy has saturated, simple analytic
models
\citep{sedov_59,kompaneets_60,laumbach_69,klimishin_81,koo_90,matzner_99}
provide a useful qualitative description of shock propagation in
hydrodynamical simulations
\citep{woosley_95,kifonidis_03,wongwathanarat_15}.  These underlying
models typically rely on the assumption of self-similarity and/or
exponential or power-law approximations for the envelope, neglect the
effect of gravity, do not account for continuous energy input into the
ejecta, and have been derived under the assumption of spherical
symmetry. During the first seconds covered in our simulations, all
these conditions are violated. It is remarkable that the approximate
formula of \citet{matzner_99},
\begin{equation}
\label{eq:matzner}
v_\mathrm{sh,MM} = 0.794 \sqrt{\frac{E_\mathrm{expl}}{m}} \left(\frac{m}{\rho_\mathrm{pre} r^3}\right)^{0.19},
\end{equation}
nonetheless provides a reasonable estimate for the shock velocity
$v_\mathrm{sh}$ already a few hundreds of milliseconds after shock
revival if it is evaluated using appropriate definitions for the
explosion energy $E_\mathrm{expl}$, the ``ejecta'' mass $m$, and the
pre-shock density $\rho_\mathrm{pre}$: We find that
Equation~(\ref{eq:matzner}) works well if $v_\mathrm{sh}$ is taken to
be the angle-averaged shock velocity, i.e.\ the time-derivative of the
angle-averaged shock radius $r_\mathrm{sh,avg}$, if the pre-shock
density is evaluated at $r_\mathrm{sh,avg}$, if $E_\mathrm{expl}$ is
identified with the time-dependent diagnostics energy, and if the
``ejecta'' mass includes the \emph{entire} mass enclosed by the shock
from above and the gain radius from below (and thus cannot properly be
termed ``ejecta'' mass as in the original work of
\citealt{matzner_99}). This is illustrated in
Figure~\ref{fig:matzner}, which shows the ratio of the angle-averaged
shock velocity $v_\mathrm{sh}$ and the analytic estimate
$v_\mathrm{sh,MM}$ of \citet{matzner_99}. While there is considerable
scatter, the numerical models fall in a band with $1<v_\mathrm{sh}$/ $v_\mathrm{sh,MM}<1.6$ most of the time, especially at
time later than $1 \ \mathrm{s}$ after bounce.  Our models suggest
$v_\mathrm{sh}= 1.3 v_\mathrm{sh,MM}$ as a good analytic
estimate for early shock propagation in core-collapse supernovae.

\begin{figure}
\includegraphics[width=\linewidth]{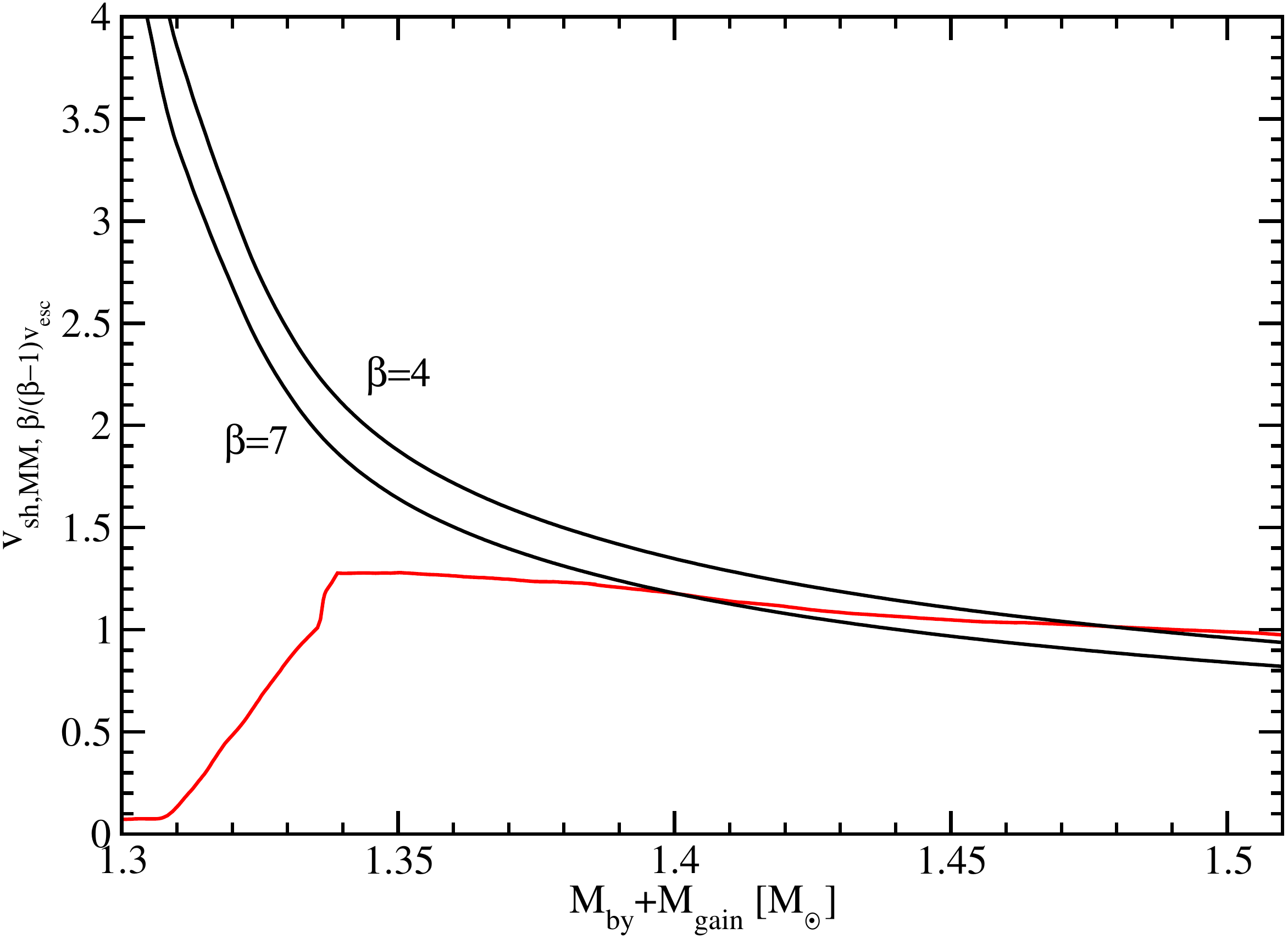}
\caption{Comparison of the post-shock velocity computed
from Equation~(\ref{eq:matzner})  (red curve) to the required
shock velocity $\beta/(\beta-1) v_\mathrm{esc}$
for the separation of outgoing and infalling mass shells
for two different values of the compression ratio $\beta$
(black curves) for model s11.2\_3D.
\label{fig:accretion}
}
\end{figure}

\subsection{Explosion Energies and Neutron Star Masses}
\label{sec:explosion_properties}
Although the explosion energy in model s11.2\_3D has not yet reached
its final value and there is still some accretion onto the
proto-neutron star, there is no doubt that the incipient explosion
will eventually expel the envelope. This is not only clear from the
steady outward movement of the shocked mass shells at late times
visible in Figure~\ref{fig:mass_shells} and in the velocity profiles
depicted in Figure~\ref{fig:velocity_profiles}; the diagnostic
explosion energy is also significantly higher than the residual
binding energy of the pre-shock matter (the ``overburden'' in the
terminology of \citealt{bruenn_13,bruenn_14}) of $5\times 10^{49}
\ \mathrm{erg}$. This is clearly different from models s11.2\_2Da and
s11.2\_2Db and the similar 2D explosion models of the same progenitor
discussed in \citet{buras_06b,marek_09,mueller_12a}.

Nonetheless, the long-time simulations of the 2D models over several
seconds show that even such supposedly tepid models eventually develop
steady shock propagation and reach sufficiently high diagnostic
explosion energies to shed the envelope (Figure~\ref{fig:comp_progs}).
For the low-mass progenitors simulated here, the outgoing and
infalling mass shells inevitably separate as the post-shock velocities
behind the entire shock front become positive once it reaches the edge
of the relatively small C/O core ($1.7 \ldots 1.8 M_\odot$), which
happens already after a few seconds in these models. The
acceleration of the shock at the steep density gradient between the
C/O core and the He shell and the small binding energy of the He shell
then result in a steady outward movement of the shocked matter. At
that point, the overburden of the unshocked envelope becomes almost
negligible (e.g.\ $10^{49} \mathrm{erg}$ for s11.2\_2Db, $5\times
10^{48} \mathrm{erg}$ for s11.6\_2D), and we can determine relatively
firm lower limits for the final explosion energy.

Continuous accretion over several seconds provides for sufficient
neutrino heating to power outflows and pump additional energy into the
ejecta over this long time-scale, albeit at a rather modest rate. As a
result, models that appear woefully underenergetic during the first
second can still develop appreciable explosion energies, the best
example being the $11.6 M_\odot$ model, where $E_\mathrm{expl}$ grows
from $3.5 \times 10^{49} \mathrm{erg}$ at $1 \ \mathrm{s}$ to $2.0 \times 10^{50} \ \mathrm{erg}$ after $11 \ \mathrm{s}$.
The explosion energies obtained after several seconds are
comparable to the 3D case and compatible with supernova
explosion energies at the lower end of the observed
spectrum (see, e.g., \citealt{pejcha_15}).

 The fact that the diagnostic explosion energies increase
  more or less steadily over several seconds in the 2D models (except
  for transient phases where the explosion geometry changes because a
  neutrino-driven outflow is shut off as discussed in
  Section~\ref{sec:constriction}) has important implications for the
  usefulness of the diagnostic energy as a predictor of the final
  explosion properties. \citet{perego_15} have recently pointed out on
  the basis of artificial 1D explosions of \citet{perego_15} that the
  diagnostic energy \emph{overshoots} the final explosion energy and
  only approaches its asymptotic value very slowly on a time-scale of
  seconds, and suggest that a better estimate for the final explosion
  energy can be obtained by subtracting the overburden
  $E_\mathrm{ov}$, i.e.\ the binding energy of the mass shells outside
  the shock, from $E_\mathrm{expl}$. As illustrated by the comparison
  of $E_\mathrm{expl}$ and $E_\mathrm{expl}-E_\mathrm{ov}$ in
  Figure~\ref{fig:shock_prog}, our 2D models show a somewhat different
  behaviour; similar to the simulations of \citet{bruenn_14}, there is
  no overshooting of $E_\mathrm{expl}$ above its prospective final
  value for which it appears to furnish a lower bound rather than an
  upper bound. This is the result of a fundamentally different way to
  power the explosion in multi-D compared to 1D: Once an explosion is
  triggered in 1D, the outflow rate quickly drops and only a weak
  neutrino-driven wind can still pump energy into the ejecta over
  time-scales of seconds. The accumulation of shocked material with
  negative total energy therefore quickly dominates the total energy
  budget of the ejecta region and $E_\mathrm{expl}$ decreases, while
  $E_\mathrm{expl}-E_\mathrm{ov}$ remains roughly constant by virtue
  of total energy conservation. The case for
  $E_\mathrm{expl}-E_\mathrm{ov}$ as a more compelling predictor of
  the final explosion energy is weaker in multi-D, however, where
  neutrino-driven outflows can continuously pump energy into the
  ejecta at a high rate, and a considerable part of the shocked
  material with negative total energy is channeled onto the
  proto-neutron star instead of being swept along by the ejecta and
  reducing the diagnostic energy.  $E_\mathrm{expl}$ may still
  decrease somewhat on time-scales longer than $5\ldots 10
  \ \mathrm{s}$ as the shock propagates through the helium shell, and
  this introduces a residual uncertainty of up to $\mathord{\sim} 15
  \%$ in the final explosion energy, which we expect to lie in the
  range bracketed by $E_\mathrm{expl}$ and
  $E_\mathrm{expl}-E_\mathrm{ov}$.  It is also noteworthy that
  $E_\mathrm{expl}-E_\mathrm{ov}$ does not appear to be a good
  predictor for the final explosion energy at early times simply
  because its rise phase is much more drawn out than in artificial 1D
  explosions and it only becomes positive $\mathord{\sim} 1
  \ \mathrm{s}$ after bounce or later.

While our simulations reach final explosion energies of the order of $10^{50}
\ \mathrm{erg}$, this comes at the expense of rather high neutron star
  masses: Figure~\ref{fig:pns_masses} shows that the baryonic neutron
  star masses $M_\mathrm{by}$ in the 2D models all end up at values
  $\gtrsim 1.47 M_\odot$ and will definitely exceed $1.6 M_\odot$ in
  cases like s11.0\_2D and s11.6\_2D. Unless selection effects favor
  the production of less massive neutron stars in binary systems for
  some reason, this potentially presents a serious conflict with the
  inferred neutron star mass distribution. Even if the lowest-mass
  neutron stars are presumed to originate from electron-capture
  supernovae, the masses of the neutron stars in the 2D models would
  end up well above the mean value of inferred baryonic masses of $1.5
  M_\odot$ \citep{schwab_10}. Since the simulated models represent
  progenitors with relatively small cores and a relatively early onset
  of the explosion, this is highly problematic. It is interesting to
  note that the 2D models of \citet{bruenn_14} also show such a
  tendency towards high neutron star masses despite their relatively
  high explosion energies (although this tendency is less striking
  than in our long-time simulations), with $M_\mathrm{by} =1.461
  M_\odot$ for their $12 M_\odot$ model B12-WH07 and values well above
  $1.6 M_\odot$ for the three remaining simulations.

The faster rise of the explosion energy in 3D could help to resolve
this discrepancy. Although the neutron star mass has not yet converged
to a final value, the spherically-averaged velocity profiles
(Figure~\ref{fig:velocity_profiles} indicate that the final ``mass cut''
is slowly emerging. At the end of the simulation, the net
mass accretion rate onto the neutron star in s11.2\_3D is lower by
a factor of $\mathord{\sim} 2$ compared to the corresponding 2D
models. 

To obtain a quantitative estimate for the final neutron star mass, we
follow \citet{marek_09}, who argued that accretion must subside once
the post-shock velocity $v_\mathrm{post}$ becomes comparable to the
escape velocity $v_\mathrm{esc}$. For a strongly asymmetric explosion
geometry, $v_\mathrm{post}$ is of course strongly direction-dependent.
Hence material ahead of the neutrino-heated plumes originating from a
given mass coordinate $m$ in the progenitor will be accelerated to a
higher post-shock velocity by the shock than material with the same
initial $m$ that is hit later in a direction where the shock expands
more slowly, so that the actual ``mass cut'' does not correspond to a
single mass shell $m$ in the progenitor. Instead, the dividing line in initial
mass coordinate will depend on angle. Nonetheless, one can argue that
the criterion $v_\mathrm{post}=v_\mathrm{esc}$ still yields a fairly
reliable estimate for the final mass of the neutron star if an
appropriate spherical average for $v_\mathrm{post}$ is used.

Equation~(\ref{eq:matzner}) for the average shock velocity allows us to
extrapolate the evolution of the $v_\mathrm{post}$ if necessary to
estimate a spherically-averaged ``mass
cut''.\footnote{Equation~(\ref{eq:matzner}) is also more convenient to
  use from the numerical point of view because the computation of the
  shock velocity as a numerical derivative of the shock position
  typically yields very noisy results.} If the pre-shock velocity is
assumed to be negligible, the post-shock velocity becomes
\begin{equation}
v_\mathrm{post}=\frac{\beta-1}{\beta} v_\mathrm{sh},
\end{equation}
in terms of the ratio $\beta$ of the post- and pre-shock density, and
equating this to the escape velocity yields the criterion
\begin{equation}
\frac{\beta-1}{\beta} v_\mathrm{sh}=\sqrt{\frac{2 G (M_\mathrm{by}+M_\mathrm{gain})}{r}},
\end{equation}
where we include the entire mass interior to the shock and not just
the mass of the neutron star when computing the escape velocity. At
late stages, the compression ratio $\beta$ typically drops below the
value $\beta=7$ for a radiation-dominated ideal gas with adiabatic
index $\gamma=4/3$ because of nuclear burning and/or because the
strong shock approximation is not strictly applicable over the
downflows.  We therefore compare $v_\mathrm{sh}$ in model s11.2\_3D to
the critical velocity $\beta/(\beta-1) v_\mathrm{esc}$ for two
different values of $\beta$ in
Figure~\ref{fig:accretion}. Figure~\ref{fig:accretion} suggests a
final baryonic remnant mass of $1.41\ldots 1.48 M_\odot$ 
for s11.2\_3D (to which
late-time fallback might be added). This would imply that the shock has
already passed the initial mass cut in some directions.  Estimates
along the same lines for the long-time simulations in 2D yield
baryonic remnant masses of $1.62 M_\odot$ for s11.0\_2D,
$1.63 M_\odot$ for s11.4\_2D, $1.69 M_\odot$ for s11.0\_2Db, and $1.63
M_\odot$ for s11.6\_2D assuming $\beta=4$.

Using the approximate formula of \citet{timmes_96} for the
gravitational neutron star mass $M_\mathrm{grav}$,
\begin{equation}
M_\mathrm{grav}
\approx M_\mathrm{by}-0.075 M_\odot \left(\frac{M_\mathrm{grav}}{M_\odot}\right)^2,
\end{equation}
which provides a reasonable fit across different nuclear equations of
states, the estimated baryonic neutron star mass for s11.2\_3D can be
converted to a gravitational mass of $1.34 M_\odot$, which would be
well within the range of observed neutron star masses and slightly
below the mean value of the higher-mass population of neutron stars
from iron core progenitors suggested by \citet{schwab_10}.  For the 2D
simulations the estimated gravitational masses are higher by more than
$0.1 M_\odot$.  The 3D effects responsible for the steeper rise of the
explosion energy thus improve the agreement with the observational
constraints considerably.

\begin{figure}
\includegraphics[width=\linewidth]{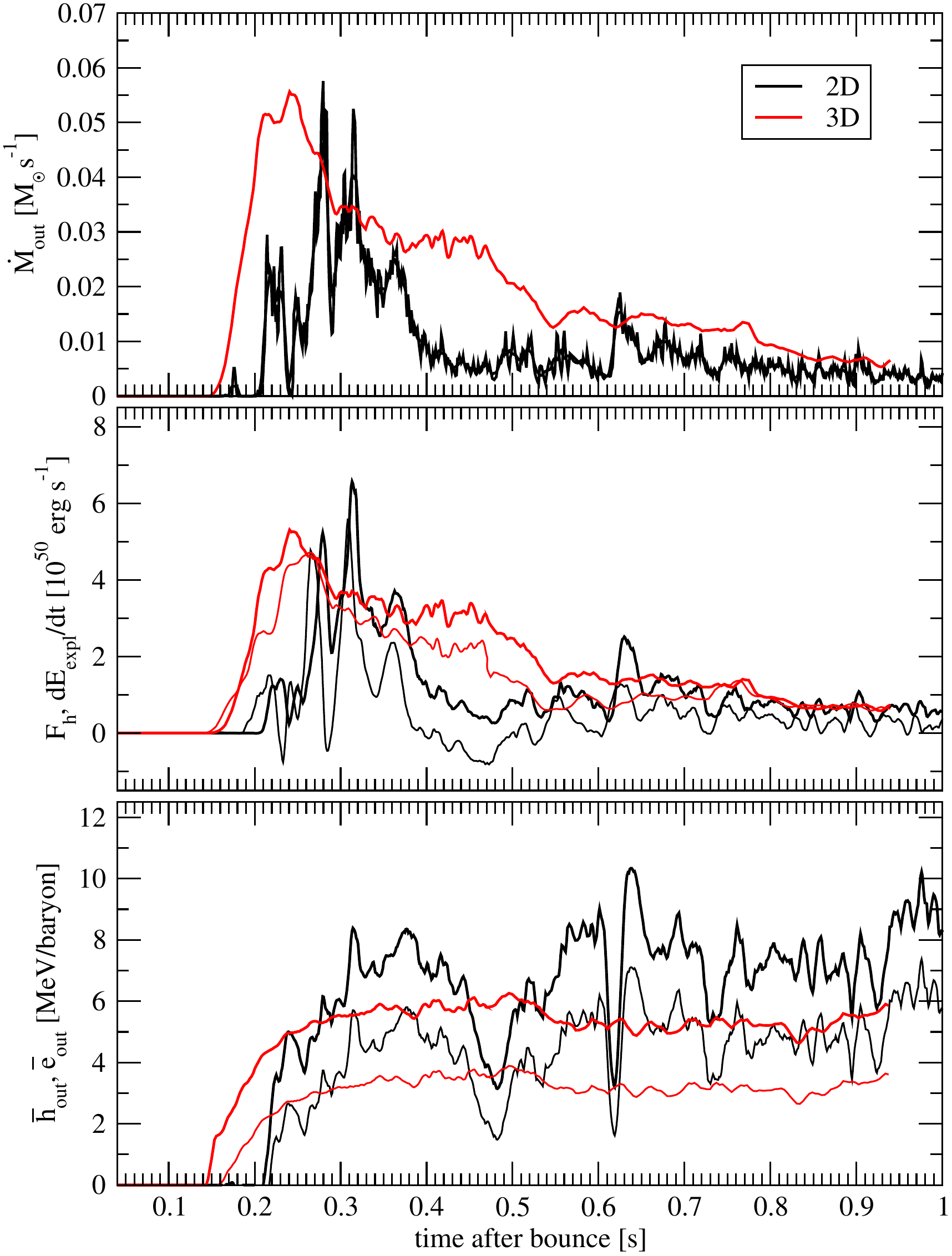}
\caption{Comparison of key properties of the neutrino-driven
outflows in 2D (black curves) and 3D (red curves) as measured
at a radius of $400 \ \mathrm{km}$. The top panel shows
the mass outflow rate $\dot{M}_\mathrm{out}$. The
middle panel shows the total enthalpy flux
$F_{h,\mathrm{out}}$ as defined in Equation~(\ref{eq:ent_flux})
(thick lines) alongside the time derivative
$\ud E_\mathrm{expl}/\ud t$ of the explosion energy (thin lines). The
average total enthalpy $\bar{h}_\mathrm{out}$ (thick lines) and total energy
$\bar{e}_\mathrm{out}$ in the outflows are shown in the bottom panel.
\label{fig:outflow_rate}}
\end{figure}

\section{Assessment of Shock Revival in the 3D Model}
\label{sec:assessment}
In the remaining part of the paper, our main thrust will be to explain
the physical mechanisms behind the pronounced differences between 2D
and 3D simulations in the explosion phase presented in
Section~\ref{sec:overview}. We do not investigate the differences in
the pre-explosion phase, because the numerical methodology used in
this study only allows limited conclusions concerning the problem of
shock revival in 3D for reasons detailed below.  Nonetheless, a few
remarks about shock revival in model s11.2\_3D are in order, if only
to motivate why the remainder of this paper focuses completely on the
explosion phase, and why simulations with a more rigorous treatment of
the neutrino transport and the neutrino rates are needed to decide the
fate of this particular progenitor model.

Superficially, our results for the $11.2 M_\odot$ star in 3D may
appear to be at odds with the recent core-collapse supernova
simulations with multi-group neutrino transport find either no shock
revival at all in 3D or only delayed shock revival compared to 3D
\citep{hanke_12,hanke_phd,tamborra_14a,lentz_15,melson_15b}.
Specifically, the $11.2 M_\odot$ model failed to explode in 3D
\citep{tamborra_14a} in a simulation using the
\textsc{Vertex-Prometheus} code \citep{rampp_02,buras_06a}. However,
one should not attach undue importance to the different outcomes of
the \textsc{Vertex-Prometheus} and \textsc{CoCoNuT-FMT} models:
Although reasonably close agreement with the more rigorous transport
scheme in \textsc{Vertex} can be reached with the FMT scheme, the
differences noted by \citet{mueller_15} are sufficiently large to
matter for a marginal model like s11.2. The fact that we find an
explosion merely underscores how close the $11.2 M_\odot$
\textsc{Vertex} model of \citet{hanke_phd} and \citet{tamborra_14a}
comes to an explosive runaway, and that the extreme sensitivity of the
supernova problem to the neutrino transport treatment and the
microphysics \citep{lentz_12a,lentz_12b,melson_15b} requires highly
accurate first-principle models in order to reliably decide whether
  an individual progenitor close to the explosion threshold explodes
  or fails (although ``imperfect'' simulations may already unearth
  much of the relevant physics from such cases).  Considering that
the comparison between the FMT scheme and \textsc{Vertex} revealed
slightly better heating conditions for a $15 M_\odot$ progenitor at
early times, and that the average shock radius initially expands
somewhat further in 3D than in 2D in the simulations of the $11.2
M_\odot$ progenitor with \textsc{Vertex}, the different outcome of the
FMT and \textsc{Vertex} runs is by no means unexpected.

Potentially, the emergence of large-scale bubbles aligned with the
coordinate axis (Figures~\ref{fig:snap_2d_3d_1},
\ref{fig:snap_2d_3d_2} and \ref{fig:volume}) also helps in pushing the
3D model above the explosion threshold at an early time
(cp.\ \citealp{thompson_00,dolence_13,fernandez_15}
for the role of the bubble size in the development of
runaway shock expansion). The alignment
is clearly a consequence of our coordinate choice and may also be
connected to the coarsening procedure for the polar region. Such
artifacts are unavoidable for standard spherical polar, cylindrical,
or Cartesian coordinates (where they manifest themselves as a
preferred excitation of $m=4$ modes instead), because the grid
geometry and spacing dictates the effective numerical diffusivity and
viscosity of a code, and physical instabilities will grow
preferentially in directions where they are least suppressed (or even
aided) by numerical dissipation. If there is sufficient time for
instabilities like convection and the SASI to reach saturation and go
through several overturn time-scales or oscillation periods, these
initial artifacts from the growth phase are eventually washed out, but
in a situation where the growth of certain modes accelerates rapidly
(e.g.\ after the infall of the Si/SiO interface) and then freezes out
they can subsist throughout the simulation. Nonetheless, we do not
view this a concern; the convective flow does not show any grid
alignment prior to the infall of the interface, and outflows
eventually develop in the equatorial plane as well.  Moreover, we
found no grid alignment of sloshing/spiral motions in SASI-dominated
models (which will be reported elsewhere).  In more realistic
simulations, the explosion geometry will be dictated by anisotropies
in the initial model, e.g.\ due to convective nuclear burning
\citep{arnett_94,bazan_94,bazan_98,asida_00,kuhlen_03,meakin_06,meakin_07,meakin_07_b,arnett_11,couch_15}
or rotation. In a sense, the alignment of the most prominent
high-entropy bubbles with the axis in model s11.2\_3D is even
fortunate for our further analysis because it eliminates the
unavoidable grid alignment of 2D explosion models as a potential cause
for the different energetics in 2D and 3D.

\begin{figure}
\includegraphics[width=\linewidth]{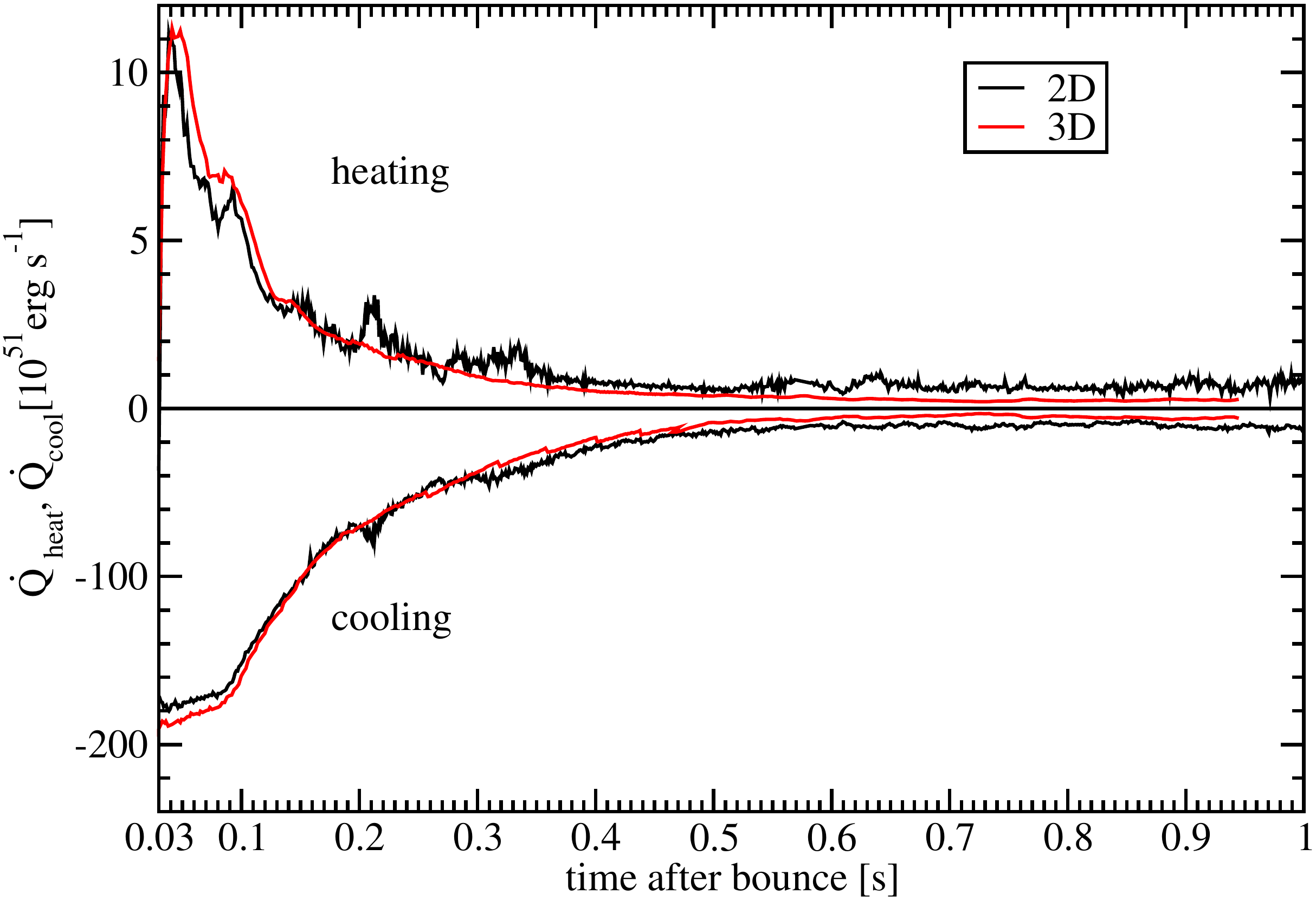}
\caption{Volume-integrated heating/cooling rates
$\dot{Q}_\mathrm{heat}$ and $\dot{Q}_\mathrm{cool}$
in the gain and cooling region for
models s11.2\_2Da and s11.2\_3D. The inner boundary of
the cooling region is defined (somewhat arbitrarily)
by a density of $10^{13} \ \mathrm{g} \ \mathrm{cm}^{-3}$.
Note that a different scale is used for both rates.
\label{fig:heating_cooling}}
\end{figure}

\begin{figure}
\includegraphics[width=\linewidth]{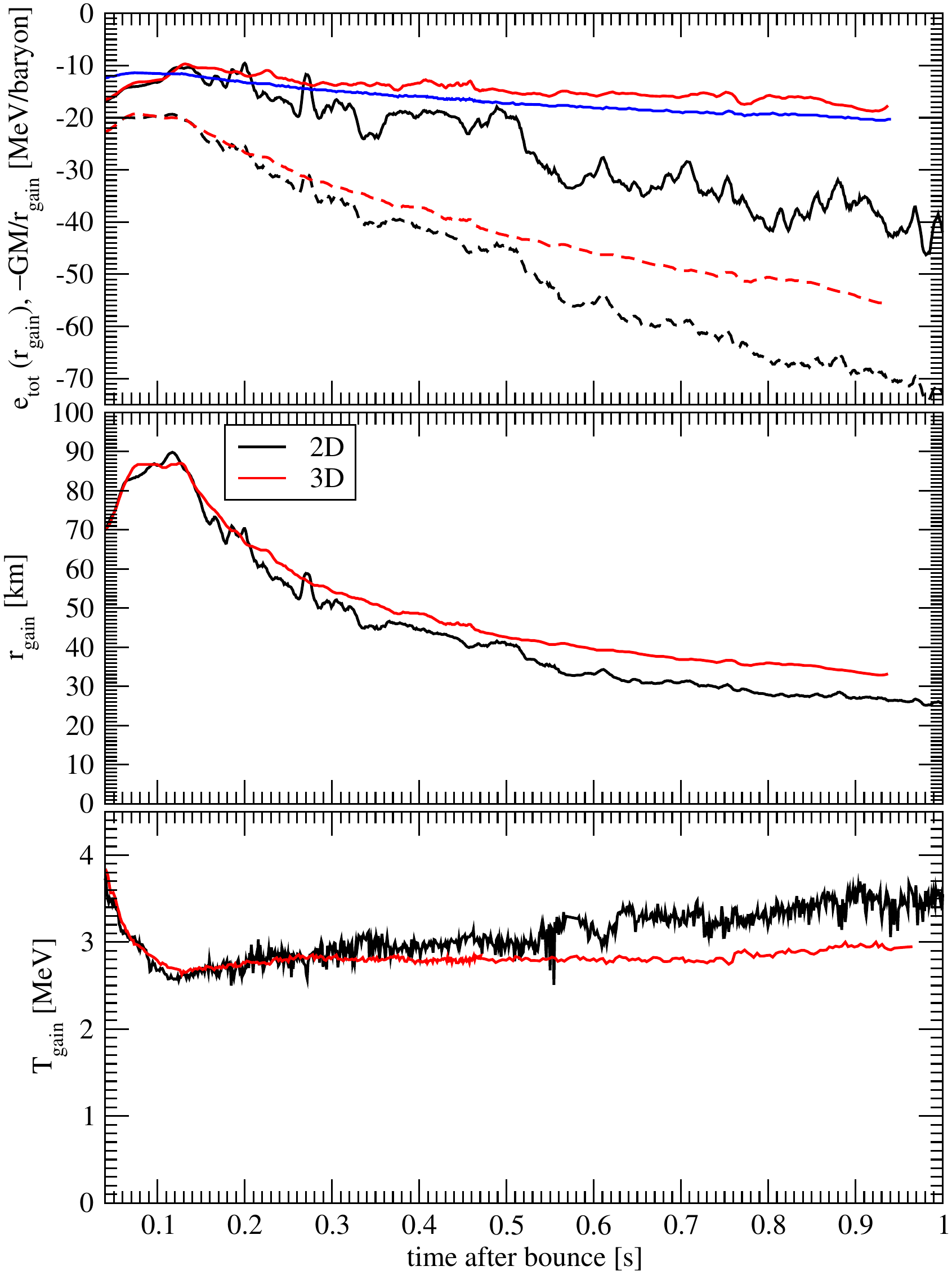}
\caption{Properties of the gain radius in 2D and 3D: The top panel
  shows the binding energy (i.e.\ the sum of the gravitational,
  kinetic, and internal energy) at the gain radius in 2D (black solid
  curve) and 3D (red solid curve) alongside the Newtonian potential
  of the neutron star $GM/\mathrm{r}_\mathrm{gain}$ (dashed lines),
  which sets the typical energy scale at the gain radius.  Here, $M$
  is the neutron star mass, and $\mathrm{r}_\mathrm{gain}$ is the gain
  radius. The estimate for $e_\mathrm{gain}$ from
Equation~(\ref{eq:egain}), which is based on the
assumption that the Bernoulli integral at the gain radius
is zero, is shown in blue for comparison.
 The middle panel shows $r_\mathrm{gain}$ itself, and the
bottom panel shows the angle-averaged temperature 
at the gain radius, $T_\mathrm{gain}$.
\label{fig:gain_radius}
}
\end{figure}

\begin{figure}
\includegraphics[width=\linewidth]{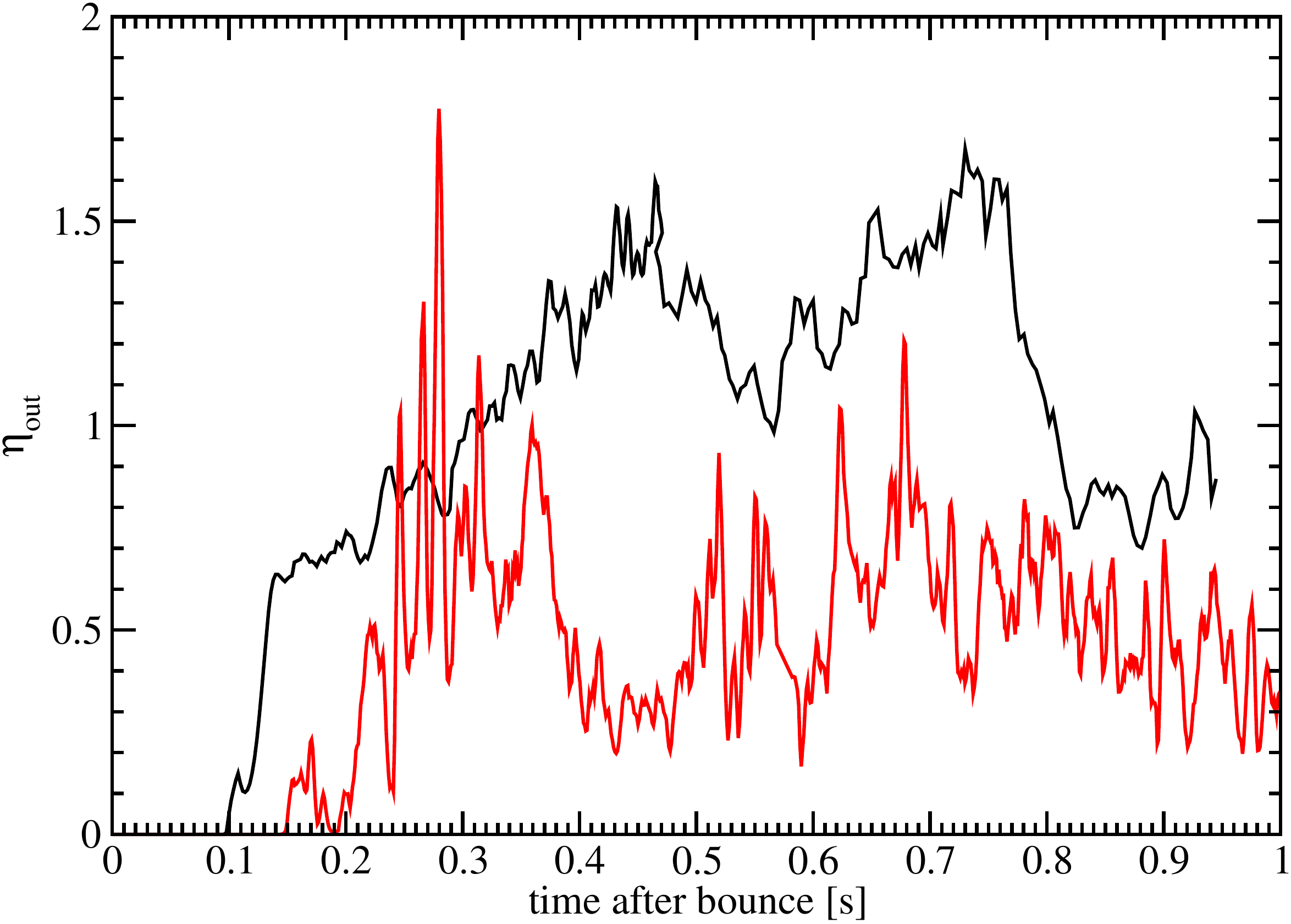}
\caption{Outflow efficiency $\eta_\mathrm{out}$,
for
models s11.2\_2Da and s11.2\_3D. $\eta_\mathrm{out}$
is defined as the ratio between the actual
mass outflow rate $\dot{M}_\mathrm{out}$
and a fiducial scale $\dot{Q}_\mathrm{heat}/|e_\mathrm{gain}|$
for the mass loss rate, see Equation~(\ref{eq:eta_out}).
Note that $\eta_\mathrm{out}$ is not limited to
values $\eta_\mathrm{out} \leq 1$ because fresh matter
for the neutrino-heated outflows is also supplied
by lateral mixing with the downflows above the gain
radius (where the matter is less tightly bound
than at the gain radius) and because recombination
also partly contributes in lifting the material
out of the gravitational potential well.
\label{fig:eta_out}}
\end{figure}

\section{Analysis of 2D/3D Differences}
\label{sec:analysis}
We now turn to the underlying physical mechanism responsible for the
different evolution of the 2D and 3D models during the explosion
phase.  The first step towards understanding the different dynamics of
the 2D and 3D models consists in considering the outflows and
downflows separately (in the vein of \citealt{melson_15a}) to analyze
the injection of mass and energy into the ``ejecta region'' with
positive binding energy (similar to \citealt{bruenn_14}).  We
partition the computational domain into two regions with positive
radial velocity ($v_r>0$) and negative radial velocity ($v_r<0$) and
then compute total fluxes and averages of several hydrodynamic
quantities.  In order not to detract the reader from the physics, we
work with Newtonian definitions here, and the generalization to the
relativistic case is discussed in Appendix~\ref{app:gr_quantities}
instead. Unless explicitly stated otherwise, the analysis is based on
models s11.2\_2Da and s11.2\_3D, i.e.\ we always refer to model
s11.2\_2Da and not to model s11.2\_2Db when talking about the 2D case.

\subsection{Mass and Enthalpy Flux into the Ejecta Region}
The first quantities to consider are the mass
fluxes $\dot{M}_\mathrm{in}$ and $\dot{M}_\mathrm{out}$
in the downflows and outflows,
\begin{equation}
\label{eq:dotm_in_out}
\dot{M}_\mathrm{in/out}
=
\int_{v_r \lessgtr 0} 
\rho v_r r^2
\ud \Omega,
\end{equation}
where $\rho$ is the density, and the
less and greater signs refer to downflows and outflows respectively.
Furthermore, we define total enthalpy\footnote{The quantity
$h_\mathrm{tot}=\epsilon+P/\rho+\mathbf{v}^2/2+\Phi$, which we shall usually
designate in this paper as ``total enthalpy'' is also referred
to as Bernoulli integral or stagnation enthalpy in other contexts
(including the case without gravity).
In this paper, we prefer the term ``total enthalpy'' to 
keep the terminology compact and stress its close relation
to the total energy per unit mass.} and energy fluxes $F_h$
and $F_e$,
\begin{equation}
\label{eq:ent_flux}
F_{h,\mathrm{in/out}}
=
\int_{v_r \lessgtr 0} 
\left[\rho (\epsilon+\mathbf{v}^2/2+\Phi) +P \right] v_r r^2\ud \Omega
\end{equation}
\begin{equation}
\label{eq:ene_flux}
F_{e,\mathrm{in/out}}
=
\int_{v_r \lessgtr 0} 
\rho (\epsilon+\mathbf{v}^2/2+\Phi) v_r r^2\ud \Omega
\end{equation}
where $\epsilon$ is the mass-specific internal energy,
$\mathbf{v}$ is vectorial fluid velocity, and $\Phi$ is the
Newtonian gravitational potential. The rationale for
including the gravitational potential in these fluxes is that
there is a conservation law for the
total energy density $e_\mathrm{tot}=\epsilon+\mathbf{v}^2/2+\Phi$,
\begin{equation}
\label{eq:energy_equation_newtonian}
\frac{\pd }{\pd t}
\left[\rho \left (\epsilon+\frac{\mathbf{v}^2}{2}+\Phi\right)
\right]+
\nabla \cdot
\left[\rho \left (\epsilon+\frac{\mathbf{v}^2}{2}+\Phi\right) \mathbf{v}
+P \mathbf{v}
\right]=0,
\end{equation}
if the gravitational potential is time-independent and the conversion
of rest mass energy is either disregarded or nuclear rest masses
are included in the internal energy. Since the diagnostic
explosion energy is defined as an integral over the total energy of
the ejecta, the energy budget of the ejecta naturally involves
the total enthalpy flux from lower regions of the gain layer to
the ``ejecta region'' with $e_\mathrm{tot}$ if the boundary
of the ejecta region remains at a roughly constant radius.

\begin{figure*}
\centering
\includegraphics[width=0.4\linewidth]{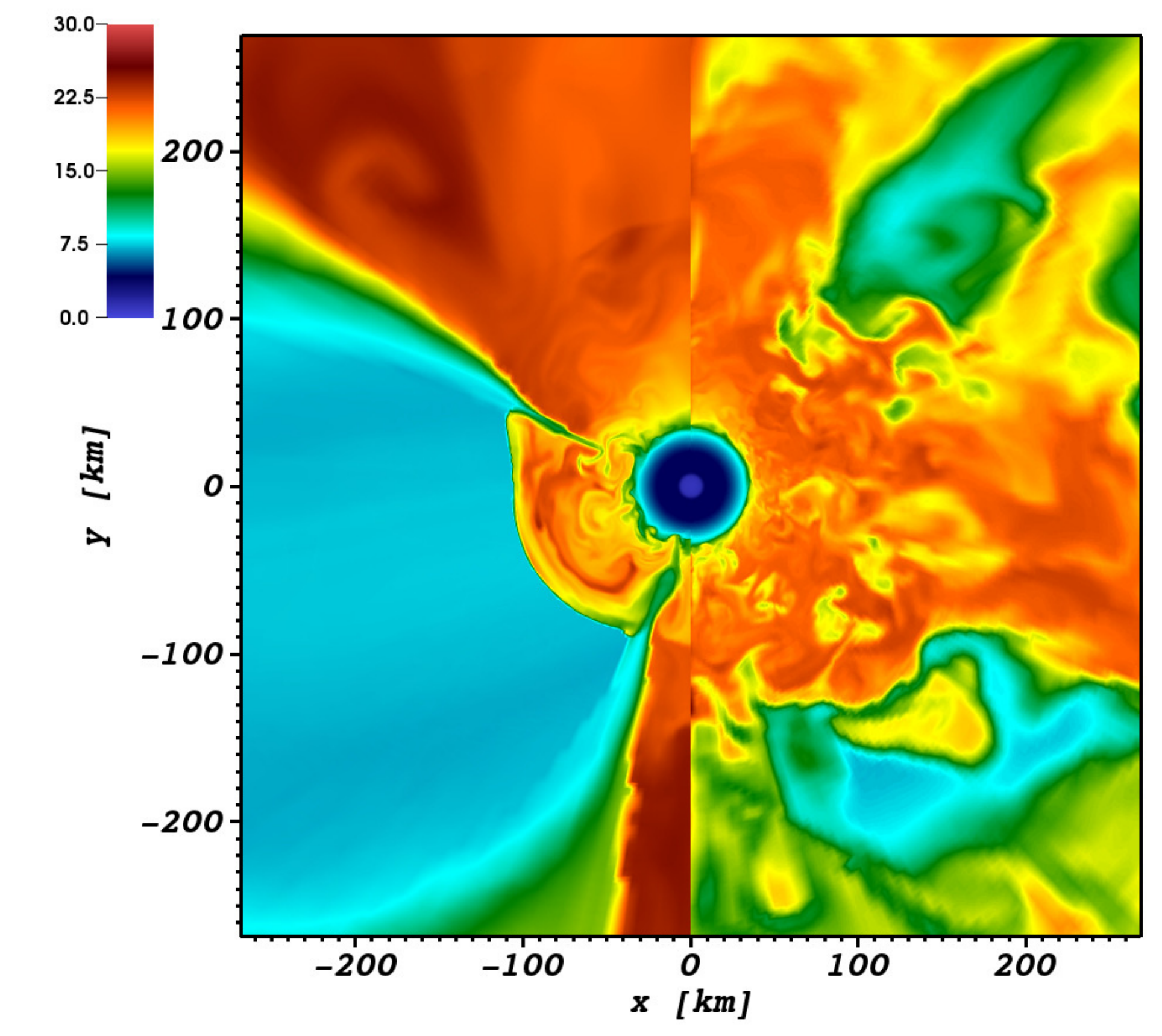}
\includegraphics[width=0.4\linewidth]{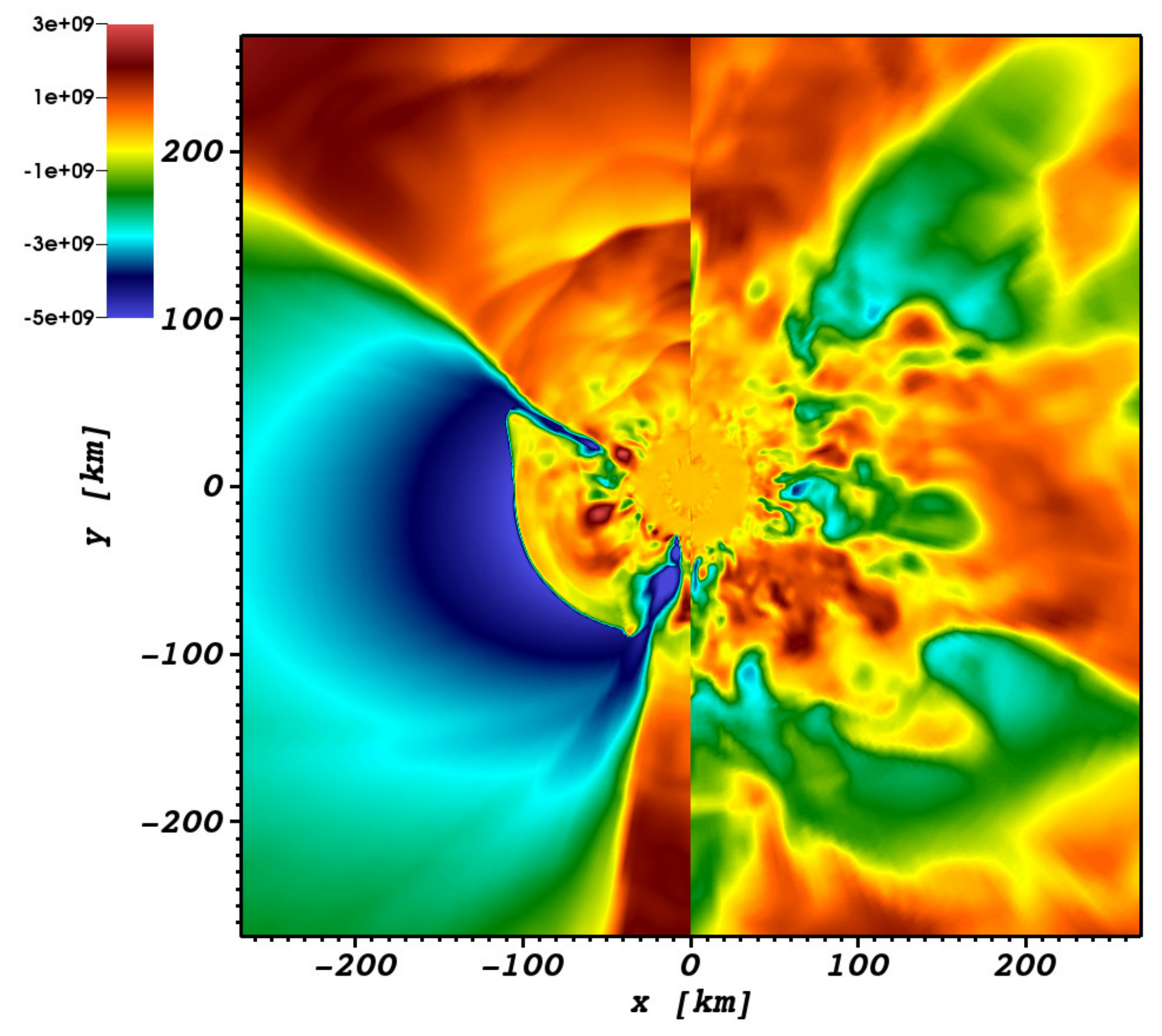}
\\
\includegraphics[width=0.4\linewidth]{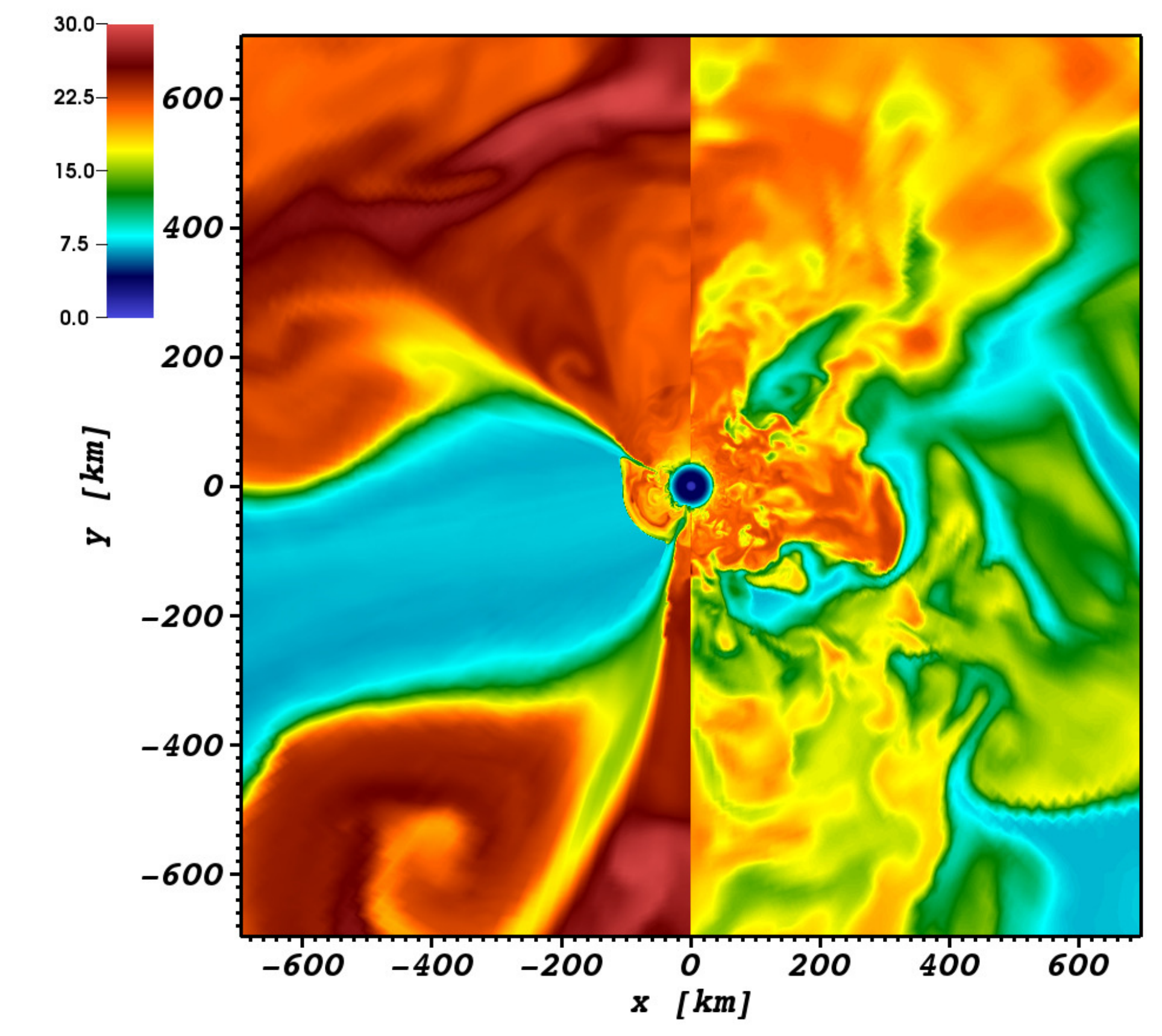}
\includegraphics[width=0.4\linewidth]{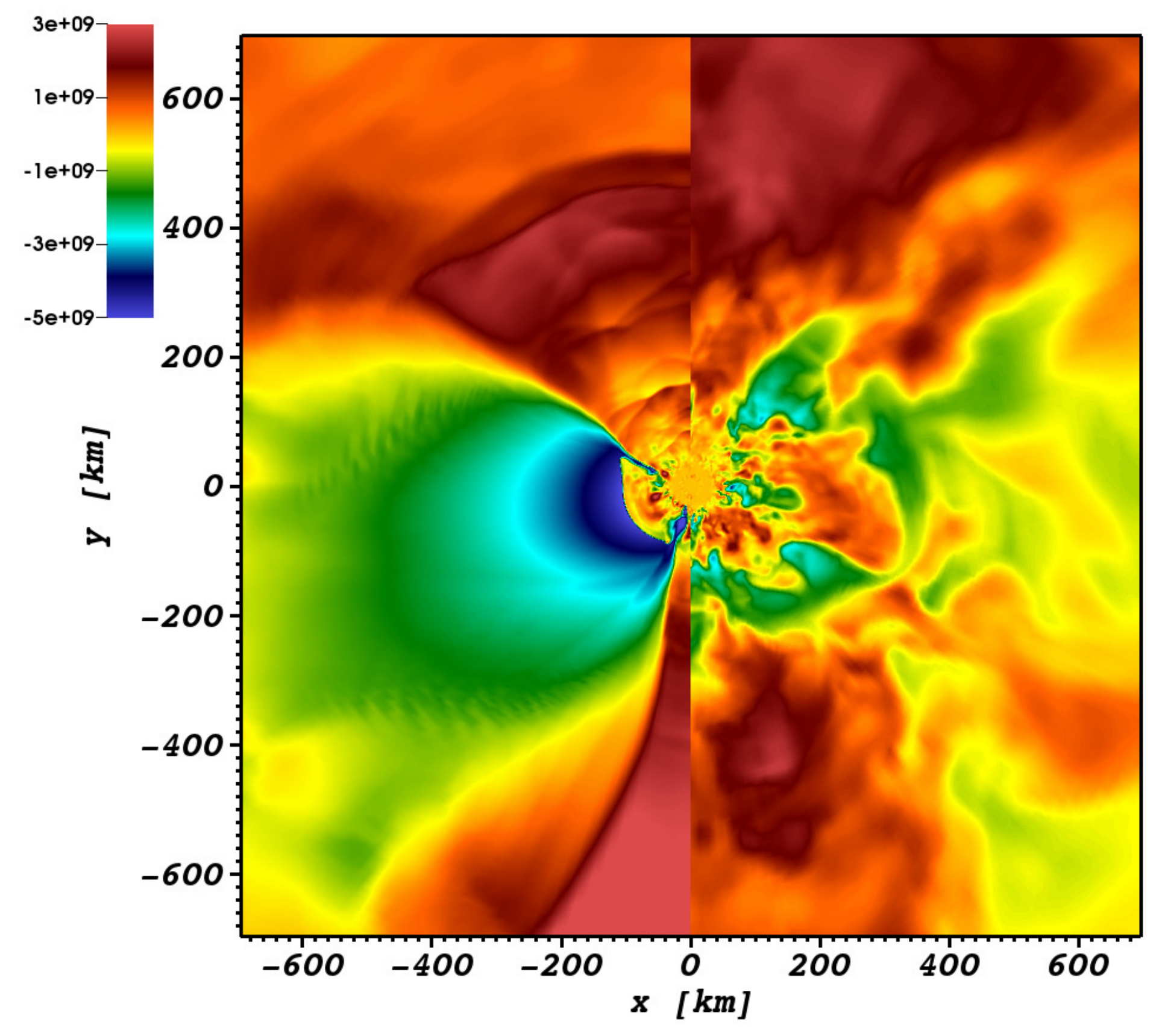}
\\
\includegraphics[width=0.4\linewidth]{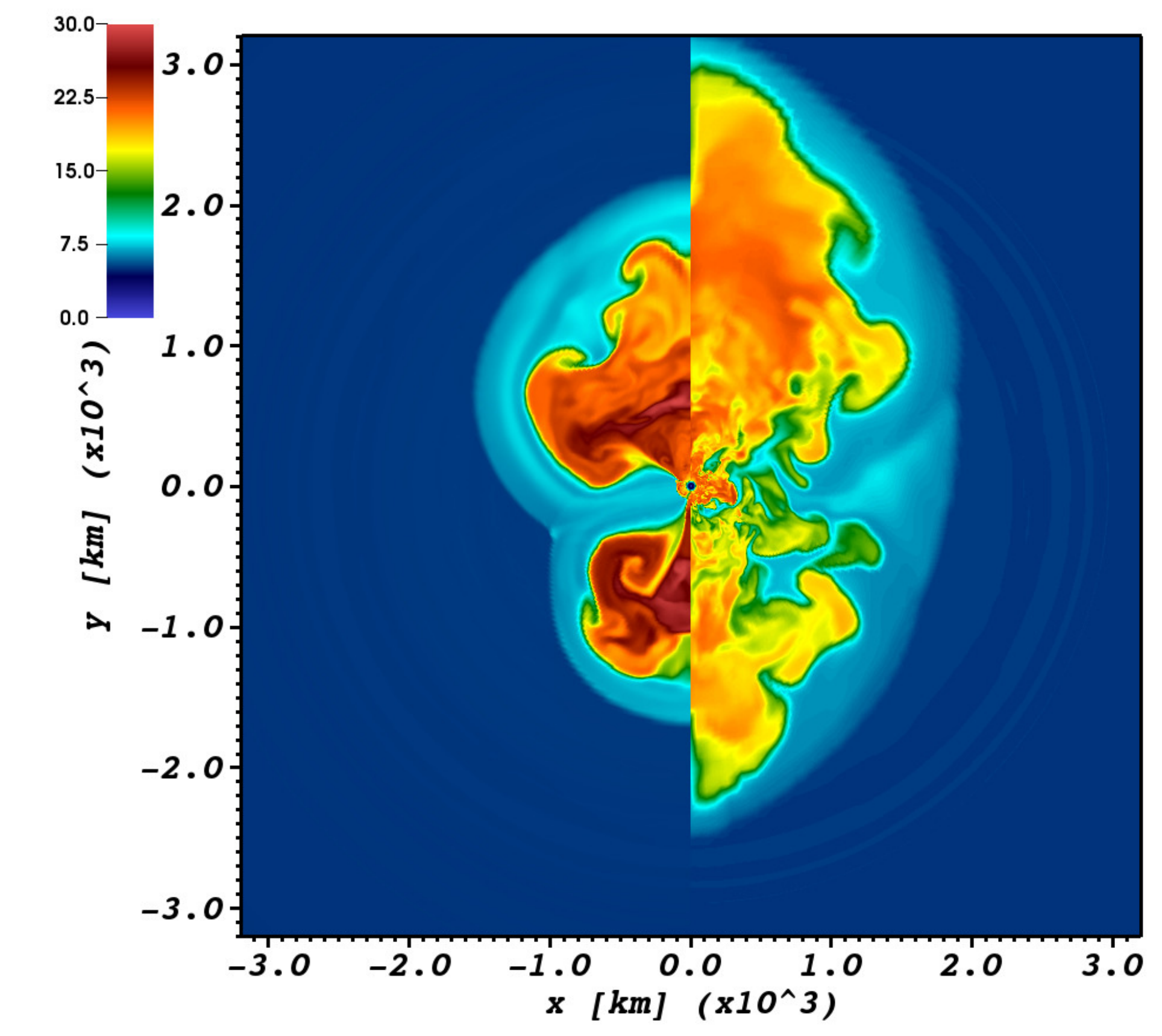}
\includegraphics[width=0.4\linewidth]{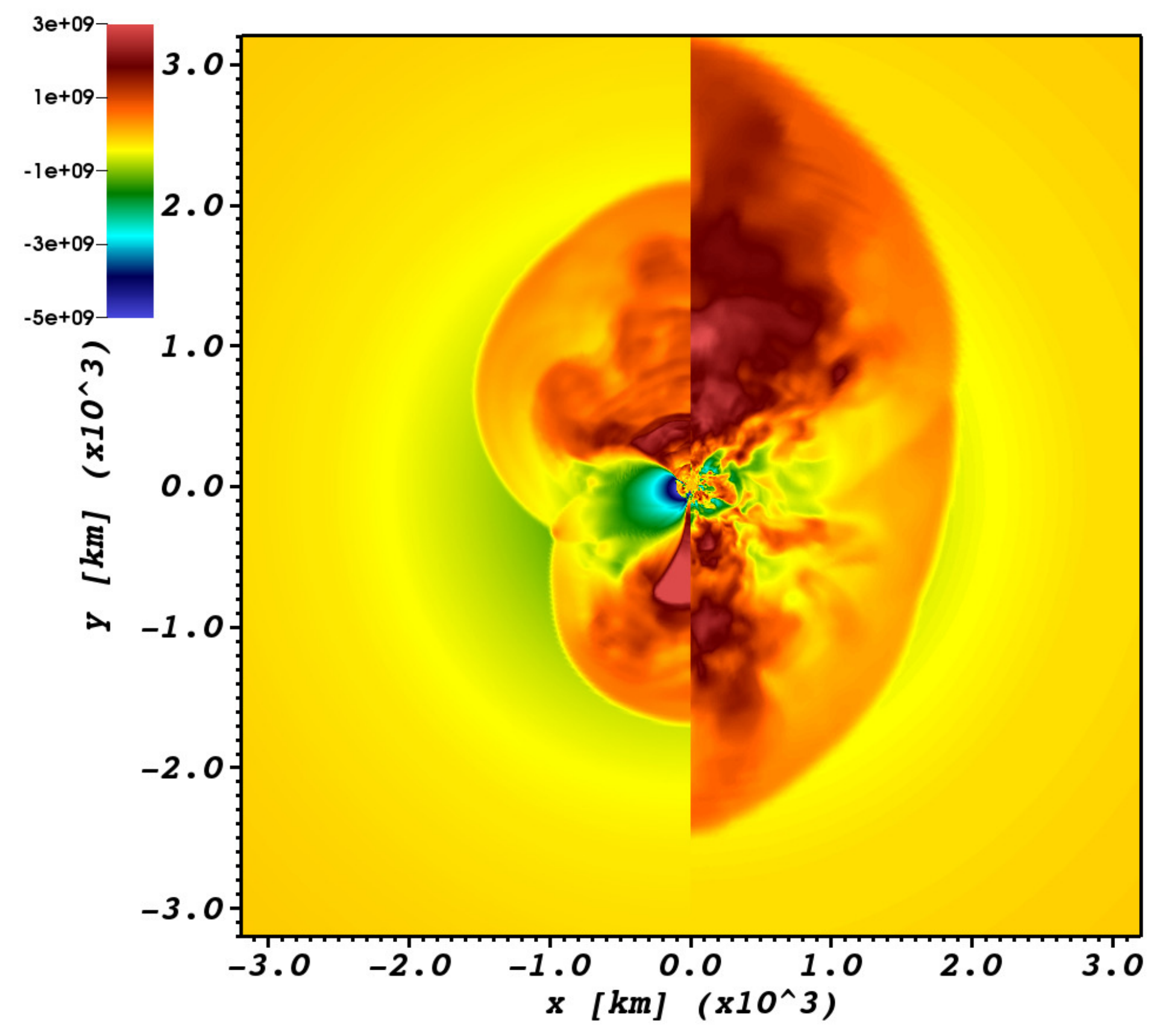}
\caption{Snapshots of 
the specific entropy $s$ (left column, measured in $k_b /\mathrm{nucleon}$)
and  the radial velocity
$v_r$ (right column, measured in $\mathrm{cm} \ \mathrm{s}^{-1}$) 
for model s11.2\_2Da (left halves of
the individual panels) and for a slice of model s11.2\_3D
(right halves) at a post-bounce time of $400 \ \mathrm{ms}$.
The same data is shown in all plots, only the zoom level is different.
Note the broad equatorial accretion downflow and the formation
of a secondary accretion shock at a radius
of $\mathord{\sim}100 \ \mathrm{km}$ in 2D.
\label{fig:snap400}}
\end{figure*}

In Figure~\ref{fig:outflow_rate}, we show $\dot{M}_\mathrm{out}$ and
the average total enthalpy and energy $\bar{h}_\mathrm{tot}$ and
$\bar{e}_\mathrm{out}$ (defined as
$\bar{h}_\mathrm{out}=F_{h,\mathrm{out}}/\dot{M}_\mathrm{out}$ and
$\bar{e}_\mathrm{out}=F_{e,\mathrm{out}}/\dot{M}_\mathrm{out}$, and
excluding rest mass contributions) in the outflows at a radius of $400
\ \mathrm{km}$ as a function of time.  This radius has been chosen
because recombination into $\alpha$-particles, which roughly sets the
final mass-specific total energy in the ejecta \citep{scheck_06}, is
already complete at this point, so that $F_{h,\mathrm{out}}$ roughly
represents the rate at which net total energy is pumped into the
ejecta region assuming steady-state conditions (i.e.\ small variations
of $F_{h,\mathrm{in}}$ with time and radius). This is indeed a very
good approximation as the comparison for $F_\mathrm{h,out}$ with the
time derivative $\ud E_\mathrm{expl}/\ud t$ of the explosion energy as
shown by the middle panel in Figure~\ref{fig:outflow_rate}.  $\ud
E_\mathrm{expl}/\ud t$ correlates extremely well with
$F_{h,\mathrm{out}}$, but is slightly smaller. The difference is due to
the accumulation of shocked material with slightly negative total
energy and energy exchange with the downflows due to turbulent
diffusion. The similarity of $\ud E_\mathrm{expl}/\ud t$ and
$F_{h,\mathrm{out}}$ also indicates that explosive nuclear burning does
not play a major role for the $11.2 M_\odot$ model in agreement with
earlier 2D simulations with the \textsc{Vertex-CoCoNuT} code
\citep{mueller_12a}.

On average, the total enthalpy flux into the ejecta region is larger
in 3D than in 2D as expected from the different evolution of the
explosion energy.  Interestingly, the relative difference between 2D
and 3D in the mass outflow rate $\dot{M}_\mathrm{out}$ is even larger
than for $F_\mathrm{h,out}$.  The smaller outflow rate in 2D is
partially compensated by a larger average mass-specific total enthalpy
and energy in the outflows, which can be \emph{larger} than the
recombination energy ($7\ldots 8.8 \ \mathrm{MeV}/\mathrm{nucleon}$).
Thus, care must be exercised in explaining differences between 2D and
3D based on the mass outflow rate or the total mass in the gain region
alone \citep{scheck_06,melson_15a} assuming the same contribution to
the explosion energy per unit mass from nucleon recombination into
$\alpha$-particles and heavy nuclei irrespective of the
dimensionality. That assumption would require similar average
enthalpies and energies in the outflows in 2D and 3D, which is clearly
not the case in general; the differences can be as large as several
$\mathrm{MeV}/\mathrm{nucleon}$. Recombination still sets the scale
for the asymptotic total energy per unit mass of neutrino-heated
ejecta, but hydrodynamic effects modify its precise value in 2D and 3D
in different directions as we shall explain below.

These differences are all the more astonishing because the
volume-integrated neutrino heating rate $\dot{Q}_\mathrm{heat}$ in the
gain region (Figure~\ref{fig:heating_cooling}) is very similar in 2D
and 3D. Especially at late times, $\dot{Q}_\mathrm{heat}$ is
consistently \emph{higher in 2D} than in 3D (as is the time-integrated
neutrino energy deposition). Assuming that the outflow rate is
determined by the total heating rate and the binding energy
$e_\mathrm{gain}$ at the gain radius as
$\dot{M}_\mathrm{out}\sim \dot{Q}_\mathrm{heat}/|e_\mathrm{gain}|$, the
lower outflow rate in 2D suggests that the material at the gain radius
is more strongly bound in this case. This is borne out by
Figure~\ref{fig:gain_radius}, which shows that the binding energy at
the gain radius is larger in 2D by up to a factor of $\mathord{\sim}2$
at late times. This is partly due a stronger recession of the gain radius
$r_\mathrm{gain}$ (bottom panel of Figure~\ref{fig:gain_radius}) as a
result of which the typical energy scale $G M/r_\mathrm{gain}$ ($M$
being the neutron star mass) at the gain radius is larger. However, it
is evident that this effect cannot fully account for the difference in
$e_\mathrm{gain}$. The small value of $e_\mathrm{gain}$ in 3D,
indicates that it is not determined by the gravitational energy scale
alone. Indeed, a much better estimate for the scale of
$e_\mathrm{gain}$ can be obtained if we suppose that the Bernoulli
integral (including rest masses) at the gain radius is roughly zero,
\begin{equation}
\epsilon+\frac{\mathbf{v}^2}{2}+\frac{P}{\rho}+\Phi=0.
\end{equation}
If we split the internal energy $\epsilon$ into a thermal component
$\epsilon_\mathrm{therm}$ and a rest mass contribution
$\epsilon_\mathrm{rm}$ (normalized to ${}^{56}\mathrm{Fe}$), and
assume vanishing velocities as well as  an ideal gas equation $P=\epsilon_\mathrm{therm}/3$ of state
for photons and non-degenerate relativistic electrons and positrons,
this leads to, 
\begin{equation}
\epsilon_\mathrm{therm}+\epsilon_\mathrm{rm}+\frac{P}{\rho}-\frac{G M }{r_\mathrm{gain}}=0,
\end{equation}
\begin{equation}
\frac{4}{3}\epsilon_\mathrm{therm}+\epsilon_\mathrm{rm}-\frac{G M }{r_\mathrm{gain}}=0,
\end{equation}
\begin{equation}
\epsilon_\mathrm{therm}=\frac{3}{4}\left(\frac{G M }{r_\mathrm{gain}}-\epsilon_\mathrm{rm}\right),
\end{equation}
where $M$ is the neutron star mass. For an electron fraction of
$Y_e=0.5$, $\epsilon_\mathrm{rm}$ would be identical to the
recombination energy of $\epsilon_\mathrm{rec} \approx 8.8
\ \mathrm{MeV}/\mathrm{nucleon}$ from protons and neutrons with equal
mass fractions into iron group elements, and for our purposes this
still provides a sufficient approximation even though we have
$Y_e<0.5$ at the gain radius. For the binding energy $e_\mathrm{gain}$
(in which we excluded rest masses), we thus obtain
\begin{equation}
\label{eq:egain}
e_\mathrm{gain}
\approx
\epsilon_\mathrm{therm}-\frac{G M }{r_\mathrm{gain}}
\approx -\frac{3}{4}\epsilon_\mathrm{rec}-\frac{GM}{4r_\mathrm{gain}}.
\end{equation}
As shown in Figure~\ref{fig:gain_radius}, this still overestimates
$|e_\mathrm{gain}|$ a bit, but accounts for the slow rise of
$|e_\mathrm{gain}|$ compared to the gravitational energy scale
$GM/r_\mathrm{gain}$ during the contraction of the neutron star.

While the different absolute value of the binding energy at the gain
radius is part of the explanation for the different mass outflow
rates, there is also an additional effect at play. In general, the
mass outflow rate will only be approximately given by
$\dot{M}_\mathrm{out}\sim \dot{Q}_\mathrm{heat}/|e_\mathrm{gain}|$,
and one can introduce an efficiency parameter
$\eta_\mathrm{out}$ to
compare the actual mass outflow rate with this fiducial
rate,
\begin{equation}
\label{eq:eta_out}
\eta_\mathrm{out}
=
\frac{|e_\mathrm{gain}| \dot{M}_\mathrm{out}}{\dot{Q}_\mathrm{heat}}.
\end{equation}
$\eta_\mathrm{out}$ is plotted in Figure~\ref{fig:eta_out}.
On average, the outflow efficiency $\eta_\mathrm{out}$
is also considerably larger in 3D (where it fluctuates
around $\eta_\mathrm{out}\approx 1$) than in 2D ($\eta_\mathrm{out}\approx 0.5$).

Our analysis of the outflows has thus revealed two reasons
for lower explosion energies in 2D: The mass loss rate,
(and hence the energy flux into the ejecta region)
is lower because the ejected material is initially
bound more tightly at the gain radius before being lifted
out of the gravitational potential well, and for a given
binding energy $e_\mathrm{gain}$ at the gain radius, the conversion
of neutrino heating into an outflow is less efficient.

\subsection{Causes for Weak Explosions in 2D}
These two effects, as well as the higher asymptotic energy per unit
mass in 2D can be traced to the constrained axisymmetric flow geometry
and a fundamentally different behaviour of the accretion downflows in
2D compared to 3D.  The different flow morphology is illustrated
qualitatively in Figure~\ref{fig:snap400}, which shows snapshots (for
different spatial scales) of the radial velocity and entropy in 2D and
3D for a representative post-bounce time of $400 \ \mathrm{ms}$.

\subsubsection{Morphology and Dynamics of Outflows and Accretion
Downflows in 2D and 3D}
\label{sec:outflows_downflows}
These snapshots reveal that the interface between the outflows and the
colder, low-entropy becomes turbulent due to the Kelvin-Helmholtz
instability in 3D, which distorts the downflows as they approach the
proto-neutron star (as already mentioned in
Section~\ref{sec:overview}), whereas Kelvin-Helmholtz instabilities
are noticeably absent at the shear interfaces between the broad
equatorial downflow and the polar bubbles in 2D (middle and bottom row
in Figure~\ref{fig:snap400}).  The tendency of the downflows to become
more turbulent in 3D has been recognized already by \citet{melson_15a}
in their $9.6 M_\odot$ model, although the morphological difference
between 2D and 3D is much more pronounced in a continuously accreting
model like the $11.2 M_\odot$ progenitor.  Moreover, although there
may be a deeper connection between the two phenomena, the stability of
the shear interfaces in 2D is likely due to a different reason than
the emergence of large-scale structures in the pre-explosion phase
\citep{hanke_12} that has been explained by the inverse turbulent
energy cascade in 2D \citep{kraichnan_76}.  Despite the inverse
turbulent cascade, subsonic shear layers/interfaces remain prone to
the Kelvin-Helmholtz instability in 2D; and 2D supernova simulations are
easily able to resolve the instability 
\citep{mueller_12b,fernandez_15} even without extraordinary
high resolution.

The situation changes, however, for the supersonic shear interfaces
between the downflows and the neutrino-heated bubbles that we
encounter during the explosion phase. Here, the classical growth rate
$\omega = k \Delta \, u/2$ (where $k$ is the wave vector, and $\Delta
u$ is the transverse velocity jump across the interface) in the vortex
sheet approximation for incompressible flow is no longer
applicable. Instead modes with a sufficiently high effective Mach
number $\mathrm{Ma} = \Delta u \cos \theta /c_\mathrm{s}$ (where $c_s$
is the sound speed and $\theta$ is the angle between the wave vector
and the vectorial velocity jump) are stabilized \citep{gerwin_68},
although the stability analysis is more complicated if finite-width
shear layers are considered
\citep{blumen_70,blumen_75,drazin_77,choudhury_84,balsa_90}.\footnote{It is noteworthy that laser-driven plasma experiments may be
  able to capture this effect and quantify the reduced growth or
  suppression of the Kelvin-Helmholtz instability in 2D
  \citep{malmud_13}.}  This implies that the Kelvin-Helmholtz
instability can be partially or completely suppressed in 2D (where
$\cos \theta =1$), while there are always unstable modes in 3D since
$\cos \theta$ can be arbitrarily small.\footnote{Loosely
  speaking, a small value of $\cos \theta$ guarantees that sound waves
  on either side of the vortex sheet can propagate in both directions along the
  wave vector $\mathbf{k}$ of a given perturbation mode to mediate the
  pressure feedback required for the growth of the Kelvin-Helmholtz
  instability: For a given vectorial velocity $\pm \mathbf{u}/2$ of
  the fluid on either side of the interface, the sound waves with
  direction $\mathbf{n}$ and velocity $c_s$ in either of the fluids
  will have a velocity component $(\pm \mathbf{u}/2+\mathbf{n}
  c_s)\cdot \mathbf{k}/|\mathbf{k}| =\pm u/2 \cos \theta + c_s
  \mathbf{n}\cdot \mathbf{k}/|\mathbf{k}|$ along the direction of
  $\mathbf{k}$ in the rest frame.  If $\cos\theta$ is sufficiently
  small, this velocity component can take on either sign depending on
  $\mathbf{n}$ in both fluids. In 2D, we always have
$\pm \mathbf{u}/2 \cdot \mathbf{k}/|\mathbf{k}|=u/2$, and sound waves
cannot propagate in both directions any longer for sufficiently large
$u$.  Note that this is only a heuristic
  explanation that cannot predict the critical Mach number correctly;
  Section~B in \citet{gerwin_68} and the other aforementioned
  references should be consulted for a rigorous derivation of the
  dispersion relation.  } In principle, it is conceivable that
numerical diffusivity and viscosity further help to suppress the
Kelvin-Helmholtz instability more strongly and earlier than the
physics might dictate, but the fact that the instability evidently
operates in the 3D model provides evidence that the numerical
resolution cannot be faulted for the behavior of the 2D
models. Furthermore, the stability of the accretion downflows is a
persistent feature even in high-resolution 2D models with continuing
accretion \citep{bruenn_14}; it is thus without doubt physical in
origin.

\begin{figure}
\includegraphics[width=\linewidth]{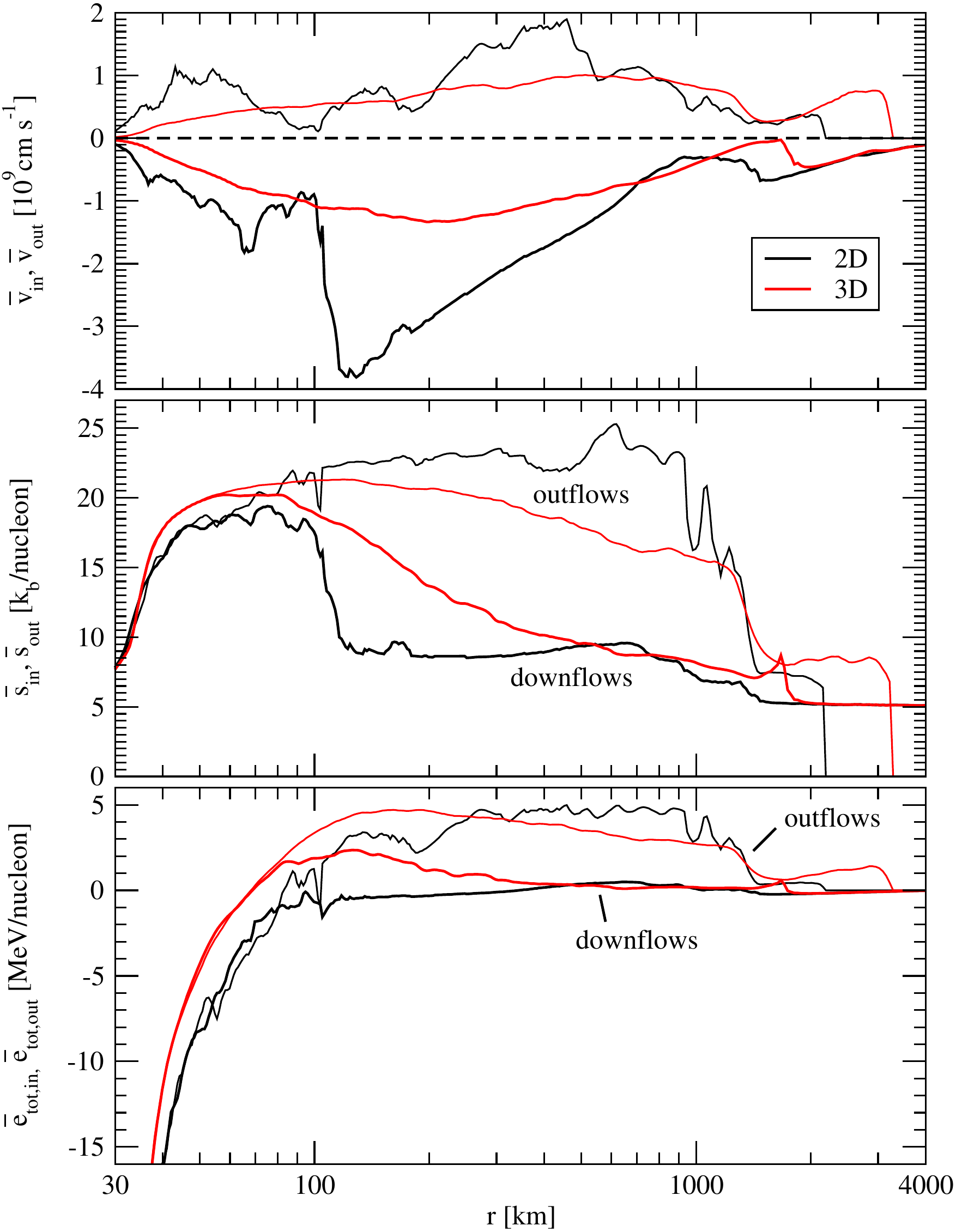}
\caption{Radial profiles of the average velocity (top panel), entropy
(middle panel), and total
energy (bottom) per nucleon in the outflows (thin lines) and downflows
(thick lines)
in 2D (black) and 3D (red) at a post-bounce
time of $400 \ \mathrm{ms}$. Note that rest mass contributions
are included in the total energy here.
\label{fig:more_profiles}}
\end{figure}

\begin{figure}
\includegraphics[width=\linewidth]{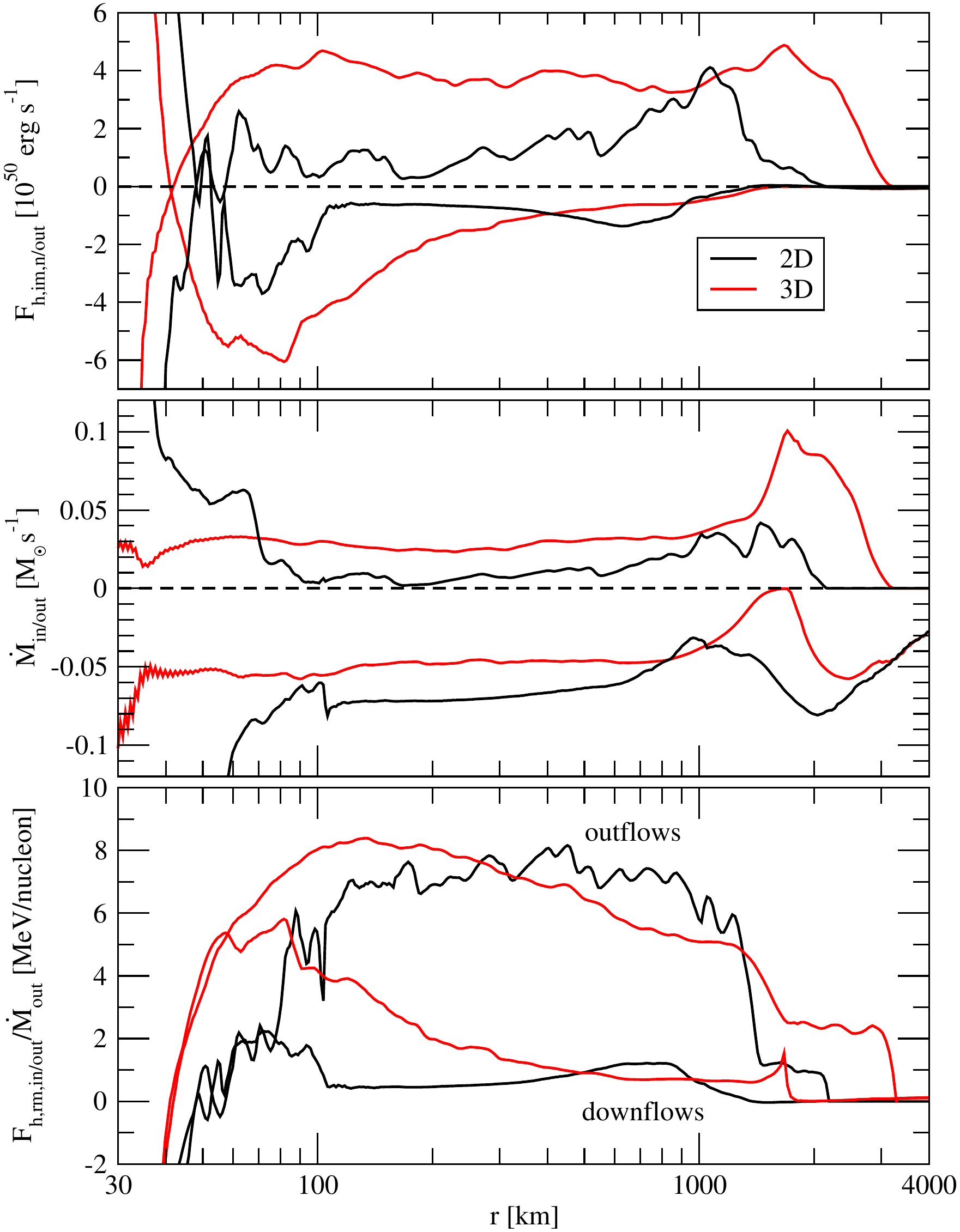}
\caption{Overview of the mass and energy fluxes in the outflows and downflows in 2D. The top panel shows the total enthalpy fluxes
$F_{h,\mathrm{rm},\mathrm{in/out}}$ in the outflows
(positive values) and downflows (negative values)
in 2D (black) and 3D (red) at a post-bounce time of $400 \ \mathrm{ms}$.
The middle panel shows the mass inflow/outflow rates $\dot{M}_\mathrm{in/out}$,
and the bottom panel shows the average flux-weighted
total enthalpies $F_{h,\mathrm{rm},\mathrm{in/out}}/\dot{M}_\mathrm{in/out}$.
Note that rest mass contributions are included in the
total entropy here, unlike in Figures~\ref{fig:outflow_rate}
and \ref{fig:more_profiles}.
\label{fig:profile_fluxes}
}
\end{figure}

The ``turbulent braking'' of the downflows in 3D is reflected
quantitatively in radial profiles of the average velocity
$\bar{v}_\mathrm{in/out}$, entropy $\bar{s}_\mathrm{in/out}$ and
mass-specific total energy $\bar{e}_\mathrm{tot,rm,in/out}$ of the downflows
and outflows. We define these quantities as density-weighted
averages (denoted by bars) of the radial velocity $v_r$, the
specific entropy $s$, and the total energy
$e_\mathrm{tot}$ as
\begin{equation}
\bar{v}_\mathrm{in/out}
=
\frac{\int_{v_r \lessgtr 0}  \rho v_r \,  \ud \Omega}
{\int_{v_r \lessgtr 0}  \rho \, \ud \Omega}
\end{equation}
\begin{equation}
\bar{s}_\mathrm{in/out}
=
\frac{\int_{v_r \lessgtr 0}  \rho s\,  \ud \Omega}
{\int_{v_r \lessgtr 0}  \rho \, \ud \Omega}
\end{equation}
\begin{equation}
\bar{e}_\mathrm{tot,in/out}
=
\frac{\int_{v_r \lessgtr 0}  \rho e_\mathrm{tot,rm}\, \ud \Omega}
{\int_{v_r \lessgtr 0}  \rho \, \ud \Omega},
\end{equation}
and show the results for models s11.2\_2Da and s11.2\_3D at a
post-bounce time of $400 \ \mathrm{ms}$ in
Figure~\ref{fig:more_profiles}. Moreover, we consider radial profiles
of the mass and total enthalpy fluxes $\dot{M}_\mathrm{in/out}$ and
$\dot{F}_\mathrm{h,rm,in/out}$ in both streams in
Figure~\ref{fig:profile_fluxes}; these are computed according to
Equation~(\ref{eq:dotm_in_out}) and (\ref{eq:ent_flux}).
However, for computing radial profiles we include rest mass
contributions in the total energy and the total enthalpy flux (as
denoted by the additional subscript $\mathrm{rm}$). This definition
has the advantage that both $\dot{M}_\mathrm{in,out}$ and
$\dot{F}_\mathrm{h,rm,in/out}$ are constant in the limit of stationary
streams without mass, energy, and momentum exchange,
so that changes in these fluxes serve as useful indicators for lateral
mixing between the outflows and downflows.

 Due to turbulent braking (i.e.\ by an
effective turbulent eddy viscosity), the average infall velocity in
the downflows reaches only $1.4 \times 10^9 \ \mathrm{cm}
\ \mathrm{s}^{-1}$ in 3D, and decreases in magnitude once the
downflows penetrate further down than a radius of $\mathord{\sim} 200
\ \mathrm{km}$.  By contrast, the downflows reach a sizable fraction
of the free-fall velocity in 2D before they are abruptly decelerated
at a secondary accretion shock at $r \approx 100 \ \mathrm{km}$.
However, the outflow velocities are also higher in 2D.

During the phase considered here, the entropy of the downflows does
not vary considerably in 2D between $r \approx 100 \ \mathrm{km}$ and
$r \approx 1000\ \mathrm{km}$ (where the equatorial downflow forms
from two converging lateral flows). Similarly, the ``flux-averaged
enthalpy''
$F_\mathrm{h,\mathrm{rm},\mathrm{in/out}}/\dot{M}_\mathrm{in/out}$
(bottom panel of Figure~\ref{fig:profile_fluxes}) does not change
appreciably in this region.  This is a further indication for a lack
of dissipation by turbulent eddy viscosity and of lateral mixing
between the downflows and outflows. By contrast, the slope in
$\bar{s}_\mathrm{in}$ and $\bar{s}_\mathrm{out}$ and the
``flux-averaged'' total enthalpy
$F_\mathrm{h,\mathrm{rm},\mathrm{in/out}}/\dot{M}_\mathrm{in/out}$
points to lateral mixing between the downflows and outflows in 3D.
The increase of
$F_\mathrm{h,\mathrm{rm},\mathrm{in/out}}/\dot{M}_\mathrm{in/out}$ in
the downflows during the infall from the shock is mirrored by a
decrease of
$F_\mathrm{h,\mathrm{rm},\mathrm{in/out}}/\dot{M}_\mathrm{in/out}$
with radius in the ouflows due to turbulent mixing of the hot ejecta
with colder matter from the downflows: This explains why the
neutrino-heated ejecta contribute only $\mathord{\sim} 5 \ldots 6
\ \mathrm{MeV}/\mathrm{nucleon}$ to the diagnostic explosion energy,
instead of the $7\ldots 8.8 \ \mathrm{MeV}/\mathrm{nucleon}$ available
from nucleon recombination.

There is thus ample evidence that turbulent viscosity and diffusion
brake the accretion funnels and transfer energy from the
outflows. It is tempting to invoke this as an explanation for the
lower binding energy at the gain radius in 3D
(Figure~\ref{fig:gain_radius}). At first glance, the fact that
\emph{both} the downflows and outflows are less strongly bound in 3D at
the bottom of the gain layer (bottom panel of
Figure~\ref{fig:more_profiles}) may seem to conflict with this
assumptions, but the lower binding energies of the outflows at small
radii is to be expected because the supply for outflow comes from
freshly accreted matter that has undergone turbulent braking in the
downflows. Turbulent braking and turbulent diffusion are perfectly
acceptable explanations for \emph{internal} energy distribution within
the gain region (but not the higher enthalpy
flux into the ejecta region, see below), and may thus account for the lower
$|e_\mathrm{gain}|$ in 3D and hence for the higher mass outflow rates.

Moreover, the turbulent braking in 3D may have implications for the
final neutron star masses. In Section~\ref{sec:explosion_properties},
we assumed that the ``mass cut'' occurs roughly when the shock accelerates the
newly swept-up material to the escape velocity. Without efficient
lateral momentum transfer and without pressure support from an
expanding hot bubble from below, it seems inevitable that material
over existing downflows must eventually fall onto the neutron star if
this condition is not met. Since the free-fall time scale at radii of
several thousands of kilometers (where the initial mass cut estimated
in Section~\ref{sec:explosion_properties} is located) is of the order
of seconds, accretion must necessarily last over a correspondingly long
duration. On the other hand, if the downflows are braked by a
turbulent eddy viscosity below a certain radius in 3D, slowly moving
matter in the downflows may instead be entrained by the high-entropy
bubbles in regions where the angular-averaged velocity is already
positive, so that the residual accretion onto the neutron star may
cease earlier and the total mass accreted during the explosion phase
may be considerably lower than estimated in
Section~\ref{sec:explosion_properties}. Longer 3D simulations will be
necessary to investigate this hypothesis.

However, the internal redistribution of energy within the gain region
 in 3D \emph{cannot} account for the significantly higher
total enthalpy flux in the outflows and the faster rise of the
explosion energy: The higher outflow rate will come at the expense of
energy loss from the outflows to the downflows -- the overall
conservation law cannot be cheated. Consequently, there must be
additional mechanisms that remove energy from the gain region in 2D
and reduce the outflow efficiency $\eta_\mathrm{out}$
(Figure~\ref{fig:eta_out}) compared to 3D. The different dynamics of
the outflows and downflows nonetheless remains a crucial element of
the explanation because it provides the basis for three mechanisms
discussed in the subsequent sections.

\subsubsection{Energy Loss by Wave Excitation}
\label{sec:waves}
\begin{figure}
\includegraphics[width=\linewidth]{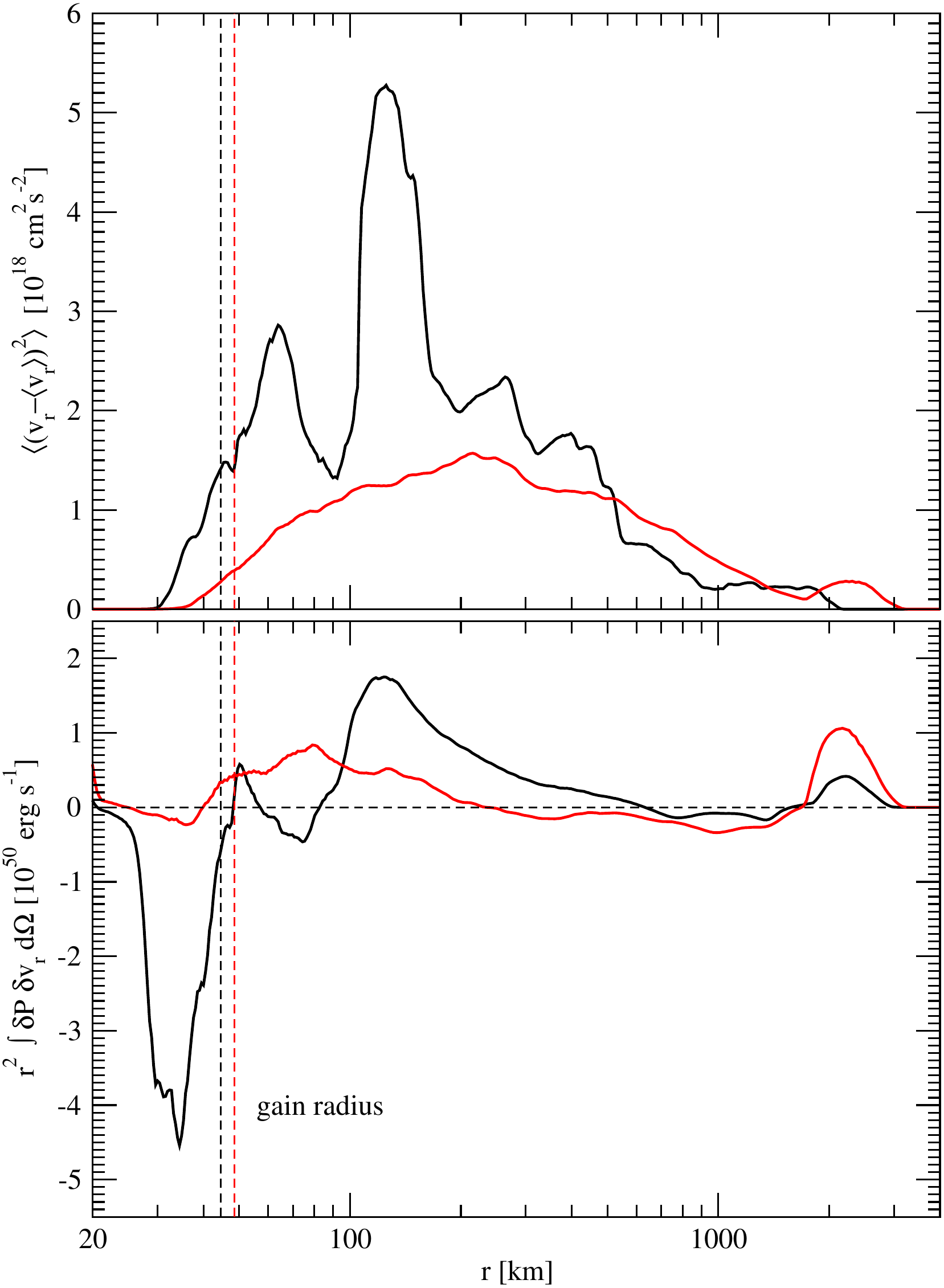}
\caption{Top: Radial velocity dispersion 
$\langle (v_r-\langle v_r\rangle)^2 \rangle$
in 2D and 3D at a post-bounce time of $400 \ \mathrm{ms}$.
Bottom: Radial profiles of the ``acoustic'' energy flux
$r^2\int \delta P \, \delta v_r \, \ud \Omega$
in 2D and 3D at a post-bounce time of $400 \ \mathrm{ms}$.
The curves show temporal averages over several time steps.
\label{fig:wave_excitation}}
\end{figure}

The lack of turbulent braking in 2D
implies that the accretion funnels either hit the neutron star
directly with a high impact velocity or are decelerated abruptly in a
secondary accretion shock (top row of Figure~\ref{fig:snap400}).  Even
in the latter case, thin accretion funnels still penetrate the hot,
neutrino-heated matter all the way down to the gain radius, cutting
off a confined high-entropy bubble from the outflows. In the snapshots
shown in Figures~\ref{fig:snap400} and \ref{fig:pns_close_up} (with an
even higher zoom level), these narrow downflows strike the
proto-neutron star surface with velocities of up to $6 \times 10^{9}
\ \mathrm{cm} \ \mathrm{s}^{-1}$. As a result they overshoot
considerably into the convectively stable cooling layer and excite
strong wave activity.  The emission of strong acoustic waves that
steepen into shocks and then dissipate is immediately evident from the
top right panel of Figure~\ref{fig:snap400}, but the deceleration of
the downflows also excites strong $g$-modes in the neutron star
surface region (a phenomenon that has been thoroughly analyzed in the
context of gravitational wave emission,
cp.\ \citealt{marek_08,murphy_09,mueller_13}). In 3D, overshooting is
much less pronounced, and so is the excitation of acoustic waves and
$g$-modes. This can be seen from the conspicuous absence of
sawtooth-like features in the velocity and by considering the radial
velocity dispersion $\langle (v_r-\langle v_r\rangle)^2 \rangle$,
which is significantly smaller in 3D below the gain radius (top panel
of Figure~\ref{fig:wave_excitation}).  This result is in agreement
with other 3D simulations of supernova explosions using
self-consistent neutrino transport \citep{melson_15a} and parameterized
neutrino heating \citep{murphy_12,handy_14} and linear theory, which
suggests that the excitation of waves ($g$-modes in particular)
at convective boundaries
is strongly sensitive to the convective Mach number $\mathrm{Ma}$
and becomes very efficient for $\mathrm{Ma}\sim 1$ 
\citep{goldreich_90,lecoanet_12}.

What has been missed so far, however, is that the excitation of
$g$-modes constitutes a \emph{non-advective energy drain} in 2D; it
transports energy from the lower layers of the gain regions deep into
the cooling region \emph{without} the need to transport mass. If the
$g$-mode energy flux is sufficiently high, it provides a very likely
explanation for the permanently higher binding energy at the gain
radius in 2D. Unfortunately, the $g$-mode energy flux in our
simulations cannot readily be quantified; this would not only require
performing a full spherical Reynolds decomposition, but also detailed
knowledge about the dispersion relation of the $g$-modes, which is
beyond the scope of this paper. However, since the transfer of the
kinetic energy from the downflows into $g$-modes involves $P\ \ud
V$-work onto the neutron star surface and
any turbulent energy flux into deeper layers should show
up in correlated pressure and velocity fluctuations in
a transition layer between the convective gain region and
the convectively stabilized cooling layer, we can formulate a crude
estimate for the $g$-mode flux by computing what is nominally an
acoustic luminosity, namely,
\begin{equation}
L_{P\ \ud V}=r^2\int \delta P\ \delta v_r \, \ud \Omega,
\end{equation}
where $\delta P$ and $\delta v_r$ denote the deviations of the
pressure and radial velocity from their respective angular
averages. The resulting estimates for the flux are shown in the bottom
panel of Figure~\ref{fig:wave_excitation} and point to a sizable
energy flux of the order of several $10^{50} \ \mathrm{erg} \ \mathrm{s}^{-1}$ from the
vicinity of the gain radius into the deeper layers of the
proto-neutron star surface (carried by $g$-modes) and to the outer
regions of the gain layer (carried by acoustic waves).  Such large
fluxes are comparable to the typical total enthalpy flux in the
outflows and even to the volume-integrated neutrino heating rate,
and therefore need to be accounted for in the total energy budget
of the gain region.

Incidentally, the excitation of acoustic waves also provides an
explanation for the high entropy (Figure~\ref{fig:more_profiles}) and
total enthalpy (Figures~\ref{fig:outflow_rate} and
\ref{fig:profile_fluxes}) in the outflows in 2D, which can still
increase somewhat at radii where neutrino heating is essentially
irrelevant.  The dissipation of the the acoustic waves in the
expanding hot bubble helps to increase the total energy and entropy in
the ejecta region beyond the $\mathord{\sim} 7 \ldots 8.8 \ \mathrm{MeV}$
available from nucleon recombination, but since these waves carry only
part of the energy lost due to by the downflows due to their
interaction with the convective boundary this effect cannot compensate
for the lower mass outflow rate in 2D, and the net effect of wave
excitation in 2D remains a detrimental one. 

The importance of wave excitation at the convective boundary will
inevitably vary between different 2D models depending on the explosion
geometry and the duration of accretion. If neutrino heating is strong,
the explosion energy rises steeply, and shock expansion is fast as in
the models of \citet{bruenn_13,bruenn_14}, it may play a less
prominent role. The formation of secondary accretion shocks as in our
2D models is likely to increase the energy loss by wave excitation
tremendously because the efficiency of this process also depends on
the frequency overlap between the convective forcing and the excited
modes \citep{goldreich_90,lecoanet_12}. The formation of a secondary
accretion shock provides for fluctuations with typical
frequencies inversely proportional to the short sound-crossing
time-scale (of the order of milliseconds) in the confined bubble.
Furthermore, the stochastic forcing of $g$-modes in 2D by
one or two strong downflows is presumably
also more efficient than in 3D, where there are more smaller
and uncorrelated downflows.

Finally, we comment on similarities and differences between $g$-mode
and acoustic wave excitation between our models and the
acoustically-driven explosion models of \citet{burrows_06,burrows_07},
where a strong flux acoustic waves, excited by an $\ell=1$ \emph{core}
$g$-mode with amplitudes of several kilometers, is responsible for
shock expansion in the first place.  Our 2D models are similar to
those of \citet{burrows_06,burrows_07} only in the sense that energy
deposition by acoustic waves contributes to the growth of the
explosion energy, but different from their simulations the acoustic
contributions remains subdominant compared to the volume-integrated
neutrino heating rate, which is more than four times larger at the
time shown in Figure~\ref{fig:wave_excitation} (a constellation that
\citealt{burrows_07} anticipated when postulating a ``hybrid
mechanism'' with combined heating by neutrinos and acoustic
waves). Moreover, the excitation mechanism for acoustic waves is
genuinely different in our case; they are excited \emph{directly} by
the interaction of the downflows with the convectively stable neutron
star surface layer without the need to channel the accretion power
through an $\ell=1$ core $g$-mode as a ``transducer'' as in the models
of \citet{burrows_06,burrows_07}. Such a large-amplitude core $g$-mode
is not found in our simulations (and could not have arisen simply
because of the spherically symmetric treatment of the neutron star
core), and the outer $g$-modes of rather modest amplitude excited in
our models could not act as an efficient transducer in the vein of
\citet{burrows_06,burrows_07} due to neutrino losses (see below).  It
is interesting to note that direct excitation at the convective
boundary still allows acoustic waves to contribute (albeit at a minor
level) to the explosion energy in 2D even without such a transducer.

Acoustic energy deposition also remains a secondary effect insofar as
this direct excitation mechanism works efficiently only \emph{after}
the onset of the explosion once the typical Mach number of the
downflows is sufficiently high. Moreover, contrary to
\citet{burrows_06,burrows_07} the net effect of wave excitation in our
models is still harmful because the power pumped into \emph{outer}
$g$-modes constitutes an energy drain that outweighs the rate of
energy deposition by acoustic waves by far. Different from their
models where the energy in the core $g$-mode is eventually
``recycled'' into an acoustic energy flux that drives shock expansion,
the energy pumped into the outer $g$-mode is manifestly lost due to
neutrino cooling in our case, and the mode coupling analysis of
\citet{weinberg_08} suggests that this should also happen if the core
$g$-mode excited due to non-linear mode coupling.  

\subsubsection{Steric Hindrances}
\label{sec:steric}
\begin{figure*}
\includegraphics[width=0.49\linewidth]{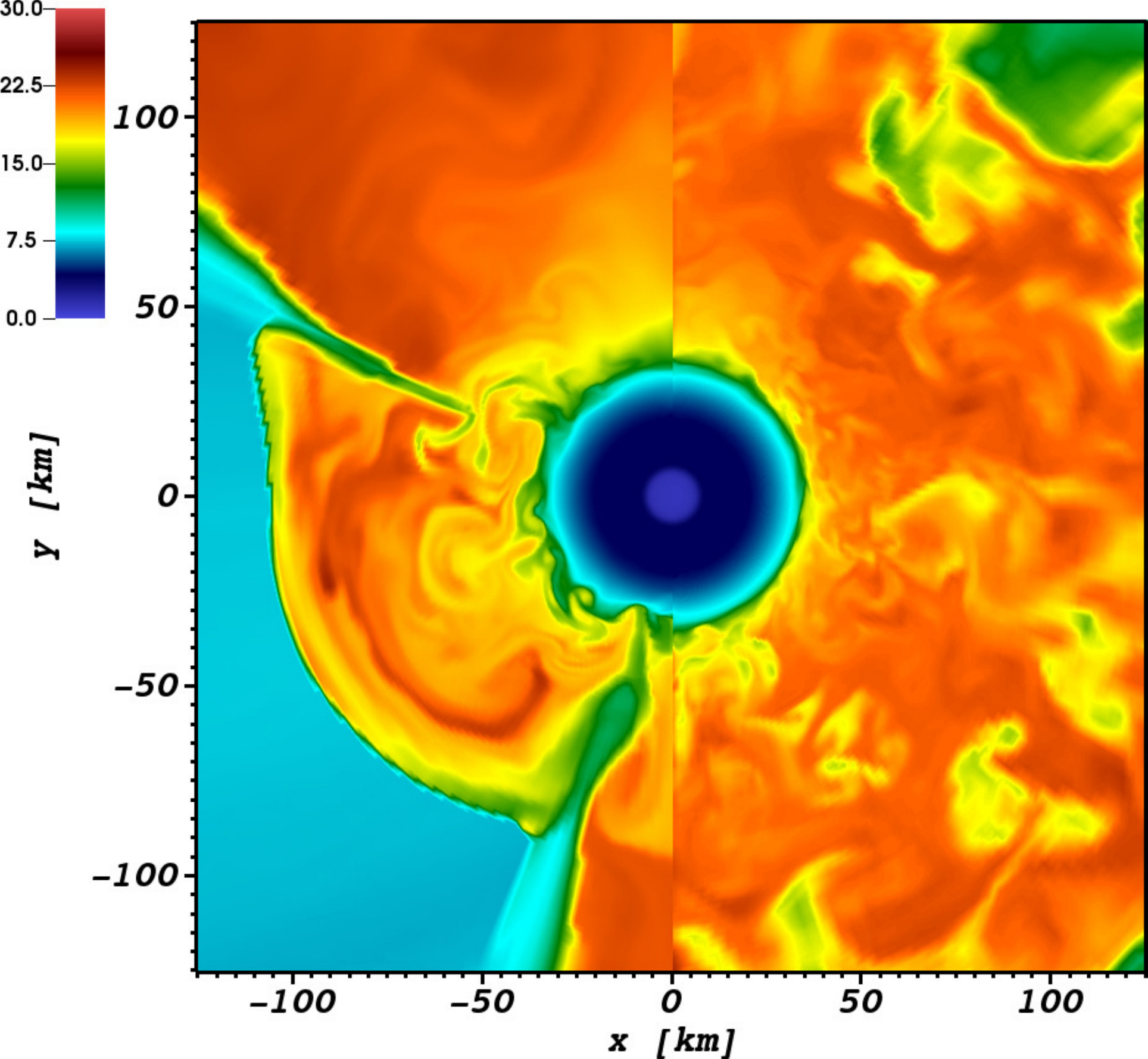}
\hfill
\includegraphics[width=0.49\linewidth]{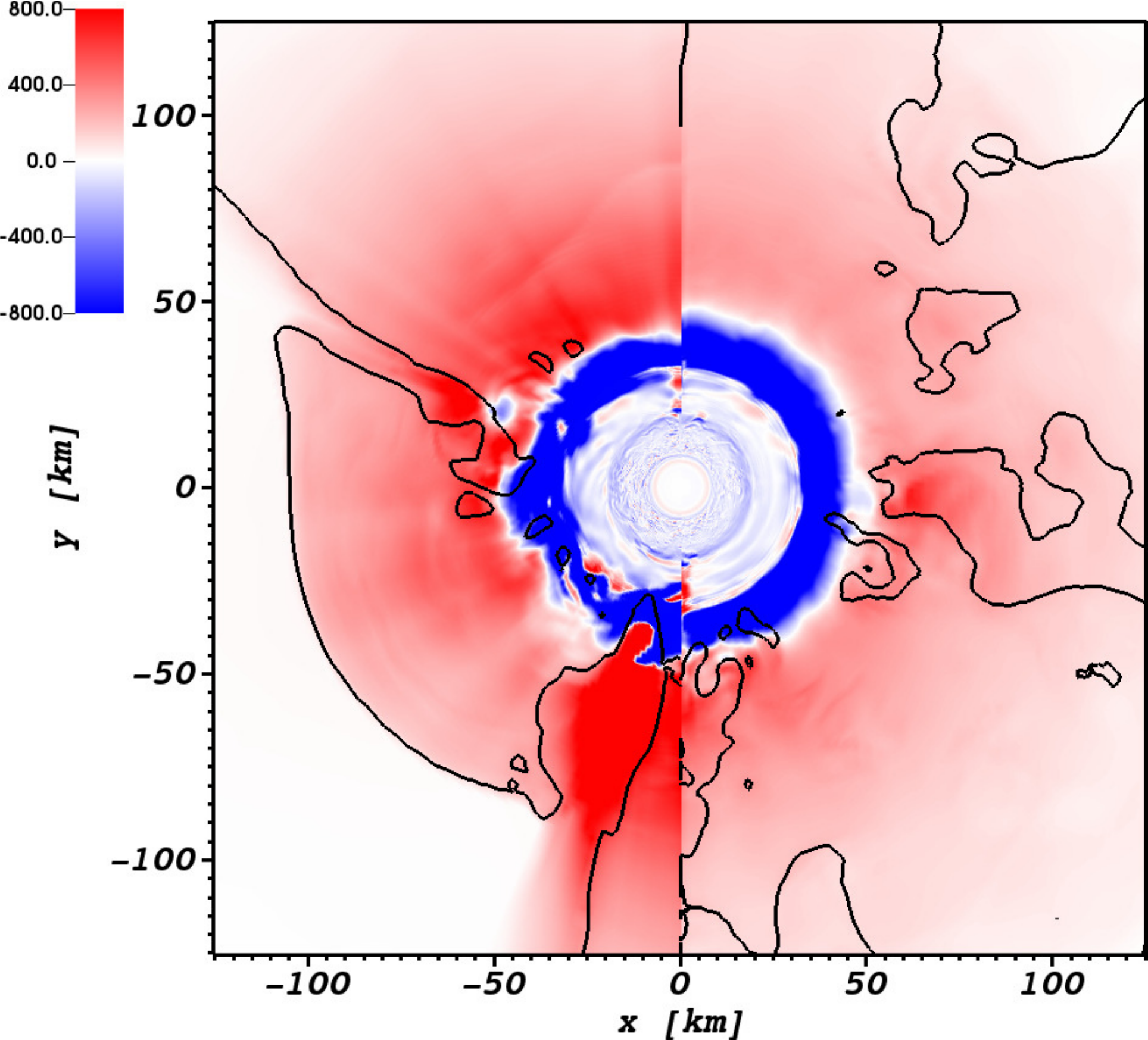}
\caption{Left: Entropy $s$ in $k_b/\mathrm{nucleon}$ in the vicinity
  of the proto-neutron star in 2D (left half of panel) and 3D (right
  half of panel) at a post-bounce time of $400 \ \mathrm{ms}$
  (identical to Figure~\ref{fig:snap400} except for the zoom level).
  Right: Heating/cooling rate in $\mathrm{MeV}/\mathrm{nucleon}$
in 2D (left half of panel) and 3D (right half of panel). Iso-velocity
contours for a radial velocity of $v_r=-10^{9} \ \mathrm{cm} \ \mathrm{s}^{-1}$
are shown to indicate the location of the accretion downflows. Note
that much of the neutrino heating occurs in the downflows and the
confined high-entropy bubble in the equatorial region and hence does not
drive an outflow.
\label{fig:pns_close_up}}
\end{figure*}

In addition to energy loss by wave excitation, which contributes to
the higher binding energy at the gain radius, the growth of the
explosion energy in 2D is further hampered by the fact that much of
the neutrino energy deposition occurs in regions where the heated
matter cannot directly escape in an outflow, i.e.\ either directly in
the accretion funnels or in high-entropy bubbles confined by downflows
and a secondary accretion shock like the equatorial bubble in
Figures~\ref{fig:snap400} and \ref{fig:pns_close_up}, a phenomenon for
which we borrow the term ``steric hindrance'' from chemistry.  In
principle, such bubbles could eventually push the secondary accretion
shock out by undergoing ``secondary shock revival'', but as long as the amount of heating is insufficient, the
bubble cannot break through the surrounding and overlying downflows.

In the snapshot shown in the right panel of \ref{fig:pns_close_up},
the surface fraction covered by the confined bubble and the downflows
exceeds $50 \%$, and the heating rate per unit mass is also largest in
the downflows. Since the fast downflows generally occupy a surface
fraction of $\mathord{\gtrsim} 50\%$ in 2D outside the typical location of a
secondary shock at $\mathord{\lesssim} 100  \mathrm{km}$
(see~Figure~\ref{fig:surfrac}), roughly half of the neutrino heating
is not used to power outflows in 2D, and consequently the outflow
efficiency oscillates around $\eta_\mathrm{out} \sim 0.5$ with some
excursions to higher values during the early explosion phase
(Figure~\ref{fig:eta_out}). By contrast, the neutrino-heated material
can escape unhindered in any direction in 3D apart from some limited
turbulent energy and momentum loss to the downflows, and the resulting
outflow efficiency is of order $\eta_\mathrm{out} \sim 1$.

\subsubsection{Constriction of Outflows and Vertical Mixing}
\label{sec:constriction}
\begin{figure*}
\includegraphics[width=0.48 \linewidth]{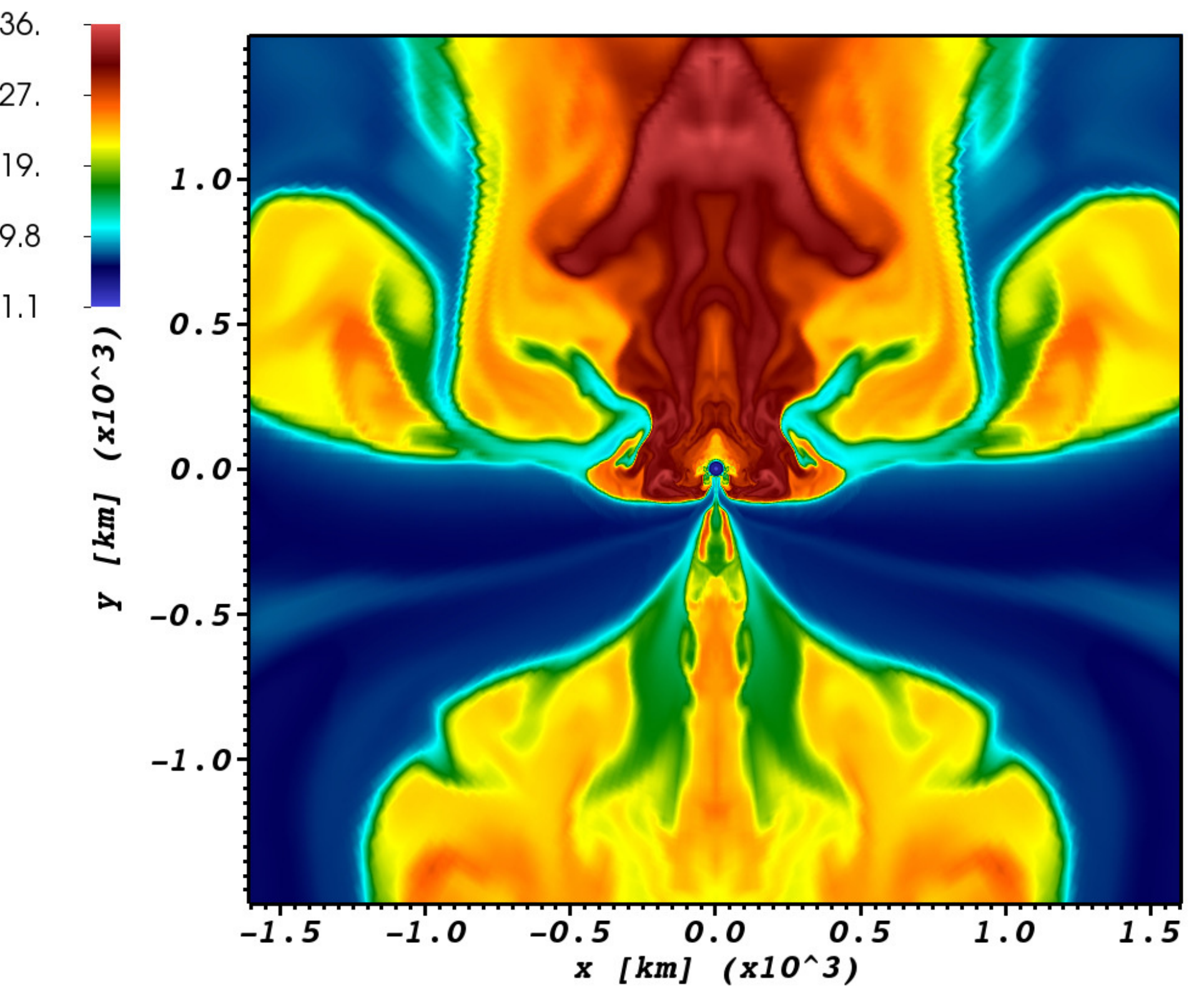}
\hfill
\includegraphics[width=0.48 \linewidth]{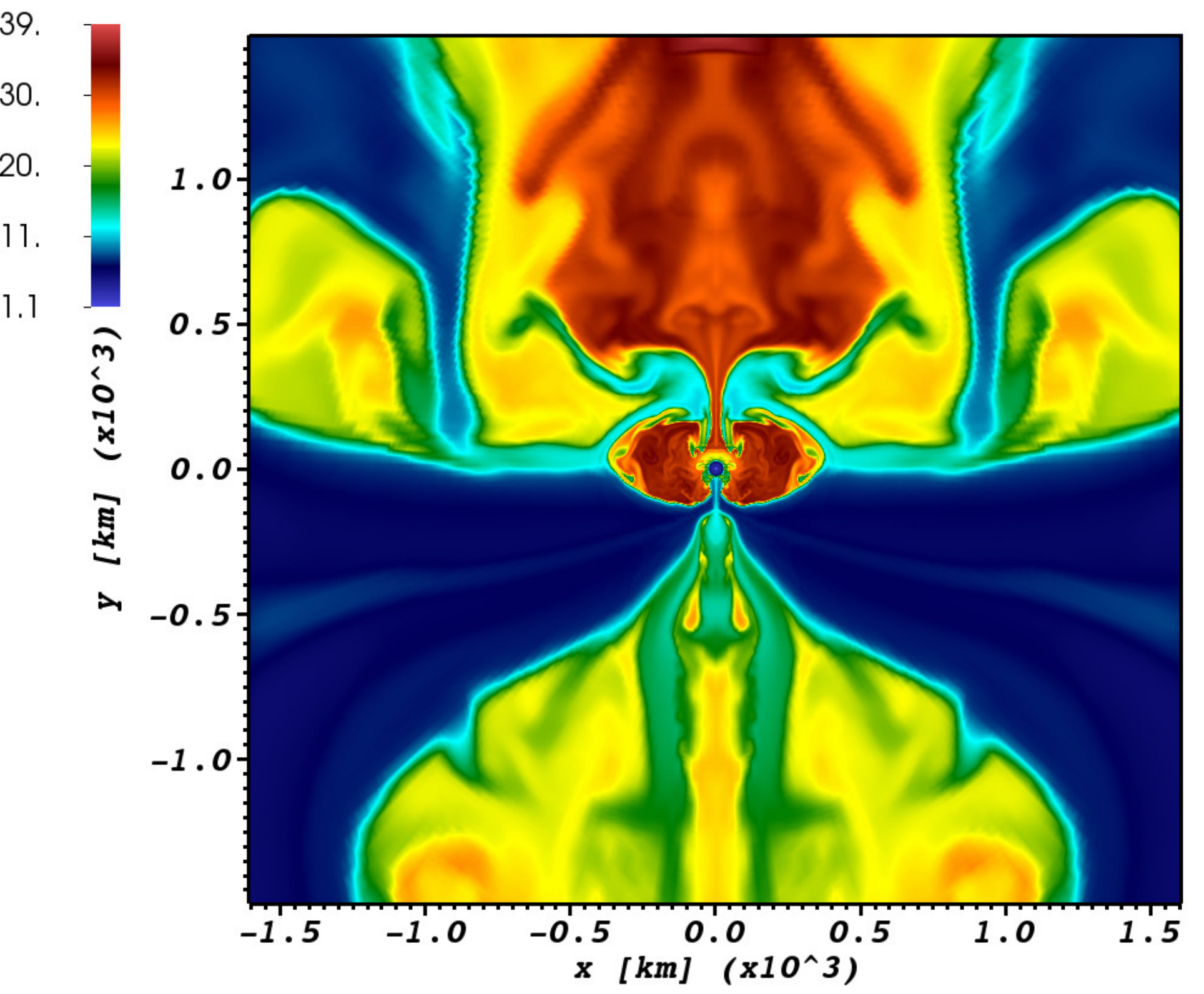}
\\
\includegraphics[width=0.48 \linewidth]{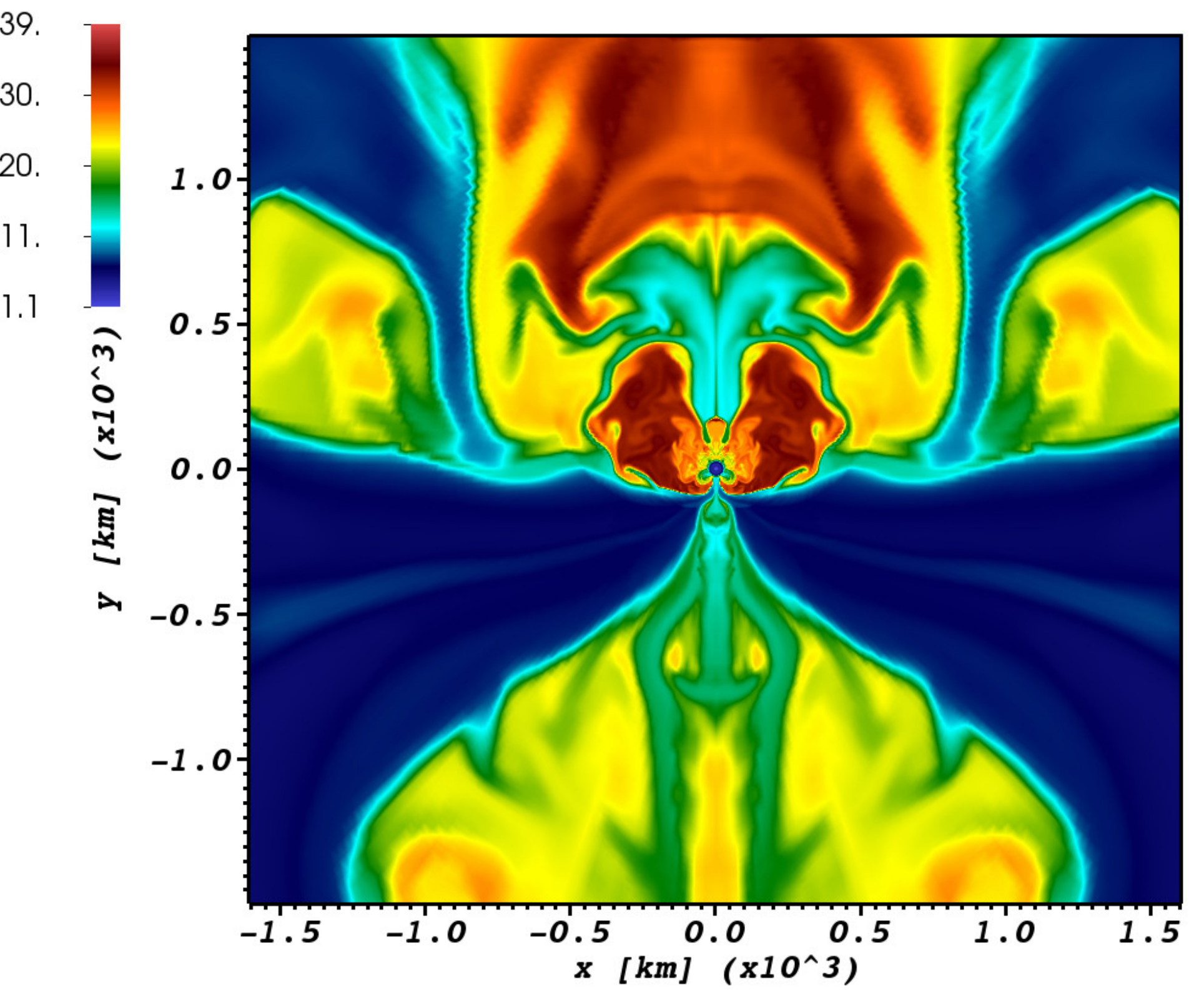}
\hfill
\includegraphics[width=0.48 \linewidth]{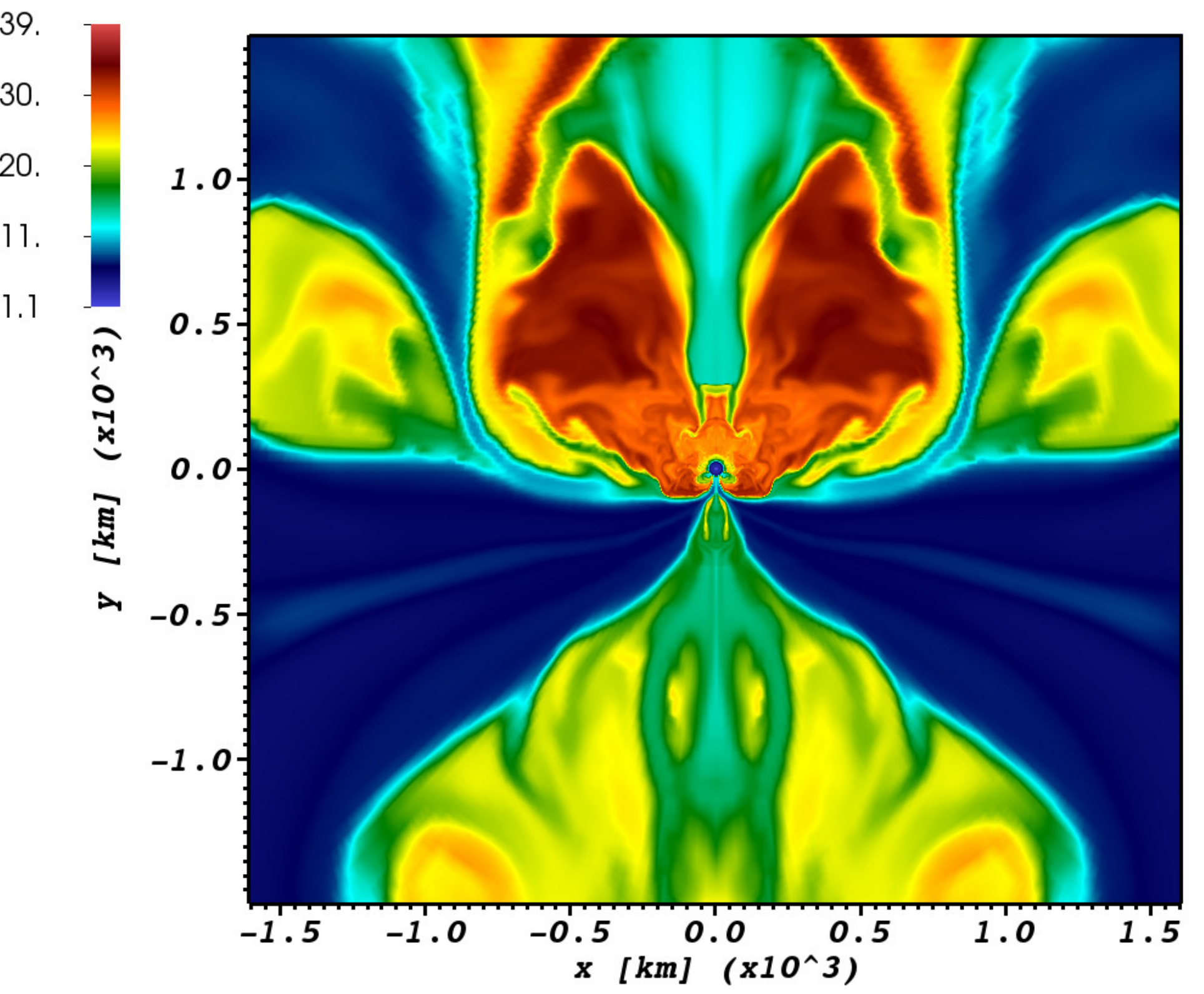}
\caption{Constriction and partial
shredding of an outflow in model s11.2\_2Da, shown by 
snapshots of the entropy at post-bounce times
of $592 \ \mathrm{ms}$,
$611 \ \mathrm{ms}$,
$628 \ \mathrm{ms}$, and
$655 \ \mathrm{ms}$.
A downflow (cyan) originating from a Rayleigh-Taylor plume
of cold matter penetrates the hot-neutrino heated bubble
in the northern hemispheres (top left), constricts
the neutrino-heated bubble to a tenuous outflow
as it approaches the axis (top right),
and eventually a considerable amount of cold material
is mixed into the outflow (bottom left). While the ejection
of matter continues (bottom right), the mixing event
lowers the average total energy per unit mass in the ejecta.
\label{fig:outflow_shredding}
}
\end{figure*}

Finally, the outflows in axisymmetric 2D simulations are less
``stable'' than in 3D in other respects as illustrated in
Figure~\ref{fig:outflow_shredding}: While the Kelvin-Helmholtz
instability between the downflows and outflows is largely suppressed
in 2D, this also implies that plumes of cold material can penetrate
far into the neutrino-heated high-entropy bubbles provided that they
develop in the first place. Because of the symmetry of the system,
these plumes are actually toroidal structures, and can therefore
completely constrict an outflow if they reach the symmetry axis
(Figure~\ref{fig:outflow_shredding}). Similarly, a downflow that
wanders toward the pole can also constrict a polar outflow and cut it
off from fresh supply of neutrino-heated material.

These events typically reduce the surface fraction covered by outflows
at the recombination radius (where they start to contribute to the
diagnostics explosion energy) for a considerable amount of time and
thus reduce the rate of increase of $E_\mathrm{expl}$. Very often the
explosion geometry changes dramatically after such an event and the
surface fraction of the outflows remains small permanently. In some
cases, a high-entropy bubbles is cut off completely from the supply of
neutrino-heated matter from below (Figure~\ref{fig:bubble_separation})
for several seconds. Even if an outflow is eventually reestablished in
the same direction, or if it is strong enough to survive because the
cold plumes reach the axis at a relatively large radius
(Figure~\ref{fig:outflow_shredding}), the ejecta will then typically
contain a large amount of cold material whose total net energy is
barely positive, and the growth of the explosion energy will still be
delayed. Such events explain excursions or even a permanent drop of
the the average total enthalpy $\bar{h}_\mathrm{tot}$ in the outflows
to low values $< 0.3$ in 2D (bottom panel of
Figure~\ref{fig:outflow_rate}).

In 3D, the lack of symmetry as well as the Kelvin-Helmholtz
instability prevent the constriction of outflows by cold plumes. While
the Kelvin-Helmholtz instability provides for some level of energy
and momentum exchange between the accretion funnels and the expanding
high-entropy bubbles as discussed in
Section~\ref{sec:outflows_downflows}, it also prevents cold plumes
from penetrating overly far into the neutrino-heated bubbles.

Mixing between downflows and outflows is thus not completely absent in
2D, it merely takes on a different guise and occurs only episodically,
but with a more catastrophic effect than in 3D. Interestingly, there
even appears to be an effect that compensates somewhat for the
 suppression of the Kelvin-Helmholtz instabilities in 2D due to the
supersonic velocities in the downflows: Rayleigh-Taylor instabilities
between the high-entropy bubbles and the cold overlying post-shock
matter develop more readily in 2D. This is a natural consequence of
higher entropies in the neutrino-heated bubbles in 2D (middle panel of
Figure~\ref{fig:more_profiles}), which implies a higher Atwood
number
between the bubbles and the colder post-shock matter. Thus, the lack of mixing by Kelvin-Helmholtz instabilities in
2D and the entropy boost due the dissipation of acoustic waves can
also have a detrimental side effect on the robustness of the
explosion.

\subsubsection{Absence of a Spherically Symmetric Neutrino-Driven Wind in 2D}

\begin{figure}
\includegraphics[width=\linewidth]{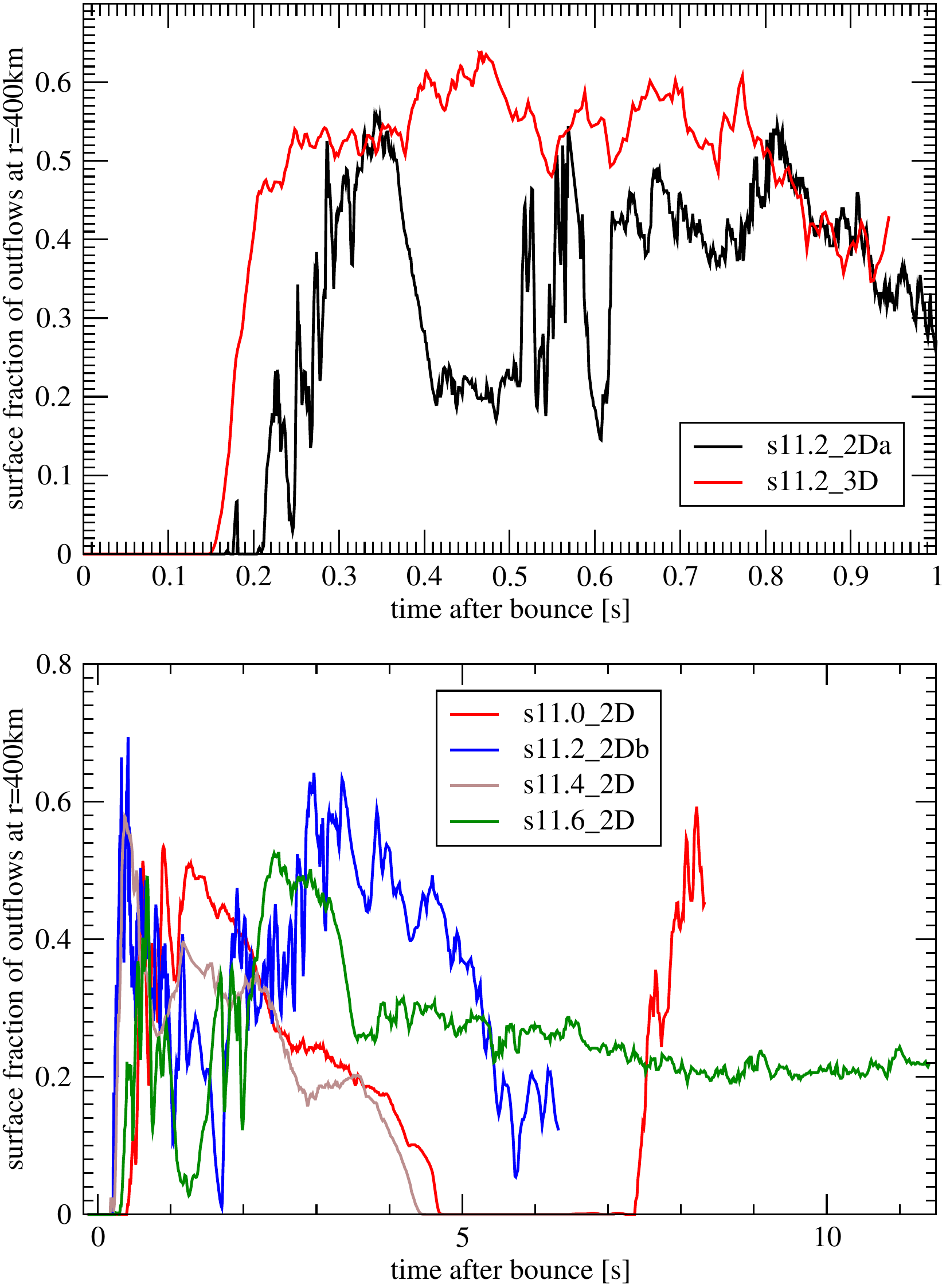}
\caption{Top: Surface fraction occupied by the outflows in models
  s11.2\_2Da (black) and s11.2\_3D (red) at a radius of $400
  \ \mathrm{km}$. The surface fraction is relatively stable with some
  fluctuations around $0.5$ in 3D. In 2D, it reaches similar values
  while the outflows are stable, but occasionally drops to
  significantly smaller values as a result of
  outflow constriction. Bottom: Long-time evolution of the outflow surface fraction for
the 2D models s11.0\_2D, s11.2b\_2D, s11.4\_2D, and s11.6\_2D.
\label{fig:surfrac}}
\end{figure}

\begin{figure*}
\includegraphics[width=0.48 \linewidth]{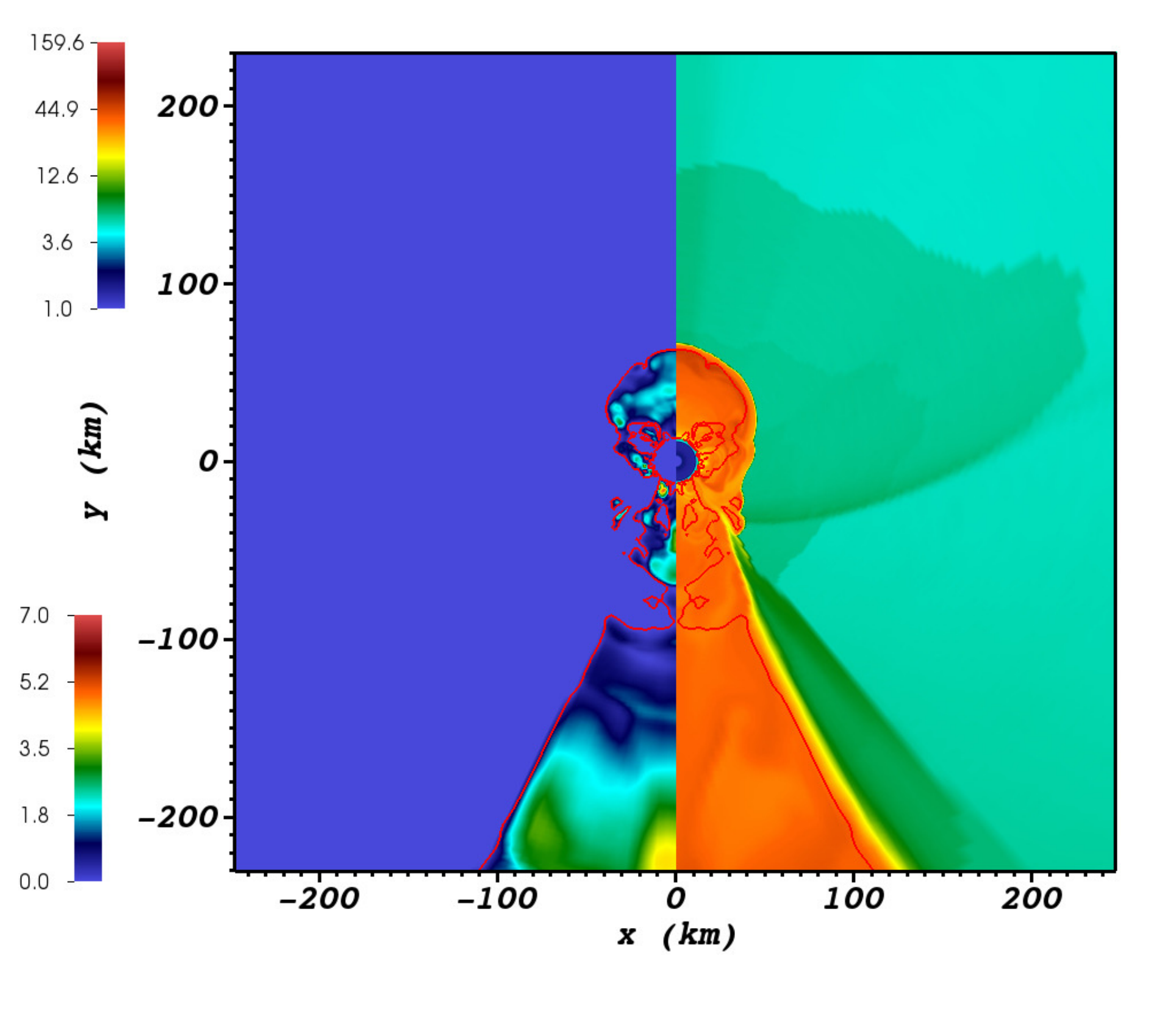}
\hfill
\includegraphics[width=0.48 \linewidth]{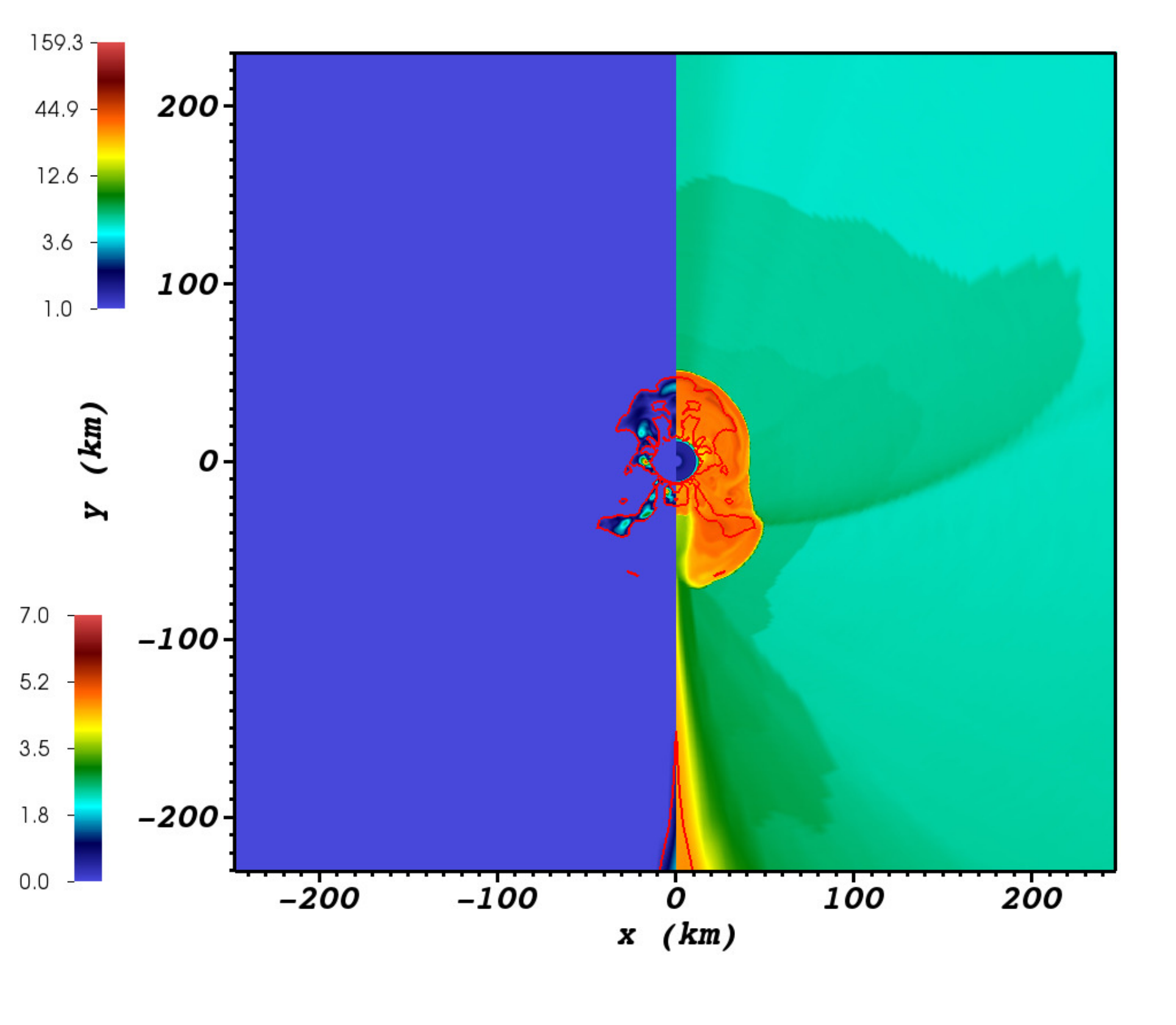}
\\
\includegraphics[width=0.48 \linewidth]{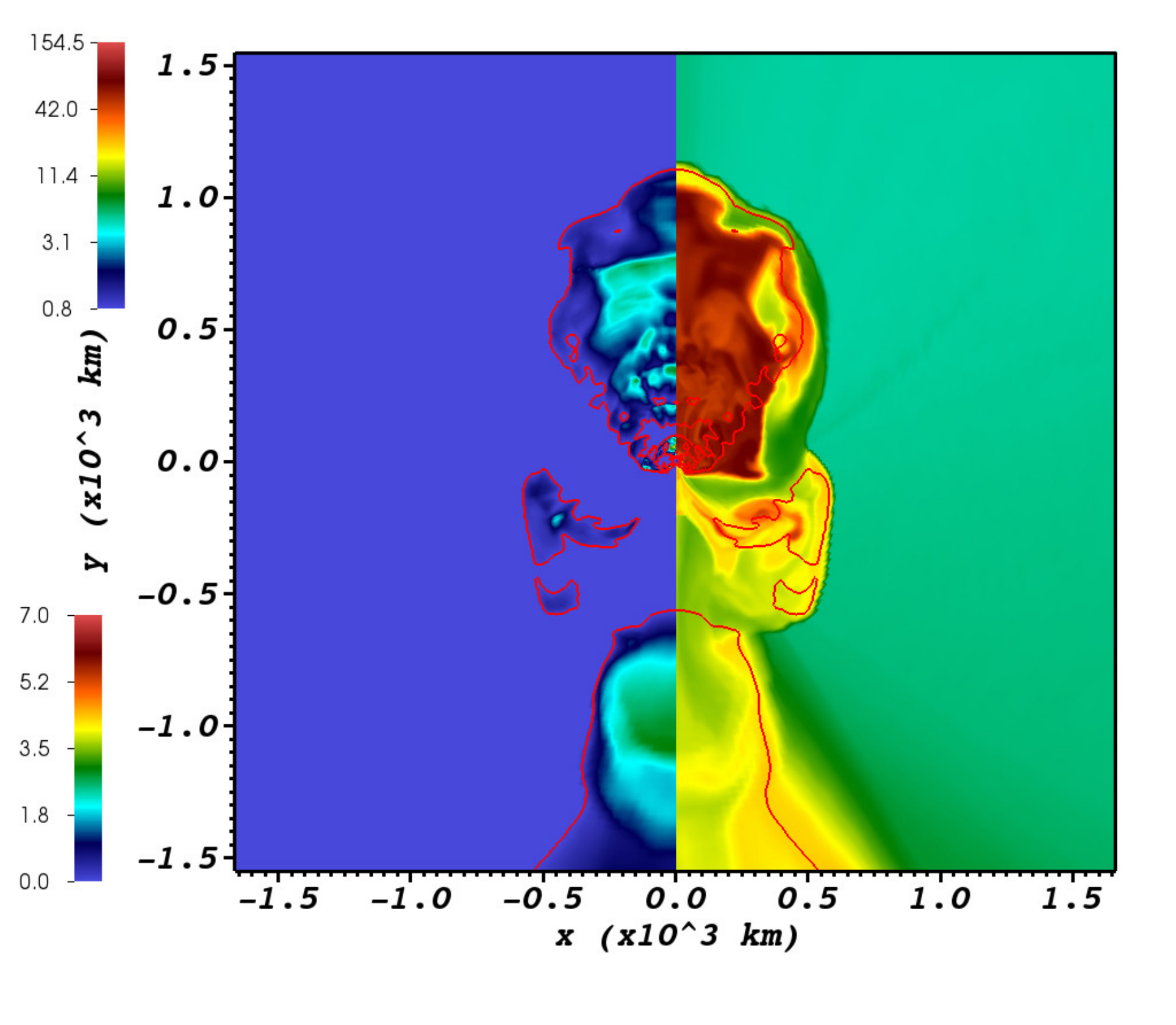}
\hfill
\includegraphics[width=0.48 \linewidth]{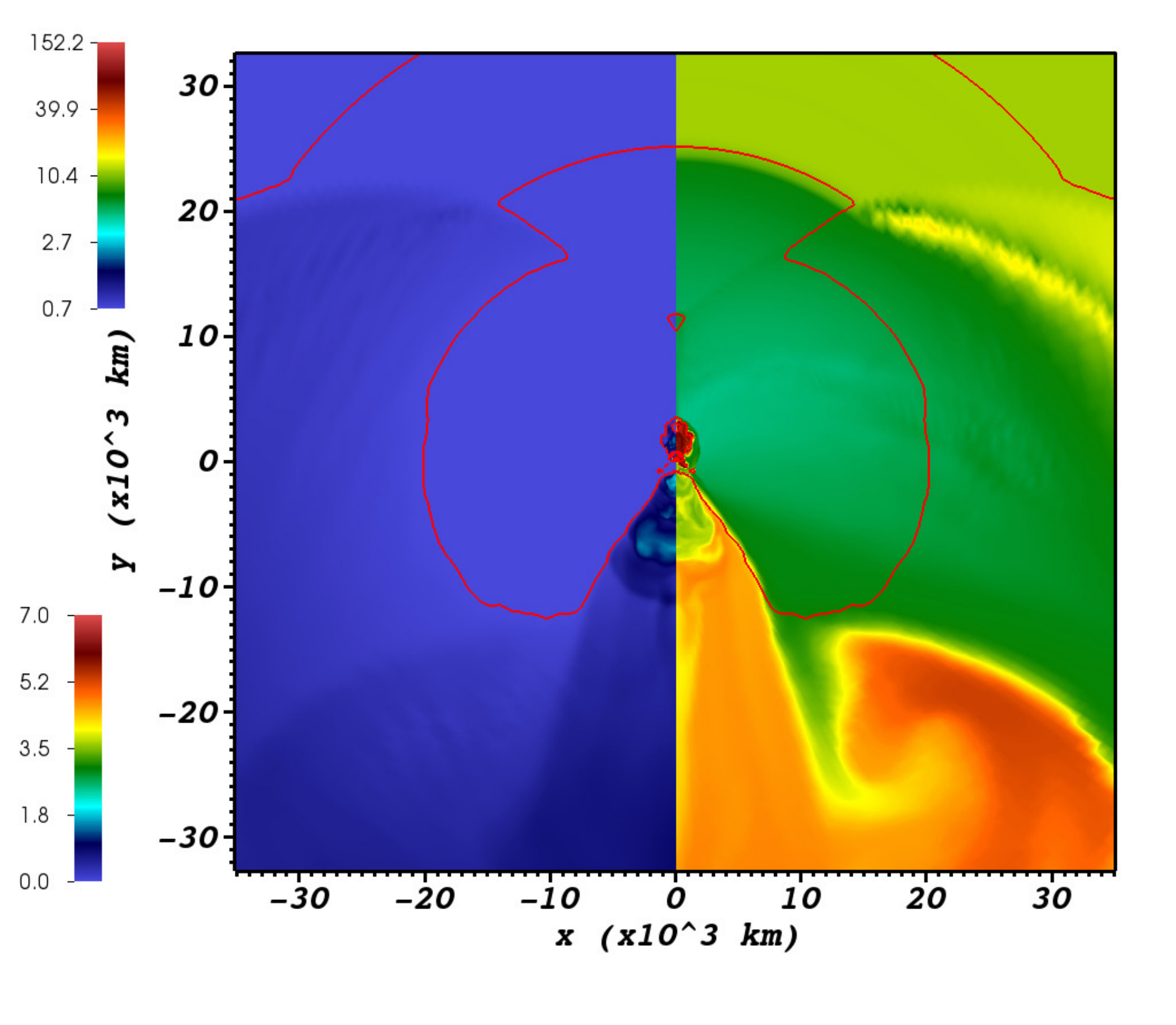}
\caption{ Snapshots of the radial velocity in units of $10^9
  \ \mathrm{cm}$ (left half of panels, lower color bar) and the
  specific entropy $s$ in $k_b/\mathrm{nucleon}$ (right half of
  panels, upper color bar) depicting the separation of the
  high-entropy outflow from the gain region in model s11.6\_2D and the
  re-establishment of an outflow after ``secondary shock revival'' in
  model s11.0\_2D. Red isovelocity contours are used to separate
  outward-moving matter with radial velocities larger than $10^7
  \ \mathrm{cm} \ \mathrm{s}^{-1}$ from infalling matter.  At $4.4
  \ \mathrm{s}$ (top left), matter is still being ejected in the
  southern hemisphere, but at $4.5 \ \mathrm{s}$ (top right) the
  outflow has become extremely thin, and matter in its wake starts
  falling back onto the neutron star.  By
$7.5 \ \mathrm{s}$ (bottom left) a new outflow has developed in
the northern hemisphere, and the expansion of the secondary accretion
shock in the southern hemisphere stops further fallback from
the outflow that was cut off earlier.
By the end of the simulation at $8.2  \ \mathrm{s}$ (bottom right),
the newly formed bubble has expanded further to several
thousands of kilometers in diameter. The post-shock velocities have become
positive in all directions at this point, and only $0.035 M_\odot$ in
the downflows are still falling toward the proto-neutron star.
\label{fig:bubble_separation}
}
\end{figure*}

In the most extreme cases of outflow constriction in 2D, the outflows
are shut off altogether, and the outflow surface fraction drops to
zero permanently, or at least over several seconds
(bottom panel of
Figure~\ref{fig:surfrac} and Figure~\ref{fig:bubble_separation}). This does not imply, however, that the
explosion has failed; it only implies that neutrino heating is not
strong enough to establish a wind that prevents the fallback of slowly
moving matter in the wake of the shock. The pockets of cold, slowly
moving matter from the C/O shell in the 2D models that will undergo
this kind of ``early fallback'' (bottom right panel of
Figure~\ref{fig:bubble_separation}) only contain a few hundredths of a
solar mass by the end of the simulations, and therefore will not
change the neutron star mass considerably. Moreover, the mass
accretion rate onto the secondary accretion shock is so low at late
times that it can start to expand after ``secondary shock revival'',
thus re-establishing an outflow (bottom right panel of
Figure~\ref{fig:bubble_separation}).

While not indicative of a failure of the explosion, the small or
vanishing outflow surface fraction in the long-time simulations nonetheless
indicates (like the models of \citealt{bruenn_14}) that the
separation of outgoing and infalling mass shells in 2D works
differently from the usual picture where a high-entropy
neutrino-driven wind with an approximately spherical flow geometry
eventually develops. The polar outflows can be viewed
as a confined wind driven jointly by neutrino heating and acoustic waves,
but they never cover the entire sphere, and because of their
flow geometry and the strong activity of acoustic waves, the outflow
dynamics is completely different from spherical winds driven 
purely by neutrino heating.

From the foregoing, it is clear that the inhibition of the
neutrino-driven wind in 2D is probably largely artificial; and we only
mention this peculiarity for that reason. As discussed in
Section~\ref{sec:outflows_downflows}, the presence of a larger
effective eddy viscosity in 3D could terminate accretion
earlier than in 2D (where supersonically infalling matter is hardly
decelerated by lateral momentum transfer), or at least decelarate
infalling matter sufficiently to be swept along by an incipient
spherical wind after a few seconds. Moreover, our models likely
underestimate the diffusive neutrino luminosity from the neutron star
core and hence the neutrino heating at late times because we ignore
the effect of nucleon correlations
\citep{burrows_98,burrows_99,reddy_99}, which shorten the
proto-neutron star cooling time-scale considerably
\citep{huedepohl_10}. It is conceivable that the concomitant increase
of the wind mass loss rate could still lead to a volume-filling
outflow for more realistic neutrino opacities after a few seconds even
in 2D.

\subsubsection{Reduced Cooling due to 3D Turbulence?}
Based on a successful supernova simulation of a $9.6 M_\odot$ star,
\citet{melson_15a} recently suggested that the more efficient braking
of the accretion downflows can be responsible for slightly higher
explosion energies in 3D because the less violent impact of the
downflows on the neutron star surface lead to reduced cooling.  In our
comparison of models s11.2\_2Da and s11.2\_3D we observe some of the
same symptoms noted by these authors, i.e.\ turbulent braking of the
downflows and a reduced cooling rate $\dot{Q}_\mathrm{cool}$ at late
times (Figure~\ref{fig:heating_cooling}). Obviously, this raises
the question whether the mechanism proposed by \citet{melson_15a}
also operates in our 3D simulation, and how the physical processes we
discussed so far in Sections~\ref{sec:outflows_downflows}
to \ref{sec:constriction} are related to it.
Unfortunately, a comparison with \citet{melson_15a} is not
straightforward. While they found a sizable increase of the explosion
energy of $10 \%$ in 3D compared to 2D, their explanation involved
relatively tiny differences in some quantities (e.g.\ the gain radius
and the temperature profiles in 2D and 3D) that cannot be confidently
diagnosed in simulations like ours where the 2D and 3D models
start to deviate from each other already shortly after bounce
once prompt convection develops (which was not simulated by
\citealt{melson_15a}). Nonetheless, there is sufficient
evidence that we observe some rather different phenomena
than \citet{melson_15a}.

Essentially, the mechanism proposed by \citet{melson_15a} involves a
\emph{recession} of the gain radius in 3D compared to 2D due to
reduced cooling to eject slightly more material in the explosion. 
Our simulations agree with \citet{melson_15a} in showing a smaller
volume-integrated cooling rate in 3D in the long term as accretion
slowly subsides (Figure~\ref{fig:heating_cooling}). 

However, we do no find a faster recession of the gain radius in 3D
during the explosion phase (middle panel of
Figure~\ref{fig:gain_radius}), and the situation is ambiguous for the
temperature at the gain radius $T_\mathrm{gain}$ (bottom panel of
Figure~\ref{fig:gain_radius}).  In 3D, the temperature
$T_\mathrm{gain}$ stagnates and falls below the 2D value around $250
\ \mathrm{ms}$ at a time when the explosion is already considerably
more vigorous in 3D than in 2D.  We believe that the stagnation of
$T_\mathrm{gain}$ is more indicative of the \emph{slower} recession of
the gain radius rather than of a higher cooling efficiency: The higher
values of the Bernoulli integral and the total energy of the downflows
at the gain radius in 3D (Figure~\ref{fig:more_profiles}
and\ref{fig:profile_fluxes}) imply that there is actually more energy
per unit mass available that can be radiated away in neutrinos as the
accreted matter settles down in the cooling region.

Instead, the faster decline of the accretion rate onto the
proto-neutron star in 3D is the dominant factor behind the lower
cooling rate, making the lower cooling rate a \emph{symptom rather than a
cause} of the more vigorous explosion.  Detailed comparisons would be
required to check whether this true for the $9.6 M_\odot$ model of
\citet{melson_15a} as well.  Since outflow constriction is unlikely to
happen for a model with robust shock expansion, the 2D/3D differences
found by \citet{melson_15a} as well as in the parameterize simulations
of \citet{handy_14} are probably most closely related to the different
outflow efficiency in 2D and 3D, i.e.\ a more efficient ``rerouting''
of freshly accreted matter into outflows. This tallies with their
finding of a smaller surface filling factor of the downflows in 3D,
which implies that a smaller fraction of the neutrino heating is wasted 
in regions where it cannot directly power an outflow. It also
accounts for reduced mass accretion into the gain region and hence a
recession of the gain radius in mass coordinate.  While this mechanism
is similar to the one discussed in Section~\ref{sec:steric} for our
models, the effect is apparently smaller in the simulations of
\citet{melson_15a} because the accretion subsides fast enough to avoid
the formation of secondary shocks and confined high-entropy bubbles in
2D, which can reduce the outflow efficiency by a factor
of $\mathord{\sim}2$ in 2D.

Potentially, wave excitation at the convective boundary could also
contribute to the 2D/3D differences in the simulations of
\citet{melson_15a}. While they take reduced convective overshooting in
3D as an indication for reduced wave excitation, the effect probably
plays a minor role in their case. The relatively small average speeds
of the downflows at the gain radius and
($\mathord{\sim}10^8\ \mathrm{cm}\ \mathrm{s}^{-1}$ compared to
$\mathord{\sim} 10^9 \ \mathrm{cm}\ \mathrm{s}^{-1}$ in our model) are
bound to make the excitation of $g$-modes rather inefficient and thus
rule them out as a major energy drain on the gain region in 2D.  This
is also suggested by the fact they find similar internal energies (and
hence binding energies) in the outer regions of the cooling layer in
2D and 3D despite the stronger recession of the gain radius in 3D,
which is quite different from what we discussed in
Sections~\ref{sec:outflows_downflows} and \ref{sec:waves}.

\begin{figure*}
\includegraphics[width=0.48 \linewidth]{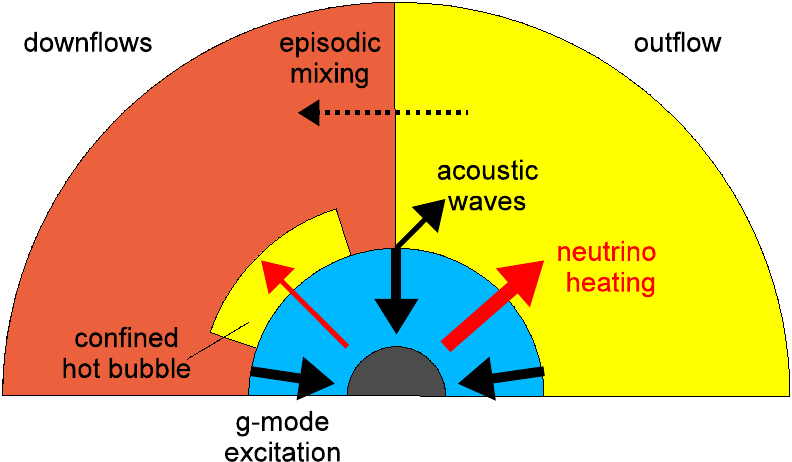}
\hfill
\includegraphics[width=0.48 \linewidth]{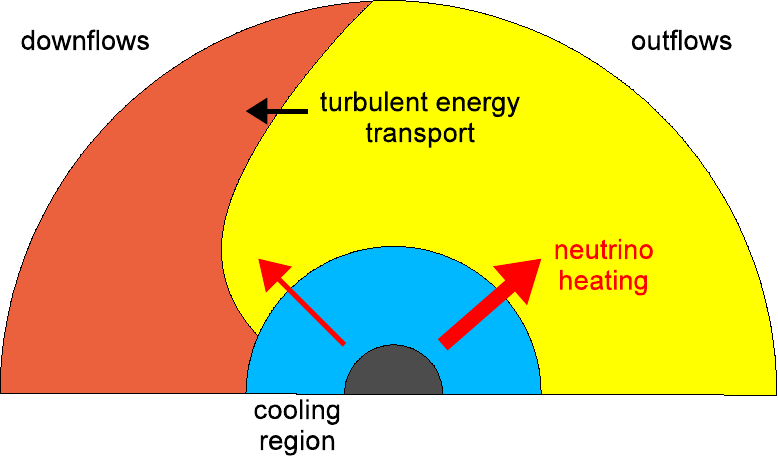}
\caption{Sketch of  the different energy budget between the outflows (yellow),
the downflows (red) and the cooling region (blue) in 2D (left) and 3D
(right). In 3D, turbulent eddy diffusivity leads to a persistent,
small energy flux (short solid arrow) from the neutrino-heated outflows to the downflows,
whereas mixing between the outflows and downflows only occurs
episodically in 2D, but has more dramatic consequences
in this case because it leads to large-scale mixing
of cold matter into the outflows (long dotted arrow). There is
a considerable energy transfer from the gain region to the cooling
region due to wave excitation in 2D and an indirect
transfer of energy from the downflows to the gain region by the
excitation of acoustic waves in 2D (which can lead to larger asymptotic
energies per unit mass in the ejecta); this is absent in 3D. Moreover, 
a considerable amount of neutrino heating (red arrows) is wasted in
2D because it is deposited in confined bubbles, whereas almost
the entire neutrino heating in 3D is used to lift matter out of the
gravitational well.
\label{fig:sketch}
}
\end{figure*}

\section{Summary and Conclusions}
\label{sec:conclusions}

We have presented a successful 3D GR simulation of the explosion of an
$11.2 M_\odot$ star using the \textsc{FMT} multi-group transport
scheme of \citet{mueller_15}. The model has been evolved to almost $1
\ \mathrm{s}$ after bounce, and has reached a diagnostic explosion
energy of $1.3 \times 10^{50} \ \mathrm{erg}$ at that point, which is
still growing by the end of the simulation. The baryonic neutron star
mass $M_\mathrm{by}$ at the end of the simulations has reached $1.33
M_\odot$ and estimates of the final neutron star mass yield
$M_\mathrm{by}\approx 1.41 \ldots 1.48 M_\odot$ and a gravitational
mass not exceeding $1.34 M_\odot$, which is compatible with the
measured neutron star mass distribution \citep{schwab_10}.  The fact
that we obtain an explosion for this progenitor with a relatively
accurate multi-group transport scheme further illustrates that even
the non-exploding state-of-the-art models
\citep{hanke_13,tamborra_14a,tamborra_14b} with the best available
neutrino transport and microphysics are apparently very close to shock
revival, something which is also suggested by the recent successful 3D
explosion models of the Garching \citep{melson_15a,melson_15b} and
Oakridge groups \citep{lentz_15}.

A comparison of the explosion dynamics after shock revival with 2D
long-time simulations of different progenitors with ZAMS masses
between $11.0 M_\odot$ and $11.6 M_\odot$ revealed a faster and more
stable growth of the explosion energy in 3D compared to 2D.  Because
accretion downflows and neutrino-driven outflows coexist over several
seconds in 2D, the explosion energy in the 2D models can still reach
values of up to $2\times 10^{50} \ \mathrm{erg}$, but this comes at
the expense of high neutron star masses ($M_\mathrm{by} \gtrsim 1.6
M_\odot$) that are likely incompatible with the observed neutron star
mass distribution. A detailed comparison of the 2D and 3D models
unearthed several physical mechanisms responsible for the more robust
rise of the explosion energy in 3D and the slower growth of the
proto-neutron star mass, which is a symptom of the faster subsidence
of accretion. The specific effects that we find are summarized below,
and we also provide a schematic visualization of the different flow geometry
and the energy budget between the downflows, the outflows and the gain
region in Figure~\ref{fig:sketch} to aid the reader's understanding:
\begin{enumerate}
\renewcommand{\theenumi}{\arabic{enumi}.}
\item In 2D, the interfaces between the accretion funnels and the
  neutrino-heated bubbles tend to become laminar after shock revival,
  while they are corrugated by the Kelvin-Helmholtz instability in
  3D. We ascribe the different behavior to the suppression of the
  purely two-dimensional modes of the Kelvin-Helmholtz instability in
  the supersonic regime \citep{gerwin_68}. The effect is thus distinct
  from the inverse turbulent energy cascade in 2D
  \citep{kraichnan_76}, which has been invoked as an explanation for
  the different behaviour of 2D and 3D models prior to shock revival,
  since the different energy cascade in 2D and 3D is not related to
  the Mach number of the flow.
\item As a consequence, the effective eddy viscosity and diffusivity
  between the downflows and outflows is larger in 3D than in 2D during
  most phases, i.e.\ there is more exchange of energy and momentum
  between the outflows and downflows. On the one hand, this implies
  that the outflows contribute only $\mathord{\sim} 6
  \ \mathrm{MeV}/\mathrm{nucleon}$ to the explosion energy in 3D, as
  some of the net total (i.e\ thermal+kinetic+potential) energy gained
  from nucleon recombination of $\mathord{\sim} 8.8
  \ \mathrm{MeV}/\mathrm{nucleon}$ is lost to the downflows by
  turbulent diffusion. On the other hand, the turbulence effectively
  ``brakes'' the downflows, and they arrive at the gain radius with
  smaller velocities but higher total energy per unit mass than in 2D.
\item The higher impact velocities of the downflows and the formation
  of secondary accretion shocks at small radii in 2D lead to a more
  efficient excitation of $g$-modes and acoustic waves at the gain
  radius that transport energy into deeper regions of the cooling layer
  and into the neutrino-heated ejecta respectively. Our analysis
  suggests that the energy loss from the gain region by wave
  excitation becomes comparable to the volume-integrated neutrino
  heating rate at late times, and by increasing the absolute
value of the binding energy
  $|e_\mathrm{tot}|$ at the gain radius significantly reduce the mass
  outflow rate that can be sustained by neutrino heating. In 3D, the
  turbulent energy flux into the gain region is small, and the binding
  energy at the gain radius is smaller by factor of $\gtrsim 2$ at
  late times, which allows for a higher mass outflow rate than in 2D.
  The dissipation of acoustic waves in the outflows in 2D provides
  only for a partial ``recycling'' of the energy lost by wave
  excitation, but can increase the total energy per unit mass in the
  outflows to values larger than the recombination energy of $8.8
  \ \mathrm{MeV}/\mathrm{nucleon}$.
\item In addition, the outflow efficiency $\eta_\mathrm{out}=
  \dot{M}_\mathrm{out}/(\dot{Q}_\mathrm{heat}/|e_\mathrm{gain}|)$, is
  also higher in 3D ($\eta_\mathrm{out}\sim 1$ with strong
  fluctuations) than in 2D ($\eta_\mathrm{out}\lesssim 0.5$ at late
  times), i.e.\ for a given amount of neutrino heating and a given
  binding energy at the gain radius, more mass is channelled into
  outflows and contributes to the explosion energy in 3D. The low outflow
  efficiency in 2D stems from the large surface fraction occupied by fast
  downflows and ``confined bubbles'' between downflows whose expansion
  is inhibited by the formation of secondary accretion shocks. 
\item Episodic mixing between the outflows and downflows still occurs
  in 2D, e.g.\ by the formation of new downflows as a result of the
  Rayleigh-Taylor between the cold shocked material and the
  neutrino-heated high entropy bubbles. While mixing only occurs
  sporadically in 2D, the consequences of these mixing events are more
  catastrophic than in 3D.  Not only do they slow down the rise of the
  explosion energy by mixing cold material into the outflows; the
  penetration of accretion funnels into the high-entropy bubbles can
  also lead to the constriction of outflows, sometimes shutting them
  off completely and permanently decreasing the outflow surface
  fraction to values of $\lesssim 0.3$.
\end{enumerate}  
  
Our simulations thus provide ample evidence that 3D effects can play a
beneficial role in core-collapse supernova explosions \emph{after}
shock revival. However, since our current 3D model has only been
evolved to $\mathord{\sim} 1 \ \mathrm{s}$ after bounce and does not
yet permit us to deduce the final explosion and remnant properties
directly because of continuing accretion (and forced us to resort to
indirect arguments about the final neutron star masses), a number of
open questions remain and invite speculation. Moreover, limited
conclusions can be drawn from a single 3D simulation of one
progenitor. Given the recent progress on other fronts
in supernova theory, the questions and perspectives for
future research on the role of 3D effects during the explosion phase
can be summarized as follows:
\begin{enumerate}
\renewcommand{\theenumi}{\arabic{enumi}.}
\item Longer 3D simulations with higher resolution are necessary
  to determine final explosion energies, Nickel masses and neutron
  star masses precisely for comparison with observations without
  recourse to indirect methods. Our estimate for the final baryonic
  neutron star mass of $1.48 M_\odot$ for the 3D model of the $11.2
  M_\odot$ progenitor still assumes the accretion of an additional
  $0.15 M_\odot$ solar masses, which is much more than a visual
  inspection of Figure~\ref{fig:pns_masses} suggests given the very
  slow rise of the neutron star mass in 3D. If the turbulent braking
  of the downflows terminates accretion before the post-shock velocity
  equals the escape velocity as speculated in
  Section~\ref{sec:outflows_downflows}, the final baryonic and
  gravitational neutron star mass might be as low as $\mathord{\sim}
  1.35 M_\odot$ and $\mathord{\sim} 1.24 M_\odot$, respectively. This
  would indicate that a plausible distribution of neutron star masses
  spanning the entire range of observed values down to the lower end
  is within reach of modern multi-D simulations of neutrino-driven
  supernovae.
\item A more rigorous analysis of the turbulent multi-dimensional flow
  in the spirit of Reynolds decomposition would be highly desirable in
  order to further bolster our qualitative interpretation of 3D
  effects in the post-explosion phase. Such quantitative analysis
  methods have considerably advanced our understanding of the
  turbulent flow during the accretion phase \citep{murphy_11}. After
  shock revival, the nonstationarity of the flow presents a challenge
  for such methods, however. Our relatively crude two-stream analysis
  based on a separation of the outflows and downflows could also be
  improved in order to account more directly and rigorously for the
  turbulent exchange of mass, momentum and energy between the two
  streams, but such an analysis faces a major challenge in the form of
  the complicated flow geometry. 
\item Whether and to what extent the positive 3D effects described in
  this paper come into play obviously depends on whether shock revival
  can be accomplished in 3D in the first place and on the delay
  compared to the 2D case. If there is a significant delay in shock
  revival, 3D models may not be able to equalize the ``head start'' of
  the 2D models at least of relatively powerful explosions where the
  diagnostic energy shows first signs of levelling off after
  $\mathord{\sim} 300 \ \mathrm{ms}$ or less \citep{bruenn_14,pan_15}.
  Even in this case, the mechanism discussed in this paper could
  nonetheless help to mitigate the ``penalty'' incurred by the delay
  of the explosion in 3D and allow the models to remain compatible
  with observational constraints. Moreover, if accretion lasts longer ---
  as in the 2D simulations of
  \citet{mueller_12a,mueller_12b,janka_12b} --- the beneficial 3D effects
  in the phase after shock revival may outweigh the ``penalty'' of
  delayed shock revival. Furthermore, it is conceivable that the
  problem of missing or delayed explosion in 3D may yet be resolved by
  the inclusion of better, multi-dimensional progenitor models with
  large-scale initial perturbations that aid shock revival
  \citep{couch_13,mueller_15,couch_15}, unknown microphysics
  \citep{melson_15b}; and strongly SASI-dominated models may even
  explode easier in 3D \citep{fernandez_15}.
\item The robustness of the mechanisms described in this paper needs
  to be studied further for a wider range of progenitors.  The 2D
  simulations \citep{buras_06b,mueller_12a} of the $11.2 M_\odot$
  progenitor considered here have been particularly noteworthy
  examples for suspiciously low explosion energies and long-lasting
  accretion. This behavior is due to the specific
    characteristics of progenitors around $11 M_\odot$, including a
    relatively small silicon core and a very pronounced density jump
    at the Si/SiO interface. These properties result in
a small proto-neutron star mass $M$ immediately after shock revival
and hence low neutrino energies (cp.~the scaling
of the electron antineutrino energy with $M$ found by \citealp{mueller_14})
as well as a small accretion luminosity. Both of these factors
contribute to relatively weak neutrino heating after shock revival
and a small mass outflow rate. The tepid nature of our
2D explosions may thus hinge very much on the peculiar structure
of low-mass supernova progenitors.

It therefore remains to be seen whether 3D effects provide a
  similarly strong boost for the growth of the explosion energy in
  other progenitors. Whenever 2D models develop the characteristic
  broad downflows and secondary shocks indicative of long-lasting
  accretion during a relatively weak explosions, such as the $15
  M_\odot$ and $27 M_\odot$ models of \citet{mueller_12a} and
  \citet{janka_12b}, the physical mechanisms identified here likely
  come into play eventually. On the other hand, they may play a
  negligible role if the volume fraction of the downflows drops very
  quickly as in the models of \citet{bruenn_14} and \citet{lentz_15}
  or the parameterized models of \citet{handy_14}. Since 2D supernova
  simulations of different groups have not yet converged sufficiently
  to decide whether there is a generic problem of ``weak explosions''
  in 2D, it is impossible to judge the generic character of our
  findings. By the same token, however, it cannot be ruled out that 2D
  models \emph{should be} generically underenergetic and overestimate
  the amount of accretion after shock revival due to the mechanisms we
  identified, and that realistic explosion enegies and remnant masses
  will only be obtained in 3D. If so, prematurely confronting the 2D
  models with the observational constraints could lead to wrong
  conclusions.
\end{enumerate}
Thus, more work is necessary to substantiate the intriguing
perspective that 3D effects could help to achieve agreement between
simulations of neutrino-driven supernovae and observational
constraints such as explosion energies and neutron star masses.  Far
from offering a complete solution due to the limitations of
computational resources that have always plagued supernova theory, our
present study can only take a first step in this direction and
adumbrate some of the physical mechanisms that could help to boost 3D
explosions after shock revival. Nonetheless, even our current results
already serve an antidote against undue pessimism after initial
setbacks in 3D multi-group neutrino hydrodynamics simulations. Along
with the recent successful explosions in first-principle models, the
identification of other beneficial effects of the third dimension on
the explosion threshold and energetics, and plausible ideas for
solving the problem of missing explosions with the help of multi-D
progenitor models and/or non-standard microphysics, they are another
piece that fits well into the overall puzzle, suggesting that a
solution for the supernova problem is slowly taking shape.

\section*{Acknowledgements}

We acknowledge fruitful exchange with H.~Andressen, A.~Burrows, Th.~Foglizzo,
A.~Heger, W.R.~Hix, H.-Th.~Janka, Y.~Levin, T.~Plewa, and T.~Waters.
The author has been supported by the Australian Research Council
through a Discovery Early Career Researcher Award (grant DE150101145)
and by the Alexander von Humboldt Foundation through a Feodor Lynen
fellowship.  The computations were performed on \emph{Raijin} at the
NCI National Facility (project fh6) using computer time contingents
obtained through NCMAS, ASTAC, and a Monash LIEF top-up grant, on the
Monash eGrid Cluster, and on the IBM iDataPlex system \emph{hydra} at
the Rechenzentrum of the Max-Planck Society (RZG).

\appendix

\section{Definition of Energies and Energy Fluxes in General Relativity}
\label{app:gr_quantities}
While the computation of mass fluxes and spherically-averaged profiles
of hydrodynamic quantities can be readily generalized  to
the relativistic case just by including the correct three-volume
element, the definition of energies and energy fluxes in the relativistic
case is less straightforward, and is therefore briefly expounded in this
Appendix. We use geometric units ($G=c=1$) throughout this section.

After adopting a $3+1$ foliation of space-time and projecting
the components of the stress-energy tensor into components
orthogonal and parallel to the 3-hypersurfaces, the energy
equation in general relativistic hydrodynamics can be written
in the formulation of \citet{banyuls_97} in terms of a new conserved variable $\tau$ as
\begin{equation}
  \label{eq:hydro_3}
  \frac{\partial \sqrt{\gamma} \tau}{\partial t}+
  \frac{\partial \sqrt{-g} \left(\tau \hat{v}^i + P v^i\right)}{\partial x^i}=
  \alpha \sqrt{-g} \left(T^{\mu 0}\frac{\partial \ln \alpha}{\partial x^\mu} -T^{\mu\nu}\Gamma^0_{\mu\nu}\right) 
.
\end{equation}
Here, $\tau$ is defined in terms of the baryonic mass density $\rho$ in
the fluid frame, the Lorentz
factor $W$, the internal energy density $\epsilon$ (including
all rest-mass contributions), the pressure $P$, and the
relativistic enthalpy $h=1+\epsilon+P/\rho$ as
\begin{equation}
\tau = \rho h W^2-P-\rho W=\rho(1+\epsilon+P/\rho) W^2-P-\rho W.
\end{equation}
Furthermore, $\gamma$ and $g$ are the determinants of
the three- and four-metric, respectively, and the advection
term contains $\hat{v}^i=v^i-\beta^i / \alpha$ instead of the
Eulerian three-velocity $v^i$. By pushing the
lapse function $\alpha$ into the temporal and spatial derivatives,
it is possible to formulate a strict conservation law
(without a source term) analogous to 
in the limit of a stationary space-time
with a zero shift vector (cp.\ Equation~A35 \citealt{mueller_10}, where
the right-hand side reduces to zero in this limit):
\begin{equation}
\label{eq:rel_ene}
  \frac{\partial}{\partial t}\left[ \sqrt{\gamma} \alpha \left(\tau+D\right)
    -\sqrt{\gamma} D \right]+ \frac{\partial}{\partial x^i} \left[ \sqrt{-g}
    \left(\alpha \tau \hat{v}^i + \alpha D \hat{v}^i - D \hat{v}^i +
    \alpha P v^i\right)\right] = 0.
\end{equation}
Here, we have introduced the baryonic mass density in the Eulerian
frame $D=\rho W$ to simplify the equation.

This suggests that in the limit of a vanishing shift vector and a
stationary metric, the Newtonian expression
$e_\mathrm{tot}=\epsilon+\mathbf{v}^2/2+\Phi$ for the total
energy per unit mass (including rest-mass contributions) can be generalized to
\begin{equation}
e_\mathrm{tot,rm}=\frac{\alpha \tau}{D}+(\alpha-1),
\end{equation}
and the role of the total Newtonian enthalpy in the flux
is taken by 
\begin{equation}
h_\mathrm{tot,rm} =
    \alpha \tau/D + (\alpha-1)+\alpha P/D.
\end{equation}
It is noteworthy that the internal energy and rest-mass contributions
(which enter the equations through $\tau$) always appears in
conjunction with factors $W$ and $\alpha$. Strictly speaking, it is
therefore no longer possible to formulate the energy equation in
general relativity without including rest-mass contributions in the
conserved quantities and the fluxes by pushing them into a nuclear source
term instead (as least not in a simple form). Computing
fluxes and total energies excluding rest-mass contributions is
therefore somewhat less meaningful in general relativity. However,
since we have $\alpha \approx 1$ and $v \lesssim 0.3 c$ in the gain
region, the higher-order relativistic corrections are small
enough to be neglected, it is still reasonable to compute total energies
and enthalpies without rest-mass contributions by using
just the thermal energy contribution 
$\epsilon_\mathrm{therm}$ to the internal energy density $\epsilon$
instead of $\epsilon=\epsilon_\mathrm{therm}+\epsilon_\mathrm{rm}$.

For the other quantities, considered in this paper, the generalization
is trivial. Mass fluxes through the surface of a sphere or parts of it
are computed as
\begin{equation}
\dot{M}=\int \alpha D v_r r^2\phi^4 \,\ud \Omega,
\end{equation}
where $v_r$ is the radial velocity component measured in the
orthonormalised Eulerian frame and $\phi$ is the conformal factor in
the xCFC metric; and the computation of energy/enthalpy fluxes works
analogously. The density-weighted spherical average $\bar{X}$ of a
quantity$X$ is computed as
\begin{equation}
\bar{X}=\frac{\int D X \phi^4 \, \ud \Omega}{\int D \phi^4 \, \ud \Omega}.
\end{equation}

\bibliography{paper}

\end{document}